\documentclass[onecollarge]{svjour2}       
\smartqed  
\usepackage{graphicx}
\usepackage{amsmath}

\journalname{Journal of Statistical Physics}

\begin{document}

\title{{Region graph partition function expansion and approximate
free energy landscapes: Theory and some numerical results}
\thanks{Research partially supported by Chinese Academy of Sciences
(grant number KJCX2-EW-J02)
 and by the National Science Foundation of China
(grant numbers 10834014 and 11121403).}}

\titlerunning{Region-graph partition function expansion}

\author{Haijun Zhou \and Chuang Wang}

\institute{H. Zhou \and C. Wang \at
              State Key Laboratory of Theoretical Physics,
              Institute of Theoretical Physics, \\
		 Chinese Academy of Sciences,
              Zhong-Guan-Cun East Rd. 55, Beijing 100190, China \\
              \email{zhouhj@itp.ac.cn}
}

\date{Received: date / Accepted: date}

\maketitle

\begin{abstract}
Graphical models for finite-dimensional spin glasses and
real-world combinatorial optimization and satisfaction problems usually have
an abundant number of short loops.
The cluster variation method and its extension, the region graph method,
are theoretical approaches for treating the
complicated short-loop-induced local correlations.
For graphical models represented by non-redundant or redundant region graphs,
approximate free energy landscapes are constructed in this paper
through the mathematical framework of region graph partition function
expansion.
Several free energy functionals are obtained, each of which use a set of
probability distribution functions or functionals as order parameters.
These probability distribution function/functionals are required to satisfy the
region graph belief-propagation equation or the region graph survey-propagation
equation to ensure vanishing correction contributions of region subgraphs
with dangling edges.
As a simple application of the general theory,  we perform region graph
belief-propagation simulations on the square-lattice ferromagnetic Ising model
and the Edwards-Anderson model. Considerable improvements over the
conventional Bethe-Peierls approximation are achieved.
Collective domains of different sizes in
the disordered and frustrated square lattice are identified by the
message-passing procedure. Such collective domains and
the frustrations among them are responsible for the low-temperature
glass-like  dynamical behaviors of the system.
\end{abstract}

\section{Introduction}

Spin glass is a paradigm for basic research on physics of collective
behaviors induced by disorder and/or frustration \cite{Anderson-1990}.
It is also a tool box for applied research on hard problems originated
from fields outside the conventional domain of physics.
The mean-field theory of spin glasses was originally developed through the
replica trick and by assuming the property of replica-symmetry-breaking (RSB)
\cite{Edwards-Anderson-1975,Sherrington-Kirkpatrick-1975,Parisi-1979}.
It was later re-derived using the cavity method
\cite{Mezard-etal-1986,Mezard-etal-1987}
without introducing replicas. Another advantage of the cavity method is
its applicability to single problem instances with fixed disorder parameters.
The cavity method was successfully extended to spin glass models defined
on finite-connectivity random graphs \cite{Mezard-Parisi-2001}, confirming and
extending the earlier theoretical results obtained by the replica method
\cite{Viana-Bray-1985,Monasson-1995,Monasson-1998}. These theoretical advances
paved the way for the fruitful applications of spin glass theory to
interdisciplinary problems in computer science, information theory, and
biological sciences \cite{Hartmann-Weigt-2005,Mezard-Montanari-2009}; for
example, constraint satisfaction and combinatorial
optimization \cite{Mezard-etal-2002,Krzakala-etal-PNAS-2007},
perceptual learning with discrete synapses
\cite{Braunstein-Zecchina-2006}, signal transmission and processing
\cite{Kabashima-Saad-1999,Tanaka-2001}, network structure inference
\cite{Mezard-Mora-2009,Weigt-etal-2009,Aurell-Ollion-Roudi-2010,Roudi-Hertz-2011},
compressive sensing \cite{Donoho-Maleki-Montanari-2009,Krzakala-etal-2011}.

The most crucial simplification made in the mean-field RSB cavity method is the
Bethe-Peierls approximation \cite{Bethe-1935,Peierls-1936a}.
Roughly speaking, this approximation ignores all the possible correlations
among the set of vertices that are nearest-neighbors to a specified central
vertex. There have been several theoretical attempts to include the effects of
correlations to the cavity method
\cite{Montanari-Rizzo-2005,Parisi-Slanina-2006,Chertkov-Chernyak-2006b,Rizzo-Wemmenhove-Kappen-2007}.
A simple and general scheme of partition function expansion was introduced
in \cite{Xiao-Zhou-2011} along this research line, following the initial work
of Chertkov and Chernyak \cite{Chertkov-Chernyak-2006b} on Ising spin glasses.
By this expansion scheme, the mean-field free energy expression at each level
of replica-symmetry-breaking can be formally derived without  imposing any
physical assumptions, and the correction contributions to these mean-field
free energies are expressed as a sum over looped subgraphs.
The strong constraint that the subgraphs with dangling edges
all have completely zero correction contribution to the free energies
leads to a set of message-passing equations, such as the belief-propagation
equation and the survey-propagation equation, that must be satisfied by
the auxiliary probability distribution functions introduced in the expansion
scheme.

For readers not familiar with spin glass theories,
the partition function expansion scheme
\cite{Xiao-Zhou-2011} can serve as a straightforward mathematical approach to
the essentials of the RSB mean field theory.

Finite-dimensional spin glass models are still very challenging for
theoretical studies. One of the major reasons is the abundance of short loops
and the associated (possibly strong) local correlations. Short loops are
also presented in numerous graphical models derived from real-world
applications in optimization, inference, and constraint satisfaction problems.
For such spin glass systems, the free energy correction contributions from
loopy subgraphs may be comparable to the mean-field free energy values.
As a consequence, the prediction power of the the mean-field theory can be
severely damaged. A powerful conceptual framework for finite-dimensional
lattice models is the cluster variation method, which was invented by
Kikuchi \cite{Kikuchi-1951} and later further developed  by many authors
(reviewed in \cite{Morita-etal-1994,Pelizzola-2005}).The key insight behind
the cluster variation method is that, in a finite-dimensional system, the
correlation between two vertices decays exponentially with the distance
(as long as the system is not approaching a continuous phase transition), and
therefore short-range correlations dominate the statistical property of the
system. Short-range correlations among clusters of neighboring vertices are
considered in a variational way in  the cluster variation method
\cite{Kikuchi-1951,An-1988,Morita-etal-1994,Pelizzola-2005}.
As an extension of the cluster variation method, Yedidia and coworkers
proposed a region graph method \cite{Yedidia-Freeman-Weiss-2005}
to tackle local correlations in finite-dimensional graphical models with more
computational flexibility. A set of regions is specified in the region graph
representation of a spin glass model \cite{Yedidia-Freeman-Weiss-2005}.
Each region contains a subset of vertices and some (or all) of the
interactions within these vertices. A partial order can be defined among
these regions, which is represented by a set of directed edges between pairs of
regions. A variational free energy is defined on the region graph, the
minimization of this free energy leads to a set of generalized
belief-propagation equations \cite{Yedidia-Freeman-Weiss-2005}. Recently,
the generalized belief-propagation was applied to the two-dimensional (2D)
Edwards-Anderson model by Ricci-Tersenghi, Mulet and co-workers
\cite{Dominguez-etal-2011,Rizzo-etal-2010,LageCastellanos-etal-2012}.

As demonstrated in a very recent paper \cite{Zhou-etal-2011}, the partition
function expansion scheme of \cite{Xiao-Zhou-2011} can also be applied to
spin glasses in the region graph representation. In the present work we give
a detailed discussion on how to construct a region graph RSB mean-field theory
for spin glass models and also to obtain the correction expressions to such
a theory. Two sets of message-passing equation, namely the
region graph belief-propagation equation and region graph survey-propagation
equation, are derived through the partition function expansion.
We also prove that, for non-redundant region graphs, the region graph
Bethe-Peierls free energy as derived through the partition function expansion
is equivalent to the variational free energy functional used in
\cite{Yedidia-Freeman-Weiss-2005}. The region graph belief-propagation (rgBP)
equation is applied to the 2D square-lattice Ising model
and the 2D square-lattice Edwards-Anderson (EA) spin glass model, and our
numerical results are compared with the results obtained by the
conventional belief-propagation (BP) procedure.

The prediction on the Ising model's Curie temperature by the rgBP is
better than the prediction of the conventional BP. The exactly known
transition temperature can be approached if region graphs with sufficiently
large maximal regions are used in the rgBP. When applying the rgBP to the
square-lattice EA model at sufficiently low temperature,
a most interesting numerical
observation for us is that, the rgBP identifies many small collective
domains of vertices in the square lattice (see figure~\ref{fig:magevolution}).
Vertices in each of these collective domains are strongly correlated
(and probably change states collectively), while the remaining vertices
outside of all these small localized domains serve as paramagnetic background.
Such highly heterogeneous patterns are in some respect similar to the
patterns of dynamical heterogeneity in dense liquid when approaching the
glass transition (reviewed in \cite{Ediger-2000,Glotzer-2000}).
A key difference is the collective domains in the square lattice do not
move in space, only their areas increase as temperature decreases. These
collective domains bring in many new time scales to the system's
low-temperature dynamics, they should be essential to understand many of
the fascinating low-temperature glass-like behaviors of this system.

For the square-lattice EA model, the existence of many localized collective
domains above the paramagnetic background naturally points to a way of
improving the power of the rgBP.
The basic idea is to use larger maximal
regions for the identified collective domains, while the sizes of the
maximal regions in the remaining parts of the lattice keep to be as small
as possible. This hybrid adaptive strategy will be explored in a
future work. We hope it will be helpful in achieving deeper
insights on the heterogeneous (but globally still
paramagnetic) behaviors of this and other 2D spin glass systems.

The next section introduces the general graphical model and its factor graph
and region graph representations, and briefly reviews the cluster variation
method. Section \ref{sec:rs} derived the region graph replica-symmetric (RS)
mean-field theory as well as the region graph belief-propagation equation;
the connection with the generalized belief-propagation of
\cite{Yedidia-Freeman-Weiss-2005} is also discussed here.
Section \ref{sec:rsb} derived the region graph first-replica-symmetry-breaking
(1RSB) mean-field theory and the associated region graph survey-propagation
equation. The RS mean-field theory (rgBP) is then applied to the 2D Ising
model and the 2D Edwards-Anderson model in Sec.~\ref{sec:2d}. We conclude
this work and  list some possible future projects
in Sec.~\ref{sec:conclude}. The paper have four appendices.

\section{The Model and Graphical Representations}
\label{sec:model}

Consider a very general model system of $N$ particles ($i=1, 2, \ldots, N$)
interacting with each other and with the environment. Each particle $i$ is
fixed in space and therefore has no translational degrees of freedom,
it is fully characterized by an internal state variable $x_i$.
For example, in Ising models, the state variable is binary, $x_i= \pm 1$;
in Potts models, the internal state can choose among $Q$
different values, $x_i \in \{1, 2, \ldots, Q\}$; in Heisenberg models, $x_i$
is a three-dimensional continuous vector of unit length. In this paper we
assume, for notational simplicity, that the state variables $x_i$ are
discrete and can take only a finite number of different values. The microscopic
configuration (state) $\underline{x}$ of the whole system is defined by the
states of all its particles, $\underline{x} \equiv \{x_1, x_2, \ldots, x_N\}$.
The configuration energy $H(\underline{x})$ is supposed to have the following
additive form
\begin{equation}
\label{eq:totalenergy}
H(\underline{x}) = \sum\limits_{i=1}^{N} E_i(x_i)
 + \sum\limits_{a=1}^{M} E_a(\underline{x}_{\partial a}) .
\end{equation}
On the right side of (\ref{eq:totalenergy}), the first term is contributed by
external forces (fields), with each energy $E_i$ depending only on a single
particle $i$ (if  $i$ is free of external forces, then $E_i(x_i) \equiv 0$).
The second term is contributed by $M$ internal interactions
($a=1, 2, \ldots, M$), each having an energy $E_a$. Let us denote by
$\partial a$ the set of particles that are involved in interaction $a$ and by
$\underline{x}_{\partial a} \equiv \{ x_i | i\in \partial a\}$ a
joint state of the particles in this set. The internal interaction energy
$E_a$ is a function only of $\underline{x}_{\partial a}$.
For example, in a two-body Ising model an internal energy
has the form $E_a = -J_a x_i x_j$ with $J_a$ being a coupling constant, then
$\partial a = \{i, j\}$ and $\underline{x}_{\partial a}=\{x_i, x_j\}$.

After equilibrium is reached in an environment of temperature $T$,
the probability of the system being in a configuration
$\underline{x}$ obeys the Boltzmann distribution
\begin{equation} \label{eq:Boltzmann}
P_B(\underline{x}) = \frac{1}{Z(\beta)} e^{-\beta H(\underline{x})} ,
\end{equation}
where $\beta \equiv 1/(k_B T)$ is the inverse temperature,
$k_B$ being the Boltzmann constant. The partition function  $Z(\beta)$ is
expressed as
\begin{equation}
 \label{eq:partitionfunction}
    Z(\beta) \equiv \sum_{\underline{x}} e^{-\beta H(\underline{x})}
  = \sum\limits_{\underline{x}} \prod\limits_{i=1}^{N} \psi_i(x_i)
 \prod\limits_{a=1}^{M} \psi_a(\underline{x}_{\partial a}) ,
\end{equation}
where $\psi_i$ and $\psi_a$ are, respectively, the Boltzmann factor for
the external field on particle $i$ and the internal interaction $a$,
\begin{equation}
 \label{eq:psi}
\psi_i(x_i) \equiv e^{-\beta E_i(x_i)}, \quad\quad
\psi_a(\underline{x}_{\partial a})
 \equiv e^{-\beta E_a(\underline{x}_{\partial a})} .
\end{equation}
From $Z(\beta)$ we can define the equilibrium free energy $F(\beta)$ as
\begin{equation}
F(\beta)  \equiv -\frac{1}{\beta} \ln Z(\beta)
= \sum\limits_{\underline{x}} P_B(\underline{x}) H(\underline{x})
 + k_B T \sum\limits_{\underline{x}}
P_B(\underline{x}) \ln P_B(\underline{x}) .
\label{eq:freeenergy2}
\end{equation}

The expression (\ref{eq:freeenergy2}) for the equilibrium free energy can
be extended to a general non-equilibrium situation. Given an arbitrary
probability distribution $\rho(\underline{x})$ on the configurations of the
model (\ref{eq:totalenergy}), the Shannon entropy functional is defined
as \cite{Cover-Thomas-1991}
\begin{equation}
\label{eq:Shannonentropy}
    S[\rho] \equiv  - k_B \sum_{\underline{x}} \rho(\underline{x})
    \ln \rho(\underline{x}) ,
\end{equation}
and the free energy functional $F[\rho]$ is
\begin{equation} \label{eq:fef}
    F[\rho ] \equiv \sum_{\underline{x}} \rho(\underline{x}) H(\underline{x})
- T S[\rho] .
\end{equation}
It is easy to check that the absolute minimum of $F[\rho]$ is equal to the
equilibrium free energy $F(\beta)$ and this minimum is achieved only when
$\rho(\underline{x})= P_B(\underline{x})$.

\subsection{The Factor Graph Representation}

The model defined by (\ref{eq:totalenergy}) can be represented conveniently
by a factor graph $G$ of variable nodes (shown as circles, denoting the $N$
particles $i,j, k, \ldots$), function nodes (shown as squares, denoting the
$M$ {\em internal} interactions $a, b, c, \ldots$), and edges between pairs of
variable and function nodes $(i,a)$.\footnote{We follow the convention in the
literature and use letters $i,j,k, l, \ldots$ to denote variable nodes and
letters $a,b,c,d,\ldots$ to denote function nodes.}
The factor graph $G$ is a bipartite graph: all the edges are between a
variable node and a function node, and an edge $(i,a)$ exists between a
variable node $i$ and a function node $a$ if and only if variable
$i$ is involved in the internal interaction represented by $a$. A
comprehensive review on factor graphs can be found
in \cite{Kschischang-etal-2001}.

\begin{figure}
\begin{center}
\includegraphics[width=0.35\linewidth]{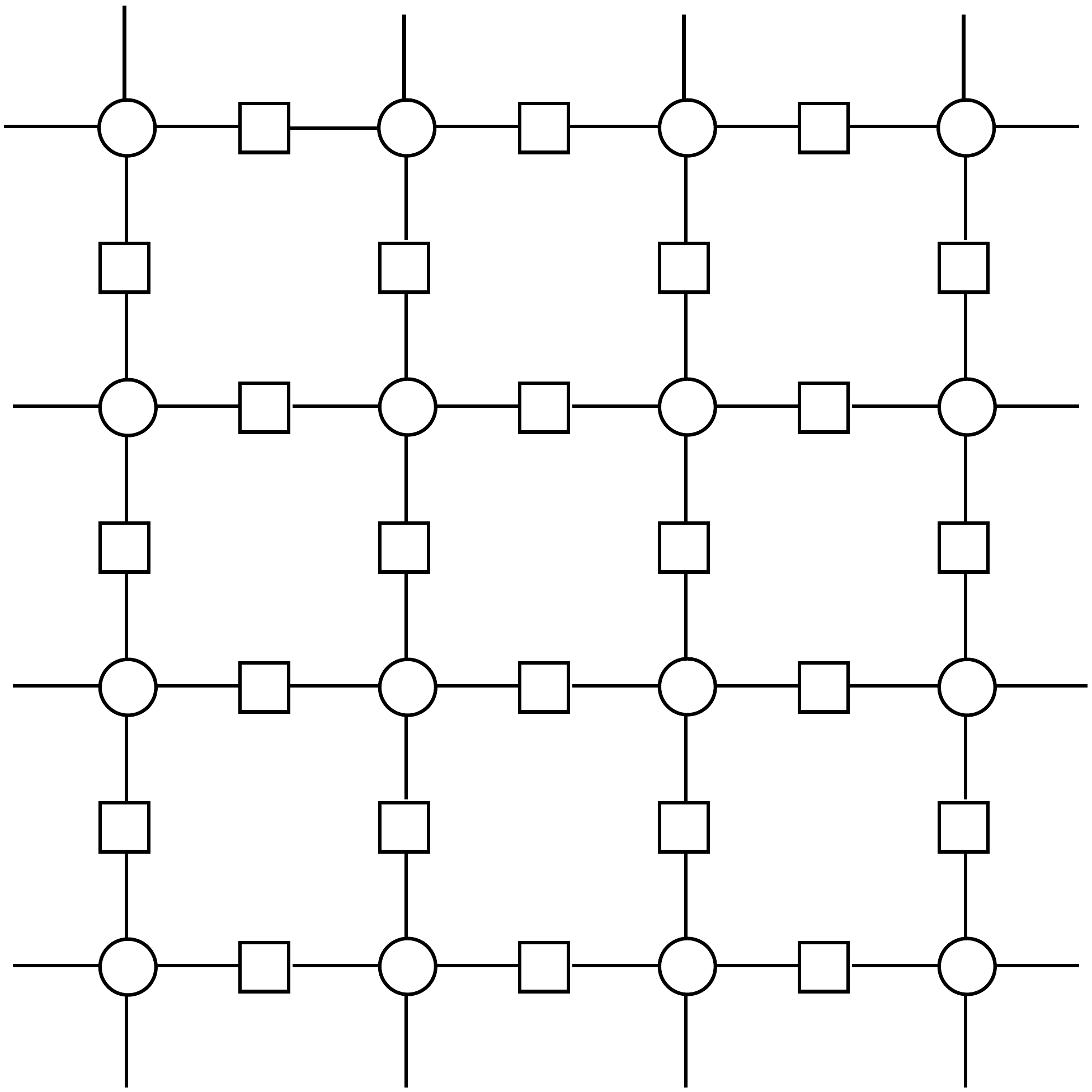}
\end{center}
\caption{\label{fig:factorgraph}
The factor graph for the two-dimensional Edwards-Anderson model.
Each variable node $i$ (circle) represents a lattice site $i$, it has a spin
state $\sigma_i$ and an energy $E_i(\sigma_i) = - h_i^0 \sigma_i$ caused by an
external field $h_i^0$; each function node $a$ (square) between two variable
nodes $i$ and $j$ represents a spin coupling with energy
$E_a(\sigma_i, \sigma_j) = - J_{i j} \sigma_i \sigma_j$.
}
\end{figure}

As a simple illustration, we show in Fig.~\ref{fig:factorgraph} part of the
factor graph for the two-dimensional Edwards-Anderson  spin glass model
\cite{Edwards-Anderson-1975} on a square lattice. The energy function of
the EA model is defined as
\begin{equation}
\label{eq:2DEA}
H(\underline{\sigma}) = - \sum\limits_{(i,j)} J_{i j} \sigma_i \sigma_j -
 \sum\limits_{i} h_i^0 \sigma_i ,
\end{equation}
where  the internal energies are due to ferromagnetic ($J_{i j} > 0$) or
anti-ferromagnetic ($J_{i j}<0$) spin couplings along the edges $(i,j)$ of the
square lattice, and $h_i^0$ is the external magnetic field on particle $i$.
We use $\sigma_i \in \{-1, +1\}$ instead of $x_i$ to denote the binary spin
state of particle $i$. The spin configuration of the whole system is
$\underline{\sigma}\equiv \{\sigma_1, \sigma_2, \ldots, \sigma_N\}$.

\subsection{A Brief Summary of The Cluster Variation Method}

The cluster variation method (CVM), as originally proposed by Kikuchi
\cite{Kikuchi-1951}, is an approximate method to calculate the Shannon
entropy (\ref{eq:Shannonentropy}). It decomposes $S[\rho]$ into contributions
from different clusters of variable nodes (see \cite{An-1988} for an
easy-to-access introduction to CVM). A cluster in the CVM is defined as a
non-empty subset of the $N$ variable nodes. The total number of possible
clusters is $2^{N}-1$. For two clusters $C_1$ and $C_2$, we say that
$C_1 \leq C_2$ (and $C_2 \geq C_1$) if and only if $C_1$ is a subset of $C_2$
(in cases that $C_1\leq C_2$ and also $C_1 \geq C_2$, then $C_1=C_2$).
If $C_1$ is a strict subset of $C_2$, we denote as $C_1 < C_2$ (and
$C_2> C_1$). This partial ordering
corresponds to the following function $\zeta(C_1, C_2)$
\begin{equation}
\zeta(C_1, C_2) = \left\{
\begin{array}{ll}
1 , & \quad {\rm for}\;\; C_1 \leq C_2 ; \\
0 , & \quad {\rm otherwise} .
\end{array} \right.
\end{equation}
This function can  be regarded as a matrix, and it is invertible:
\begin{equation}
\label{eq:inverse}
\sum\limits_{C_2} \zeta(C_1, C_2) \mu(C_2, C_3) =
\sum\limits_{C_2} \mu(C_1, C_2) \zeta(C_2, C_3) = \delta(C_1, C_3) ,
\end{equation}
where $\delta(C_1, C_2)$ is the Kronecker delta such that
$\delta(C_1, C_2) = 1$ if $C_1=C_2$ and $=0$ otherwise. The function
$\mu(C_1, C_2)$ is the M\"{o}bius inversion function \cite{Rota-1964} defined
as
\begin{equation}
\mu(C_1, C_2) = \left\{
\begin{array}{ll}
(-1)^{|C_2|-|C_1|} , \quad & {\rm for}\;\;  C_1 \leq C_2 ; \\
0 , & {\rm otherwise} .
\end{array}\right.
\end{equation}
In the above equation, $|C|$ denotes the number of elements in cluster $C$.

The configuration of a cluster $C$ is defined as the collection of states of
all its elements and is denoted as $\underline{x}_C \equiv \{x_i : i\in C\}$.
Given a probability distribution $\rho(\underline{x})$ for all the $N$
variable nodes, the marginal distribution for the configuration
$\underline{x}_C$ is
\begin{equation}
\rho_C(\underline{x}_C) \equiv
\sum\limits_{\underline{x} \backslash \underline{x}_C} \rho(\underline{x}) .
\end{equation}
Following (\ref{eq:Shannonentropy}), we can define for each cluster $C$ its
Shannon entropy functional as
\begin{equation}
S_C [\rho_C] \equiv - k_B \sum\limits_{\underline{x}_C} \rho_C
(\underline{x}_C) \ln
\rho_C (\underline{x}_C) .
\end{equation}
Let us define an entropy increment $\Delta {S}_C$ for each cluster $C$ as
(see, for example, \cite{An-1988})
\begin{equation}
\label{eq:entropydelta}
\Delta S_C \equiv \sum\limits_{C^\prime \leq C}
(-1)^{|C|-|C^\prime|}  S_{C^\prime} = \sum\limits_{C^\prime }
 S_{C^\prime} \mu(C^\prime, C) .
\end{equation}
Then from (\ref{eq:entropydelta}) and (\ref{eq:inverse}) we obtain that
\begin{equation} \label{eq:ecluster}
S_C  = \sum\limits_{C^\prime } \Delta S_{C^\prime} \zeta(C^\prime, C)
= \sum\limits_{C^\prime \leq C} \Delta S_{C^\prime} .
\end{equation}
In other words, the entropy $S_C$ of each cluster $C$ is the sum of the
entropy increments of all its sub-clusters (including $C$ itself).

If we take $C$ as the set of all the variable nodes in $G$, the above equation
(\ref{eq:ecluster}) then states that, the entropy functional $S[\rho]$ of the
whole system  is the sum of the entropy increments of all the $2^N-1$
different clusters, i.e., $S[\rho]=\sum_{C} \Delta S_C[\rho_C]$.
There are an exponential terms in this summation. The insight of Kikuchi
\cite{Kikuchi-1951} was that, the entropy increment $\Delta S_C$ often
decays quickly with the size $|C|$ of the clusters, therefore a good
approximation of $S[\rho]$ can be obtained by summing over only a subset of
small clusters. Let us choose a set of maximal clusters
$C_1^{(m)}, C_2^{(m)}, \ldots, C_p^{(m)}$. These $p$ clusters are maximal in
the sense that any of them is not a sub-cluster of any another
cluster in this chosen set. These $p$ maximal clusters and all their
sub-clusters form a set, denoted as $P$. We then have the following
approximate expression for the entropy functional:
\begin{equation}
\label{eq:S-CVM}
S\bigl[\rho \bigr] \approx
\sum\limits_{C \in P} \Delta S_C\bigl[\rho_C \bigr] .
\end{equation}
This is the CVM approximation.

Inserting (\ref{eq:entropydelta}) into the above expression, we get
\begin{equation}
S\bigl[\rho \bigr] \approx  \sum\limits_{C\in P} a_C S_C\bigl[ \rho_C \bigr] ,
\end{equation}
where the coefficient $a_C$ of a cluster $C$ is expressed as
\begin{equation}
a_C = \sum\limits_{C^\prime \in P} \mu(C, C^\prime) =
\sum\limits_{\{C^\prime | C^\prime \in P, C^\prime  \geq C\}}
 (-1)^{|C^\prime| - |C|} .
\end{equation}
It is easy to check that each of the $p$ maximal clusters (say $C^{(m)}$) has
the coefficient $a_{C^{(m)}} = 1$. For other clusters of $P$, their
coefficients can be calculated iteratively as
\begin{equation}
\label{eq:partialsum}
a_C = 1 - \sum\limits_{\{C^\prime | C^\prime \in P, C^\prime > C \}}
a_{C^\prime} .
\end{equation}

The applications of the CVM method were reviewed in
\cite{Morita-etal-1994,Pelizzola-2005}. This method can be combined with
Suzuki's coherent anomaly method \cite{Suzuki-etal-1995} to compute the
critical exponents of a given finite-dimensional system.

\subsection{The Region graph Representation}
\label{sec:regiongraph}

In the CVM approximation (\ref{eq:S-CVM}), for each maximal cluster, all its
sub-clusters are included into the cluster set $P$. This requirement was
relaxed in the work of Yedidia and co-authors, who proposed a more flexible
region graph representation \cite{Yedidia-Freeman-Weiss-2005}.
A region graph $R$ is formed by regions\footnote{In this paper, we use Greek
symbols $\alpha, \beta, \gamma, \ldots$ to denote the regions of a
region graph $R$.} and a set of directed edges between the regions.
A region $\alpha$ of a factor graph $G$
includes a set of variable nodes and a set of function nodes, with the
condition that, if a function node $a$ belongs to a region $\alpha$, then all
the variable nodes in the set $\partial a$ also belong to the region $\alpha$.
A configuration of a region $\alpha$ is defined by the states of all the
variable nodes in this region,
$\underline{x}_\alpha \equiv \{x_i | i\in \alpha\}$.

If there is a directed edge pointing from a region $\mu$ to $\nu$ in the
region graph $R$, then it must be the case that $\nu$ is a sub-region of
$\mu$ (that is, all the variable nodes and function nodes of region $\nu$
also belong to region $\mu$). We use $\mu\rightarrow \nu$ to emphasize the
directness of the edge; sometimes if the direction of an edge is
unimportant for the  discussion, the notion $(\mu, \nu)$ is also used.
It should be emphasized that, a region $\nu$ being a sub-region of another
region $\mu$ does not necessarily indicate the existence of a directed edge
from $\mu$ to $\nu$. If there is a directed edge from region $\mu$ to $\nu$,
then $\mu$ is regarded as a parent of $\nu$ and $\nu$ a child of $\mu$.
If there is a directed path from region $\alpha$ to region $\nu$, we say that
$\alpha$ is an ancestor of $\nu$ and $\nu$ is a descendant of $\alpha$.
The ancestor-descendant relationship between two regions $\alpha$ and $\nu$
is denoted by $\nu < \alpha$ and $\alpha > \nu$. (If $\nu$ is not a descendant
of $\alpha$, such a relationship does not exist even if $\nu$ is a
sub-region of $\alpha$.)

\begin{figure}
\begin{center}
\includegraphics[width=0.6\linewidth]{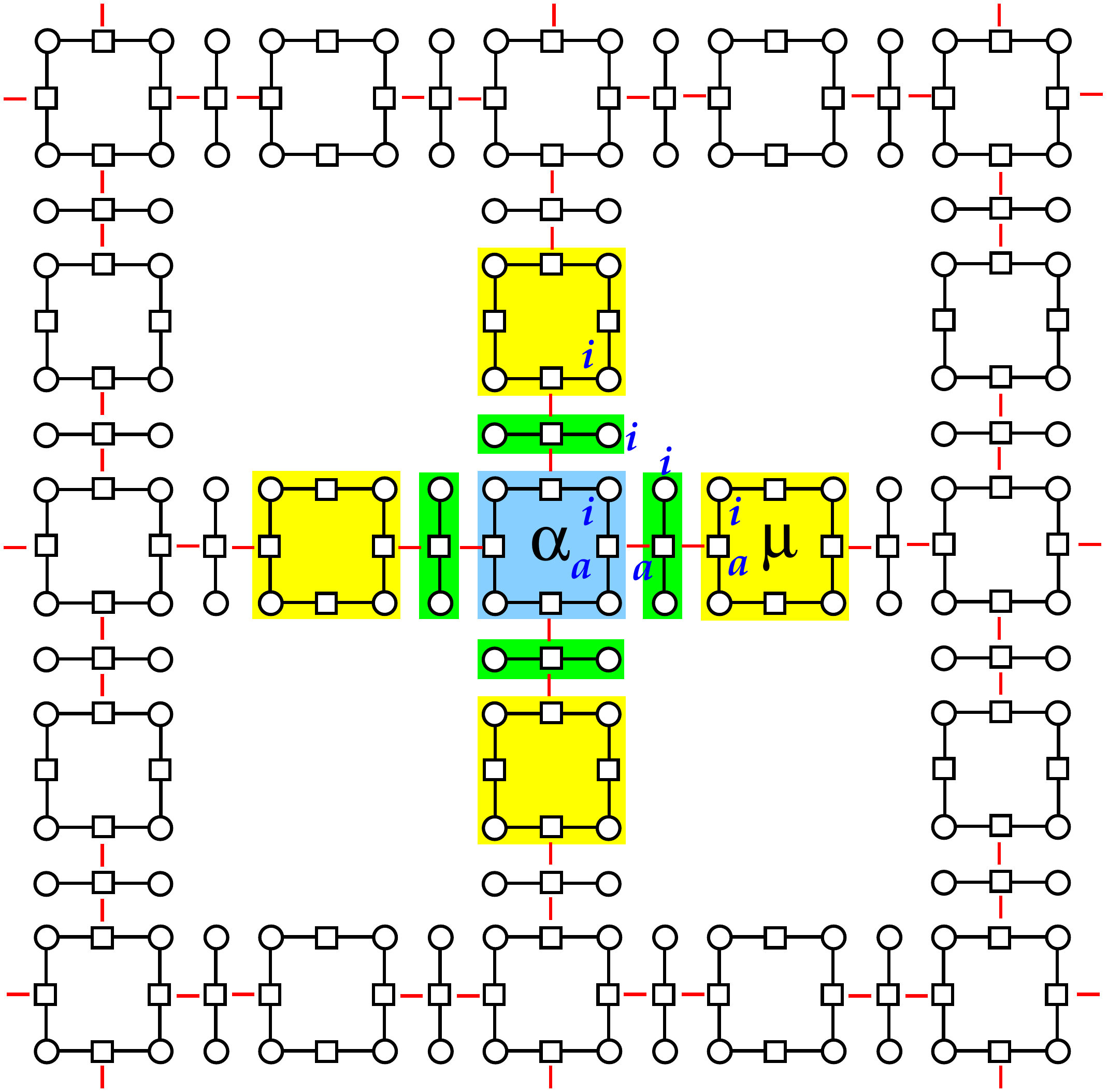}
\end{center}
\caption{ \label{fig:rgfig} A non-redundant region graph $R$ 
for the factor graph shown in Fig.~\ref{fig:factorgraph}. Only a
part of the full region graph is shown.
There are three types of
regions: each `square' region (e.g., region $\alpha$)
contains $n \times n$ variable nodes and
$2 n (n-1)$ function nodes (in this example, $n=2$), 
its counting number is $c=1$; 
each `rod' region contains $n$ variable
nodes and $n-1$  function nodes, its counting number is $c=-1$; and each
`stripe' region (e.g., region $\mu$)
contains $n \times 2$ variable nodes and $n$ 
function nodes, with counting number $c=1$. Each stripe region is
connected to two rod regions, and each rod region connects to one
stripe region and one square region. The
short red lines between two regions indicate the 
parent-child relationship (the directions of these edges are
not shown, as they are obvious). In this region graph,
variable node $i$ appears in $5$
different regions, and function node $a$ appears in $3$
different regions. }
\end{figure}

Each region $\gamma$ is assigned a counting number $c_\gamma$,
which is determined recursively by
\begin{equation} \label{eq:rgcn}
c_\gamma = 1 - \sum\limits_{\{\alpha| \alpha \in R, \alpha>\gamma\}} c_\alpha .
\end{equation}
Notice the similarity between (\ref{eq:rgcn}) and (\ref{eq:partialsum}).
In the region graph $R$, the subgraph $R_i$ induced by a variable node $i$ is
defined as the subgraph  that include all the regions containing
$i$ and all the directed edges between these regions. Similarly, $R_a$
denotes the region graph induced by function node $a$, it includes all the
regions containing $a$ and all the directed edges between these regions.
The region graph $R$ and its associated counting numbers $\{c_\gamma\}$ are
required to satisfy the following region graph conditions
\cite{Yedidia-Freeman-Weiss-2005}:
\begin{enumerate}
\item[(1)]
For any variable node $i$, the induced subgraph $R_i$ is connected,
and the sum of counting numbers within this subgraph is unity:
\begin{equation} \label{eq:cni}
    \sum\limits_{\gamma \in R_i} c_\gamma = 1 .
\end{equation}

\item[(2)]
For any function node $a$, the induced subgraph $R_a$ is connected, and the
sum of counting numbers within $R_a$ is unity:
\begin{equation}
\label{eq:cna}
\sum\limits_{\gamma \in R_a} c_\gamma = 1 .
\end{equation}
\end{enumerate}

Given a factor graph $G$, one may construct many different region graphs $R$,
all of them satisfying the region graph conditions.
A region graph $R$ is considered as being {\em non-redundant} if it
satisfies the following additional condition: the region subgraph $R_i$
induced by any variable node $i$ contains no loops
(it is a tree of regions). Region graphs which do not satisfy this
tree condition are referred to as being  redundant.
As we will discuss in detail, non-redundant region graphs
have some nice properties for partition function expansion.

A simple region graph $R$ is shown in Fig.~\ref{fig:rgfig}
for the two-dimensional EA model. Although this region graph contains
many loops at the region level, it is non-redundant as can
be easily checked.

\section{Partition Function Expansion on a Region Graph}
\label{sec:rs}

Given a region graph $R$ for a factor graph $G$, we now expand the equilibrium
partition function (\ref{eq:partitionfunction}) as a sum over the contributions
of region subgraphs of $R$. The mathematical framework of \cite{Xiao-Zhou-2011}
will be followed.

\subsection{Introducing auxiliary states of variable nodes
and removing redundancy}
\label{sec:rs01}

First of all, because of the constraints (\ref{eq:cni}) and (\ref{eq:cna}),
the partition function can be expressed as
\begin{equation}
\label{eq:Z-exp-1}
    Z(\beta) = \sum\limits_{\underline{x}}
    \prod\limits_{\alpha \in R} \biggl[
    \prod_{i\in \alpha} \psi_i(x_i) \prod_{a\in \alpha}
\psi_a(\underline{x}_{\partial a})
    \biggr]^{c_\alpha} .
\end{equation}

The microscopic configuration of each region $\alpha$ is denoted as
$\underline{x}_\alpha \equiv \{x_i^\alpha | i \in \alpha\}$, where
$x_i^\alpha$ is the state of variable node $i$ in region $\alpha$.
For a function node $a$ in  region $\alpha$,
the variable states at its neighborhood are denoted as
$\underline{x}_{\partial a}^\alpha \equiv \{ x_i^\alpha | i \in \partial a\}$.
A Boltzmann factor $\Psi_\alpha(\underline{x}_\alpha)$ is defined for
region $\alpha$ as
\begin{equation}
\label{eq:Psialpha}
\Psi_\alpha(\underline{x}_\alpha) =
 \prod\limits_{i\in \alpha} \psi_i^{c_\alpha} (x_i^\alpha) \prod_{a\in \alpha}
\psi_a^{c_\alpha}(\underline{x}_{\partial a}^\alpha) .
\end{equation}
A variable node $i$ may belong to two or more regions
(say $\alpha, \gamma, \ldots$). If this is the case, we regard the states of
node $i$ in the different regions, $x_i^\alpha, x_i^\gamma, \ldots$,
as different variables. With the introduction of these new variables,
the partition function expression (\ref{eq:Z-exp-1}) is re-written as
\begin{equation}
    Z(\beta)  =  \sum\limits_{\underline{x}}
    \sum\limits_{\{\underline{x}_\alpha\}}
    \prod\limits_{\alpha \in R} \biggl( \Psi_\alpha(\underline{x}_\alpha)
    \prod_{i\in \alpha} \delta(x_i^\alpha, x_i) \biggr)
    =  \sum\limits_{\{\underline{x}_\alpha\}}
    \prod\limits_{\alpha \in R}  \Psi_\alpha(\underline{x}_\alpha)
    \prod_{i=1}^{N} \biggl[
    \sum\limits_{x_i} \prod\limits_{\gamma\in R_i}
     \delta(x_i^\gamma, x_i) \biggr].
     \label{eq:Z-exp-2b}
\end{equation}
The Kronecker delta functions $\delta(x_i^\alpha, x_i)$ guarantee that,
the partition function is contributed only by those region configurations
$\{\underline{x}_\alpha | \alpha \in R\}$ for which the states
$x_i^\alpha, x_i^\gamma, \ldots$ of each variable node $i$ in different
regions are equal to the same value.

Let us for the moment focus on the region subgraph $R_i$ induced by variable
node $i$. Assume $R_i$ contains $n_i$ regions. The expression within the
square brackets in (\ref{eq:Z-exp-2b}) requires that the
states of variable node $i$ take the same value among the $n_i$ regions,
which is equivalent to $(n_i-1)$ constraints. These $(n_i-1)$ constraints
of each  variable node $i$ can be conveniently implemented in
the following way: (1) Attach on each edge $(\mu,\nu)$ of region graph $R$ a
variable node set, the cross-linker set $\mu\#\nu$, which is
initially empty.
(2) For each variable node $i\in \{1, 2, \ldots, N\}$, choose $(n_i-1)$
different edges of the induced region graph $R_i$ in such a way that the
$n_i$ regions of $R_i$ form a connected tree with these edges;
then for each of these $(n_i-1)$ chosen edges $(\mu, \nu)$,
add $i$ to its cross-linker set $\mu\#\nu$.
(3) After all the cross-linker sets are constructed, for each variable
node $j$ in any cross-linker set $\mu\#\nu$, a constraint
$\delta(\sigma_j^\mu, \sigma_j^\nu)$ is applied, forcing the states of
$j$ to be the same in the two connected regions $\mu$ and $\nu$.

\begin{figure}
\begin{center}
\includegraphics[width=0.45\linewidth]{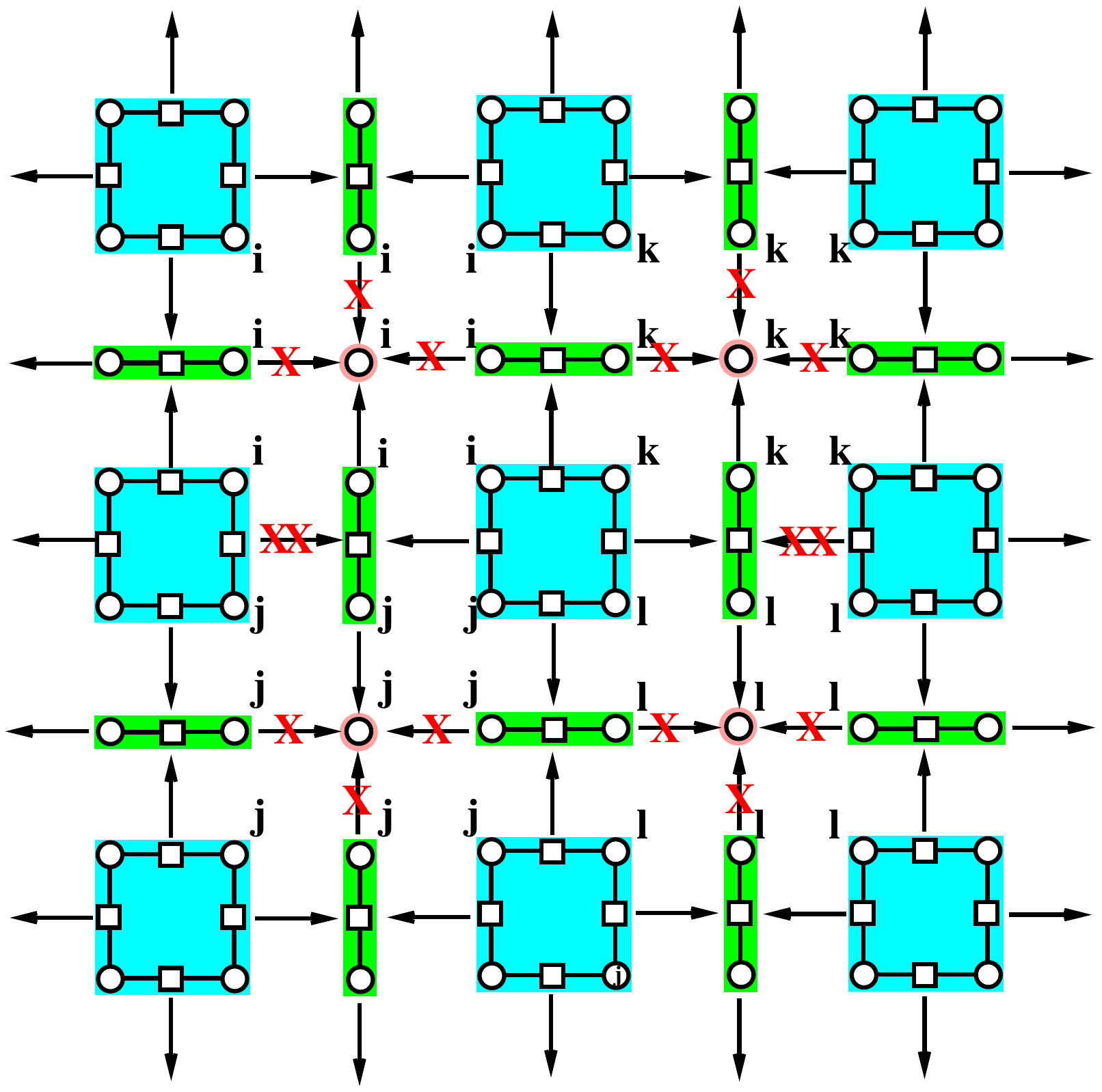}
\hspace*{1.0cm}
\includegraphics[width=0.45\linewidth]{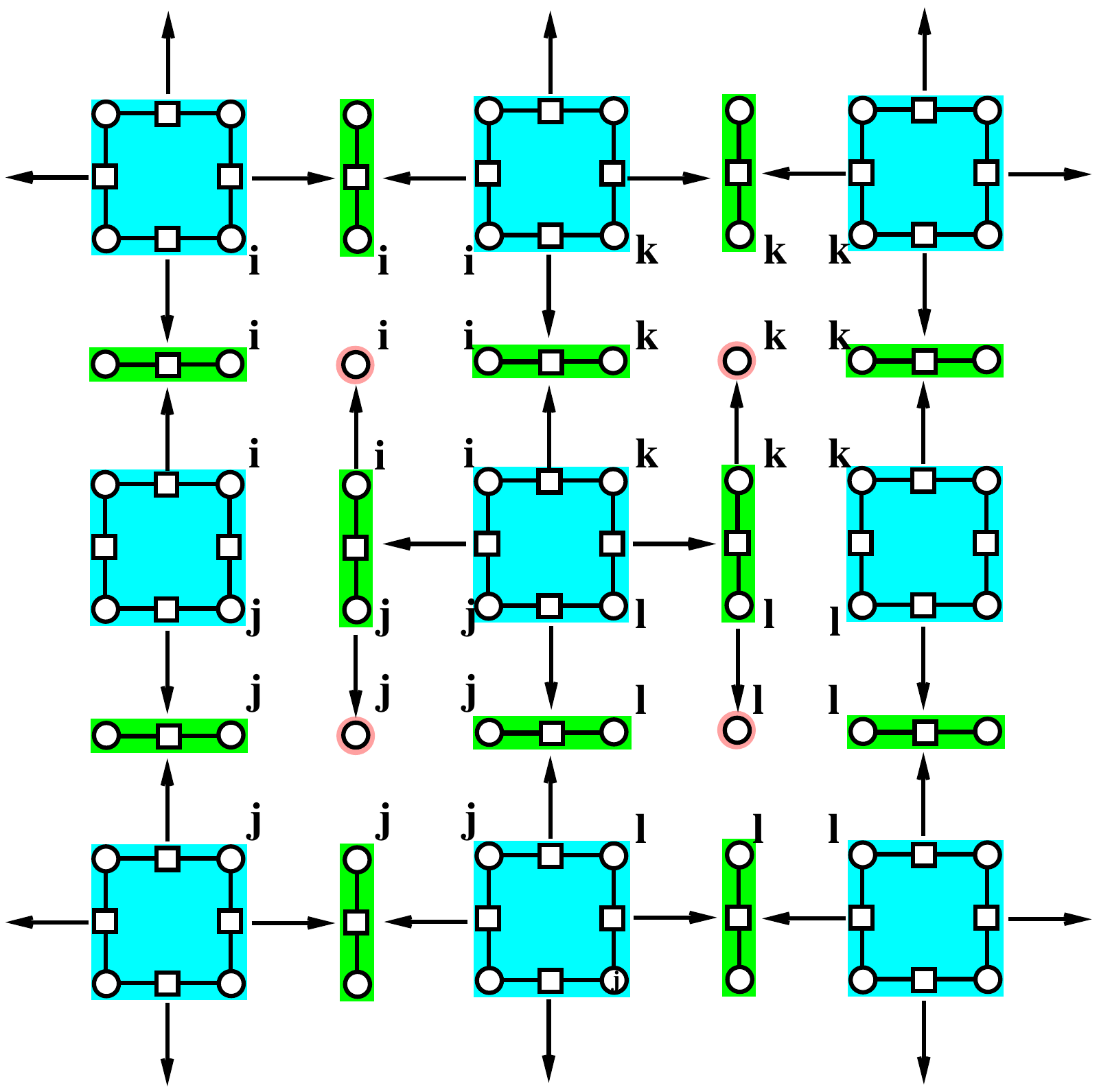}
\end{center}
\caption{ \label{fig:reduntrgfig} Removing redundancy from a region graph.
(Left panel) a redundant region graph $R$ for the factor graph 
shown in Fig.~\ref{fig:factorgraph}. There are three types of
regions in this region graph, as indicated by the three different colors.
Each variable node (e.g., $i, j, k, l$) 
appears in $9$ different regions. 
A {\tt X} symbol on a directed edge $\mu\rightarrow \nu$ (between
a parent region $\mu$ and a child region $\nu$) means that the single common
variable node (say $i$) of $\mu$ and $\nu$ is not included into the
cross-linker set $\mu\#\nu$ (that is, $\mu\#\nu=\emptyset$).
A {\tt XX} symbol on the edge $\mu\rightarrow\nu$ means that
the two common variable nodes (say $i,j$) of $\mu$ and $\nu$ are not
included into the set $\mu\#\nu$ (again, $\mu\#\nu=\emptyset$).
(Right panel) the resulting non-redundant graph $R^\#$ after
all the redundancy in $R$ has been removed.
}
\end{figure}

The region graph shown in figure \ref{fig:rgfig} is non-redundant, and
there is no need to remove redundancy. 
The left panel of figure \ref{fig:reduntrgfig} shows a redundant
region graph $R$ for the 2D Edwards-Anderson model. This region graph has
three types of regions, and each variable node (such as
node $i$) appears in $9$ different regions.  The region subgraph $R_i$
induced by variable node $i$ contains $12$ directed edges. One particular
way of removing redundancy is indicated by the {\tt X} and {\tt XX} symbols
on the directed edges of $R$. After removing redundancy, the
resulting graph $R^\#$ is shown in the right panel of
figure ~\ref{fig:reduntrgfig}. This graph $R^\#$ is actually
equivalent to the region graph shown in figure ~\ref{fig:rgfig} (we will
return to this point later). 

By this construction, it is easy to check that the states of each variable
node $i$ is associated with $(n_i-1)$ constraints. For some edges $(\mu,\nu)$
of the region graph $R$, the cross-linker set might be empty,
$\mu\#\nu = \emptyset$. We can remove all such edges from the region graph
$R$, and the resulting graph is denoted as $R^\#$
(see the right panel of figure \ref{fig:reduntrgfig} for an example).
The partition
function (\ref{eq:Z-exp-2b}) is re-expressed as
\begin{equation}
Z(\beta)  = \sum\limits_{\{\underline{x}_\alpha\}}
    \prod\limits_{\alpha \in R}  \Psi_\alpha(\underline{x}_\alpha)
   \prod\limits_{(\mu,\nu)\in R^\#} \biggl[ \prod_{i\in \mu\#\nu}
 \delta(x_i^\mu, x_i^\nu) \biggr] .
     \label{eq:Z-exp-2c}
\end{equation}

We should emphasize that the above-mentioned
procedure of removing redundancy does not change the counting numbers, it
only assigns a cross-linker set $\mu\#\nu$ (might be empty) on each
directed edge $(\mu, \nu)$ of the region graph $R$.

\subsection{Introducing auxiliary cavity probability functions}

For each edge $(\mu, \nu)$ of the region graph $R^\#$ we now introduce two
auxiliary probability distribution functions,
$p_{\mu\rightarrow \nu}(\underline{x}_{\mu\#\nu}^{\nu})$ and
$p_{\nu \rightarrow \mu}(\underline{x}_{\mu\#\nu}^\mu)$. They are non-negative
and normalized but otherwise arbitrary.
The microscopic configuration $\underline{x}_{\mu\#\nu}^\mu$
denotes the states of variable nodes of set $\mu\#\nu$
{\em in region $\mu$}, i.e.,
$\underline{x}_{\mu\#\nu}^\mu \equiv \{ x_i^\mu | i \in \mu\#\nu\}$; similarly,
$\underline{x}_{\mu\#\nu}^\nu \equiv \{x_i^\nu | i\in \mu\#\nu\}$.
We further rewrite (\ref{eq:Z-exp-2c}) as
\begin{equation}
    Z(\beta) = \sum\limits_{\{\underline{x}_\alpha\}}
    \prod\limits_{\alpha} \biggl[ \Psi_\alpha (\underline{x}_\alpha )
    \prod_{\gamma \in \partial^{\#} \alpha}
     p_{\gamma\rightarrow \alpha}(\underline{x}_{\alpha\#\gamma}^\alpha)
 \biggr] \prod_{(\mu, \nu)\in R^{\#}} \biggl[
    \frac{\prod_{i\in \mu\#\nu} \delta(x_i^\mu, x_i^\nu)}
    {p_{\mu\rightarrow \nu}(\underline{x}_{\nu\#\mu}^\nu)
    p_{\nu\rightarrow \mu}(\underline{x}_{\mu\#\nu}^\mu)}\biggr] ,
      \label{eq:Z-exp-3}
\end{equation}
where $\partial^\# \alpha$ denotes the set of nearest neighboring regions of
region $\alpha$  in region graph $R^\#$.

It is helpful to define two partition function factors as follows:
\begin{eqnarray}
     Z_\alpha  & \equiv & \sum\limits_{\underline{x}_\alpha}
     \Psi_\alpha(\underline{x}_\alpha)
      \prod\limits_{\gamma \in \partial^\# \alpha}
 p_{\gamma \rightarrow \alpha}(\underline{x}_{\alpha\#\gamma}^\alpha)  ,
                                      \label{eq:Zalpha} \\
     Z_{\mu,\nu} &\equiv &
     \sum\limits_{\underline{x}_{\mu\#\nu}^\mu}
 \sum\limits_{\underline{x}_{\nu\#\mu}^\nu}
     p_{\nu\rightarrow \mu}(\underline{x}_{\mu\#\nu}^\mu)
     p_{\mu \rightarrow \nu}(\underline{x}_{\nu\#\mu}^\nu)
 \prod_{i\in \mu\#\nu} \delta(x_i^\mu, x_i^\nu) .       \label{eq:Zedge}
\end{eqnarray}
Then (\ref{eq:Z-exp-3}) is written as the following simple form
\begin{equation}
Z(\beta) = Z_0   \sum\limits_{\{\underline{x}_\alpha\}}
    \prod\limits_{\alpha} w_\alpha(\underline{x}_\alpha)
     \prod_{(\mu, \nu)\in R^\#} \Bigl[1+ \Delta_{(\mu,\nu)}
     (\underline{x}_{\mu\#\nu}^\mu,
     \underline{x}_{\nu\#\mu}^\nu) \Bigr]  ,
\label{eq:Z-exp-4}
\end{equation}
where the coefficient $Z_0$ is equal to
\begin{equation}         \label{eq:Z0}
    Z_0 = \prod\limits _{\alpha\in R} Z_\alpha
    \prod_{(\mu,\nu)\in R^\#} \frac{1}{Z_{(\mu, \nu)}} ,
\end{equation}
and $\Delta_{(\mu,\nu)}$ is expressed as
\begin{equation}\label{eq:Dcorrection}
\Delta_{(\mu, \nu)}(\underline{x}_{\mu\#\nu}^\mu,
\underline{x}_{\nu\#\mu}^\nu)
     = \frac{Z_{(\mu,\nu)} \prod_{i\in \mu\#\nu} \delta(x_i^\mu,
    x_i^\nu)}{p_{\mu\rightarrow \nu}(\underline{x}_{\nu\#\mu}^\nu)
    p_{\nu\rightarrow \mu}(\underline{x}_{\mu\#\nu}^\mu)} - 1 ,
\end{equation}
and $w_\alpha(\underline{x}_\alpha)$ is a probability distribution defined as
\begin{equation}
w_\alpha (\underline{x}_\alpha) \equiv \frac{1}{Z_\alpha}
\Psi_\alpha(\underline{x}_\alpha)
\prod\limits_{\gamma \in \partial^\# \alpha} p_{\gamma \rightarrow \alpha}
(\underline{x}_{\alpha\#\gamma}^\alpha) .
\end{equation}

We regard $\Delta_{(\mu, \nu)}$ as small quantities and
expand the edge-product of (\ref{eq:Z-exp-4}), and finally obtain
\begin{equation}\label{eq:Zexpan}
    Z(\beta) = Z_0 \biggl[    1+ \sum\limits_{r\subseteq R^\#} L_r
    \biggr]  ,
\end{equation}
where $r$ denotes any subgraph of the region graph $R^\#$, which contains a
subset of the edges of $R^\#$ and the associated regions. The correction
contribution $L_r$ of a region subgraph $r$ is expressed as
\begin{equation}\label{eq:correction}
   L_r =  \sum\limits_{\{\underline{x}_\gamma | \gamma \in r\}}
    \prod\limits_{\gamma \in r} w_\gamma(\underline{x}_\gamma)
     \prod_{(\mu, \nu) \in r}
  \Delta_{(\mu,\nu)}(\underline{x}_{\mu\#\nu}^\mu,
     \underline{x}_{\nu\#\mu}^\nu) .
\end{equation}

\subsection{Region graph belief-propagation equation}
\label{sec:rgbp}

Many of the subgraphs of the region graph $R^\#$ contain dangling edges.
A dangling edge $(\mu, \nu)$ in a subgraph $r$ is such an edge that if it is
cut off from subgraph $r$, either region $\mu$ or region $\nu$ (or both) will
be an isolated region of $r$. For a dangling edge $(\mu,\nu)$ of subgraph $r$,
suppose the region $\mu$ has only this edge attached to
it. The correction contribution $L_r$ of subgraph $r$ is calculated to be
\begin{eqnarray}\label{eq:cdangle}
L_r &=& \sum\limits_{\{\underline{x}_\alpha |
\alpha \in r \backslash \mu \}}
  \prod\limits_{\alpha \in r \backslash \mu}
w_\alpha(\underline{x}_\alpha)
    \prod_{(\eta, \lambda) \in r\backslash (\mu,\nu)} \Delta_{(\eta,\lambda)}
    (\underline{x}_{\eta\#\lambda}^\eta,
\underline{x}_{\lambda\#\eta}^\lambda)
      \nonumber \\
& &    \times \biggl\{
    \frac{\hat{p}_{\mu\rightarrow \nu}(\underline{x}_{\nu\#\mu}^\nu)
    \sum_{\underline{x}_{\mu\#\nu}^\mu} 
\sum_{\underline{x}_{\nu\#\mu}^{\nu}} \prod_{i\in \mu\#\nu}
 \delta(x_i^\mu, x_i^\nu)
 p_{\nu\rightarrow \mu}(\underline{x}_{\mu\#\nu}^\mu)
    p_{\mu\rightarrow \nu}(\underline{x}_{\nu\#\mu}^\nu)}
     {p_{\mu\rightarrow \nu}(\underline{x}_{\nu\#\mu}^\nu)
     \sum_{\underline{x}_{\mu\#\nu}^\mu}
      \sum_{\underline{x}_{\nu\#\mu}^\nu}
 \prod_{i\in \mu\#\nu}  \delta(x_i^\mu, x_i^\nu)
 p_{\nu\rightarrow \mu}(\underline{x}_{\mu\#\nu}^\mu)
    \hat{p}_{\mu\rightarrow \nu}(\underline{x}_{\nu\#\mu}^\nu)}
    -1
    \biggr\} ,
\end{eqnarray}
where $\hat{p}_{\mu\rightarrow \nu}(\underline{x}_{\nu\#\nu}^\nu)$ is a
probability distribution function determined by
\begin{eqnarray}
 \hat{p}_{\mu\rightarrow \nu}(\underline{x}_{\nu\#\mu}^\nu)
    & \equiv & B_{\mu\rightarrow \nu}\bigl(
    \{p_{\gamma\rightarrow \mu} | \gamma \in \partial^\# \mu\backslash \nu\}
  \bigr)
    \label{eq:rgBmn}\\
 &=&    \frac{
    \sum\limits_{\underline{x}_\mu} \Psi_\mu(\underline{x}_\mu)
    \prod\limits_{\gamma\in \partial^\# \mu\backslash \nu}
    p_{\gamma\rightarrow \mu}
    (\underline{x}_{\mu\#\gamma}^\mu) \prod\limits_{i\in \mu\#\nu}
     \delta(x_i^\mu, x_i^\nu)
    }
    {
    \sum\limits_{\underline{x}_\mu}
    \Psi_\mu(\underline{x}_\mu)
    \prod\limits_{\gamma \in \partial^\# \mu\backslash \nu}
    p_{\gamma\rightarrow \mu}
    (\underline{x}_{\mu\#\gamma}^\mu)} . \label{eq:hatrho}
\end{eqnarray}
In the above expressions, $\partial^\# \mu \backslash \nu$ denotes the set of
nearest-neighboring regions of $\mu$ in the region graph $R^\#$, but with
$\nu$ being removed from this set.

From (\ref{eq:cdangle}) we know that, if the auxiliary probability
distributions $\{p_{\mu\rightarrow \nu}, p_{\nu\rightarrow \mu}\}$ are chosen
as a fixed-point of the following equation,
\begin{equation}
\label{eq:rgbp}
p_{\mu\rightarrow \nu}(\underline{x}_{\nu\#\mu}^\nu)
 =    \frac{
    \sum\limits_{\underline{x}_\mu} \Psi_\mu(\underline{x}_\mu)
    \prod\limits_{\gamma\in \partial^\# \mu\backslash \nu}
   p_{\gamma\rightarrow \mu}
    (\underline{x}_{\mu\#\gamma}^\mu)
    \prod\limits_{i\in \mu\#\nu} \delta(x_i^\mu, x_i^\nu)
    }
    {
    \sum\limits_{\underline{x}_\mu}
    \Psi_\mu(\underline{x}_\mu)
    \prod\limits_{\gamma \in \partial^\# \mu\backslash \nu}
       p_{\gamma\rightarrow \mu}
    (\underline{x}_{\mu\#\gamma}^\mu)} ,
\end{equation}
then all the subgraphs $r$ of the region graph $R^\#$ with at least one
dangling edges will have zero correction contribution. Then the free energy
of the system is expressed as
\begin{equation} \label{eq:feexpan}
F(\beta) = F_{0} -\frac{1}{\beta}
\ln \Bigl[1+ \sum\limits_{r^{loop} \subseteq R^\#}
L_{r^{loop}} \Bigr] ,
\end{equation}
where $r^{loop}$ denotes a subgraph of $R^\#$ that is free of dangling edges
(in $r^{loop}$, each region has at least two edges attached).
Equation (\ref{eq:rgbp}) is referred to as the region graph belief-propagation
(rgBP) equation. It ensures that all the subgraphs of $R^\#$ with one or more
dangling edges have zero correction contribution to the partition function
$Z(\beta)$, therefore greatly reduces the number of terms in the free energy
correction expression.

It is not difficult to check that, the expression of the rgBP equation
(\ref{eq:rgbp}) for the region graph $R^\#$ shown in 
the right panel of figure~\ref{fig:reduntrgfig} is equivalent to that 
of the rgBP equation for the non-redundant region graph shown in
figure~\ref{fig:rgfig}. In this sense, $R^\#$ of figure~\ref{fig:reduntrgfig}
(right panel) is equivalent to $R$ of figure~\ref{fig:rgfig}.

\subsection{Region graph Bethe-Peierls free energy $F_0$}

When all the loop correction contributions in (\ref{eq:feexpan}) are neglected,
the remaining term $F_0$ is an approximation to the true free energy
$F(\beta)$. We refer to $F_0$ as the {\em region graph Bethe-Peierls free energy}, its
expression is
\begin{equation}
\label{eq:f0}
F_{0} = \sum\limits_{\alpha \in R} f_\alpha - \sum\limits_{(\mu,\nu)\in R^\#}
f_{(\mu, \nu)} ,
\end{equation}
with
\begin{equation}
\label{eq:fvande}
f_{\alpha}  =    -\frac{1}{\beta}
\ln \biggl[ \sum\limits_{\underline{x}_\alpha}
\Psi_\alpha(\underline{x}_\alpha) \prod\limits_{\gamma \in \partial^\# \alpha}
 p_{\gamma\rightarrow \alpha}(\underline{x}_{\alpha\#\gamma}^\alpha)
\biggr] , \quad
f_{(\mu,\nu)}  =  -\frac{1}{\beta}
\ln \biggl[\sum\limits_{\underline{x}_{\mu\#\nu}^\mu}
     p_{\mu\rightarrow \nu}(\underline{x}_{\mu\#\nu}^\mu)
     p_{\nu \rightarrow \mu}(\underline{x}_{\mu\#\nu}^\mu)
\biggr] .
\end{equation}

The free energy $F_0$ can also be regarded as a functional of the probability
functions $\{p_{\mu\rightarrow \nu}, p_{\nu\rightarrow \mu}\}$. A nice
property of this functional is that, the partial derivative of $F_0$ with
respective to any probability function $p_{\mu\rightarrow \nu}$ vanishes
at a fixed point of the rgBP equation (\ref{eq:rgbp}). This can easily be
checked by showing that
\begin{eqnarray}
\frac{\delta ( \beta F_{0})}{\delta p_{\mu\rightarrow \nu}(
\underline{x}_{\nu\#\mu}^{\nu})}
= \frac{p_{\nu\rightarrow \mu}(\underline{x}_{\nu\#\mu}^{\nu})}{
\sum\limits_{\underline{x}_{\mu\#\nu}^{\mu}}
p_{\mu\rightarrow \nu}(\underline{x}_{\mu\#\nu}^{\mu})
p_{\nu\rightarrow \mu}(\underline{x}_{\mu\#\nu}^{\mu})}
- \frac{\hat{p}_{\nu\rightarrow \mu}(\underline{x}_{\nu\#\mu}^{\nu})}{
\sum\limits_{\underline{x}_{\mu\#\nu}^{\mu}}
p_{\mu\rightarrow \nu}(\underline{x}_{\mu\#\nu}^{\mu})
\hat{p}_{\nu\rightarrow \mu}(\underline{x}_{\mu\#\nu}^{\mu})
} ,
\end{eqnarray}
where the probability function $\hat{p}_{\nu\rightarrow \mu}$ has been
defined through (\ref{eq:hatrho}).

From the variational property of $F_0$, we know that each fixed point of the
rgBP equation (\ref{eq:rgbp}) locates a stationary point of the functional
$F_0$ and {\em vice versa}, namely there is a one-to-one correspondence
between the stationary points of the $F_0$ functional and the fixed points of
the rgBP equation. A stationary point of $F_0$ could be a minimum, it could
also be a maximum, or be a saddle point.

\begin{figure}
\begin{center}
\includegraphics[width=0.7\textwidth]{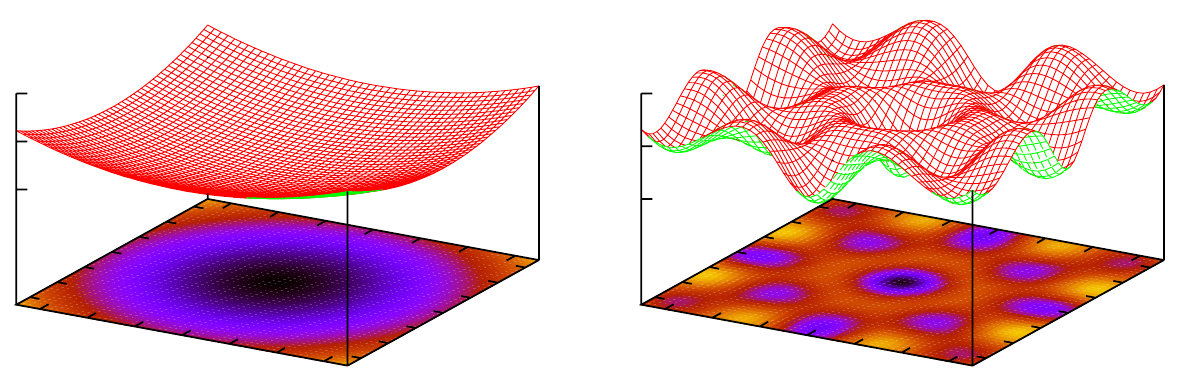}
\end{center}
\caption{\label{fig:freelandscape}
The region graph Bethe-Peierls free energy $F_0$ as a functional of
probability functions may have only a single stationary point (left panel),
or it may have multiple stationary points including minima, maxima, and saddles
(right panel). This qualitative change in functional shape is induced by
variation in environmental temperature or other control parameters.
}
\end{figure}

Using the set of probability
functions $\{p_{\mu\rightarrow \nu}, p_{\nu\rightarrow \mu}\}$ as order
parameters, the functional $F_0$ gives an approximate description of
the free energy landscape of model
(\ref{eq:totalenergy}).
If $F_0$ has only one stationary point (expected to be a minimum), then
the rgBP equation  has only one fixed point (solution). This simple situation
can be referred to as the `replica-symmetric' (RS) case, in connection to the
physics approaches based on the replica method and the cavity method
\cite{Mezard-etal-1987,Mezard-Montanari-2009}. In many non-trivial problems,
however, the $F_0$ functional may turn out to be very rugged in shape and have
many stationary points, then (\ref{eq:rgbp}) has multiple solutions. This
later complex situation can be referred to as the `replica-symmetry-broken'
(RSB) case. We will discuss this RSB case in the next section
\ref{sec:rsb}. A schematic illustration of the qualitative change in shape of
$F_0$ is shown in Fig.~\ref{fig:freelandscape}.

The free energy $F_0$ can be expressed as a sum over the contributions of all
the regions. For any region $\alpha$ of the region graph $R$, we have the
identity $c_\alpha + \sum_{\gamma > \alpha} c_\gamma = 1$. Then $F_0$ can be
re-written as
\begin{eqnarray}
F_0 & = & \sum\limits_{\alpha} \Bigl[c_\alpha+ \sum_{\gamma > \alpha}
c_\gamma\Bigr] f_\alpha -\sum\limits_{(\mu\rightarrow \nu) \in R^\#}
\Bigl[c_\mu + \sum_{\alpha>\mu} c_\alpha \Bigr] f_{(\mu,\nu)}
\nonumber \\
&=& \sum\limits_{\alpha} c_\alpha \tilde{F}_\alpha .
\label{eq:f0cluster}
\end{eqnarray}
Each region $\alpha$ contributes a term $c_\alpha \tilde{F}_\alpha$ to $F_0$.
The expression of $\tilde{F}_\alpha$ is
\begin{equation}
\label{eq:fa3}
\tilde{F}_\alpha = f_\alpha - \sum\limits_{(\alpha\rightarrow \gamma)
\in R^\#} f_{(\alpha,\gamma)} + \sum\limits_{\mu < \alpha}
\biggl[ f_\mu - \sum\limits_{(\mu\rightarrow \nu)\in R^\#}
f_{(\mu,\nu)} \biggr] .
\end{equation}
Notice that (\ref{eq:fa3}) has the same expression as (\ref{eq:f0}), the only
difference is that the summations are restricted to the subgraph containing
region $\alpha$ and all its descendants. Therefore, $\tilde{F}_\alpha$ can be
regarded as the region graph Bethe-Peierls free energy of this subgraph
$\alpha$.

\subsection{Factor graph as the simplest region graph}

It has been demonstrated in \cite{Xiao-Zhou-2011} that
the conventional belief-propagation (BP) equation can be derived from the
framework of partition function expansion.
Actually BP is just a limiting case of the more general
rgBP equation.

Given a factor graph $G$, let us consider the following simplest
region graph $R$: there are $N$ `variable' regions denoted by
$i=1,2,\ldots, N$, each of which contains a single variable node $i$, with
counting number $c_i = 1- k_i$ ($k_i\equiv |\partial i|$ being the degree of
node $i$ in $G$); and there are $M$ `function' regions denoted by
$a=1,2,\ldots,M$, each of which contains a function node $a$ and the set
$\partial a$ of all the nearest-neighboring variable nodes of node $a$, with
counting number $c_a = 1$. Such a simple region graph is
non-redundant, although it in general contains many loops.

Let us  define two new probability functions $b_{i\rightarrow a}(x_i)$ and
$b_{a\rightarrow i}(x_i)$ as
\begin{equation}
b_{i\rightarrow a}(x_i) \equiv \frac{ \psi_i(x_i) p_{i\rightarrow a}(x_i)}{
\sum_{x} \psi_i(x) p_{i\rightarrow a}(x)} , \quad\quad
b_{a\rightarrow i}(x_i) \equiv \frac{ \psi_i^{-1}(x_i)
 p_{a\rightarrow i}(x_i)}{
\sum_{x} \psi_i^{-1}(x) p_{a\rightarrow i}(x)}  .
\end{equation}
Using the rgBP equation (\ref{eq:rgbp}), one can easily check that the newly
defined probability functions satisfy the conventional BP equation:
\begin{equation}
\label{eq:fgbp}
b_{i\rightarrow a}(x_i) =\frac{\psi_i(x_i)
\prod\limits_{c\in \partial i\backslash a}
 b_{c\rightarrow i}(x_i)}{ \sum\limits_{x}
\psi_i(x) \prod\limits_{c\in \partial i\backslash a} b_{c\rightarrow i}(x)
} ,
\quad\quad
b_{a\rightarrow i}(x_i) =
\frac{\sum\limits_{\underline{x}_{\partial a}\backslash x_i}
\psi_a(\underline{x}_{\partial a}) \prod\limits_{j\in \partial a\backslash i}
 b_{j\rightarrow a}(x_j)}{
\sum\limits_{\underline{x}_{\partial a}}
\psi_a(\underline{x}_{\partial a}) \prod\limits_{j\in \partial a\backslash i}
b_{j\rightarrow a}(x_j)}.
\end{equation}

The probability $b_{i\rightarrow a}(x_i)$ can be interpreted as the
state distribution of variable node $i$ in the
absence its interaction with function node $a$;
while the probability $b_{a\rightarrow i}(x_i)$ can be interpreted as
the state distribution of $i$ if it only interacts  with function
node $a$.

The free energy $F_0$ for this simplest
region graph reduces to the conventional Bethe-Peierls free energy. Its
expression is
$F_0 = \sum_{a\in G} \tilde{F}_a - \sum_{i\in G}(k_i -1) \tilde{F}_i$.
The free energies $\tilde{F}_i$ (for a variable node $i$) and
$\tilde{F}_a$ (for a function node $a$) are expressed as
\begin{eqnarray}
\tilde{F}_i =
-\frac{1}{\beta} \ln \biggl[\sum_{x_i} \psi_i(x_i)
\prod_{a\in \partial i} b_{a\rightarrow i}(x_i) \biggr] , \quad
 \tilde{F}_a  = -\frac{1}{\beta}\ln \biggl[
\sum_{\underline{x}_{\partial a}} \psi_a(\underline{x}_{\partial a})
\prod_{i\in \partial a} \Bigl[ \psi_i(x_i)
 \prod_{c\in \partial i\backslash a} b_{c\rightarrow i}(x_i)
\Bigr] \biggr] .
\end{eqnarray}
Only the messages $b_{a\rightarrow i}(x_i)$ from the  function regions
(parents) to the variable regions (children) appear in the above two
expressions. We now proceed to demonstrate that, such a property holds for a
general non-redundant region graph.

\subsection{Parent-to-child message-passing in a non-redundant region graph}

In a non-redundant region graph $R$, the subgraph $R_i$ induced by any
variable node $i$ is a connected tree. Because of this property,  $R^\#$ is
identical to $R$ and the cross-linker set for each directed edge
$\mu\rightarrow \nu$ contains all the variable nodes in $\nu$
($\mu\#\nu= \{i | i\in \nu\}$). It is easy to see that, if there is a
directed path pointing from a region $\alpha$ to another region $\gamma$,
then such a path must be unique in $R$; furthermore, the subgraph $R_a$
induced by any function node $a$ is also a connected tree.

When we approximate the free energy $F(\beta)$ by the
region graph Bethe-Peierls free
energy $F_0$, the corresponding approximate expression for the marginal
configuration distribution of a region $\gamma$ is
\begin{equation}
\label{eq:pgamma}
p_\gamma (\underline{x}_\gamma) \propto
\Psi_\gamma (\underline{x}_\gamma)
\prod\limits_{\alpha \in \partial^\# \gamma}
p_{\alpha \rightarrow \gamma}(\underline{x}_{\gamma\#\alpha}^\gamma) .
\end{equation}
This marginal distribution has the following consistency property:
if $(\mu, \nu)$ is an edge of the region graph $R$, then
\begin{equation}
\label{eq:mncons}
\sum\limits_{\underline{x}_\mu \backslash \underline{x}_{\mu\#\nu}}
p_{\mu}(\underline{x}_\mu)
 \ = \
\sum\limits_{\underline{x}_\nu \backslash \underline{x}_{\mu\#\nu}}
p_{\nu}(\underline{x}_\nu) ,
\end{equation}
where $\underline{x}_{\mu \# \nu} \equiv \{ x_i | i \in \mu\#\nu\}$ is
the configuration of the variables in the cross-linker set $\mu\#\nu$.
Equation (\ref{eq:mncons}) ensures that the marginal configuration
distribution of the cross-linker set $\mu\#\nu$  is the same
whether it is inferred from $p_{\mu}(\underline{x}_\mu)$ or from
$p_\nu (\underline{x}_\nu)$.

\begin{figure}
\begin{center}
\includegraphics[width=0.35\textwidth]{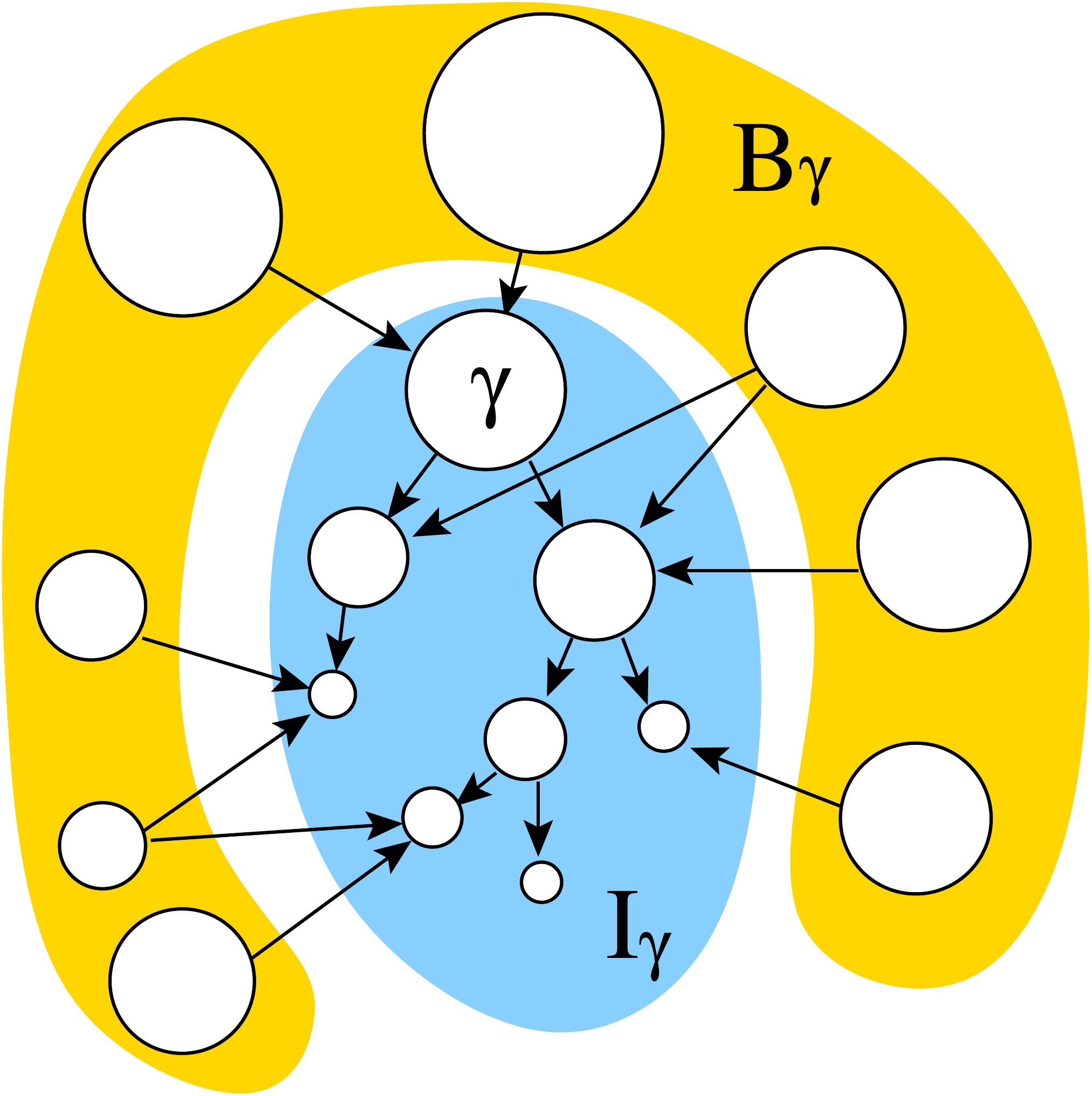}
\end{center}
\caption{\label{fig:rgptoc}
For any region $\gamma$ in a non-redundant region graph $R$, the set
$I_\gamma$ is formed by $\gamma$ and all its
descendant regions, and the set $B_\gamma$ is formed by all the other regions
that do not belong to $I_\gamma$ but are
pointing to regions of $I_\gamma$ by one or more edges. Each white
disk in this figure represents a region.
}
\end{figure}

When $R$ is non-redundant, we have
$\partial^\# \gamma =\partial \gamma$ in (\ref{eq:pgamma}), where
$\partial \gamma$ is just the
 set of nearest-neighboring regions of region $\gamma$. Then, with the
help of (\ref{eq:rgbp}), the expression (\ref{eq:pgamma}) is rewritten as
\begin{equation}
\label{eq:pgamma2}
p_\gamma (\underline{x}_\gamma ) \propto
\prod\limits_{\eta\in I_\gamma} \Psi_\eta (\underline{x}_\eta)
\prod\limits_{\{(\mu\rightarrow \nu) | \mu\in B_\gamma, \nu\in I_\gamma\} }
p_{\mu\rightarrow \nu} (\underline{x}_\nu) .
\end{equation}
In this expression, the region set $I_\gamma$ contains $\gamma$ and all its
descendants, i.e., $I_\gamma \equiv \{ \eta | \eta \leq \gamma\}$; $B_\gamma$
is another region set with the property that any region in $B_\gamma$ does
not belong to $I_\gamma$ but is parental to at least
one region in $I_\gamma$, see figure~\ref{fig:rgptoc}.
The  configuration $\underline{x}_\nu$ for any region
$\nu \in I_\gamma$ should be understood as
$\underline{x}_\nu \equiv \{x_i^\gamma | i\in \nu\}$.

For a non-redundant region graph $R$, we can prove (see the
appendix~\ref{app:proof}) that, the marginal probability distribution
(\ref{eq:pgamma2}) is equivalent to
\begin{equation}
\label{eq:pgamma4}
p_{\gamma}(\underline{x}_\gamma) \propto \biggl[
\prod\limits_{a\in \gamma} \psi_a(\underline{x}_{\partial a})
\prod\limits_{i\in \gamma} \psi_i(x_i) \biggr]
\prod\limits_{\{(\mu\rightarrow \nu) | \mu\in B_\gamma, \nu\in I_\gamma\}}
 m_{\mu\rightarrow \nu}(\underline{x}_\nu) ,
\end{equation}
where $m_{\mu\rightarrow \nu}(\underline{x}_\nu)$ is a probability
distribution that satisfies the self-consistent
equation (\ref{eq:miteration}). The message
$m_{\mu\rightarrow \nu}(\underline{x}_\nu)$ is interpreted as
the (cavity) probability that the variable set $\{i \in \nu \}$ takes
configuration $\underline{x}_\nu$ in  region $\mu$
when all the interactions that also appear in region $\nu$ are not considered
(namely, setting $\psi_i(x_i) = 1$ for any $i\in \nu$ and
$\psi_a(\underline{x}_{\partial a})=1$ for any $a\in \nu$,
while all the other interactions of region $\mu$ are not modified).
The marginal probability expression
(\ref{eq:pgamma4}) has the nice property
that only the parent-to-child messages $m_{\mu\rightarrow \nu}$ are needed,
but not the messages from children to parents.

The self-consistent equation for
$m_{\mu\rightarrow  \nu}(\underline{x}_\nu)$, derived from
(\ref{eq:rgbp}) after a lengthy process (see
Appendix \ref{app:proof}), is intuitively simple:
\begin{equation}
m_{\mu\rightarrow \nu}(\underline{x}_\nu) \propto
\sum\limits_{\underline{x}_\mu \backslash \underline{x}_\nu}
\biggl[
\prod\limits_{a\in \mu\backslash \nu} \psi_a(\underline{x}_{\partial a})
\prod\limits_{i\in \mu\backslash \nu} \psi_i(x_i)
\biggr]
\prod\limits_{\{(\alpha\rightarrow \eta) | \alpha\in B_\mu, \eta
 \in I_\mu\backslash I_\nu\}}
m_{\alpha\rightarrow \eta}(\underline{x}_\eta) .
\label{eq:miteration}
\end{equation}

For a non-redundant region graph $R$, the marginal probability distributions
$p_\gamma(\underline{x}_\gamma)$ should satisfy the following
edge-consistency condition, namely
\begin{equation}
\label{eq:rgconsistency2}
\sum\limits_{\underline{x}_\mu \backslash \underline{x}_\nu}  p_\mu (
\underline{x}_{\mu})  = p_\nu (\underline{x}_{\nu}) ,
\quad\quad \forall (\mu\rightarrow \nu) \in R .
\end{equation}
To check this consistency, plug (\ref{eq:pgamma4}) into the two sides
of the  above equation,
we get the follow equation that must be satisfied by $m_{\mu\rightarrow \nu}$:
\begin{equation}
m_{\mu\rightarrow \nu} (\underline{x}_\nu) =
C
\frac{\prod\limits_{\underline{x}_\mu \backslash \underline{x}_\nu}
\prod\limits_{a\in \mu\backslash \nu} \psi_a(\underline{x}_{\partial a})
\prod\limits_{i\in \mu\backslash \nu} \psi_i(x_i)
\prod\limits_{\{(\alpha\rightarrow \eta ) | \alpha \in B_\mu, \eta
 \in I_\mu \backslash I_\nu\}}
m_{\alpha \rightarrow \eta}(\underline{x}_\eta)}
{\prod\limits_{\{\eta \rightarrow \lambda | \eta
\in (B_\nu \cap I_\mu)\backslash \mu,
\lambda \in I_\nu \backslash \nu\}}
m_{\eta\rightarrow \lambda}(\underline{x}_\lambda)} ,
\label{eq:ptoceq}
\end{equation}
where the normalization constant $C$ is determined by
$\sum_{\underline{x}_\nu} m_{\mu\rightarrow \nu}(\underline{x}_\nu) = 1$.
For a non-redundant graph $R$, the set
$(B_\nu\cap I_\mu)\backslash \mu$ is empty, and hence the denominator of
 the expression on the right side
of (\ref{eq:ptoceq}) is equal to $1$. Since (\ref{eq:ptoceq})
is equivalent to (\ref{eq:miteration}),
the edge-consistency condition (\ref{eq:rgconsistency2}) holds.

If the region graph $R$ is redundant, the marginal probability distribution
of each region $\gamma$ can no longer
be written in the form of (\ref{eq:pgamma4}), and the edge-consistency
condition (\ref{eq:rgconsistency2}) is
no longer valid (but the weaker condition (\ref{eq:mncons}) is
still valid). In \cite{Yedidia-Freeman-Weiss-2005}, however, Yedidia and
co-authors
{\em defined} that a region $\gamma$ has a marginal probability distribution
(\ref{eq:pgamma4}), and then they required that this marginal probability
distribution should satisfy the edge-consistency condition
(\ref{eq:rgconsistency2}), which leads to the self-consistent equation
(\ref{eq:ptoceq}) for $m_{\mu\rightarrow \nu}(\underline{x}_\nu)$.
Such a treatment might be able to get good results for some problems, but in
our opinion it lacks a solid theoretical foundation.

For a non-redundant graph $R$, the free energy $\tilde{F}_\alpha$
(\ref{eq:fa3}) can also be expressed by parent-to-child messages (see
Appendix \ref{app:fgbp} for details). The final expression is
\begin{equation}
\tilde{F}_\alpha = -\frac{1}{\beta} \ln \biggl[
\sum\limits_{\underline{x}_\alpha}
\prod\limits_{a\in \alpha} \psi_a(\underline{x}_{\partial a})
\prod\limits_{i\in \alpha} \psi_i(x_i)
\prod\limits_{\{(\mu\rightarrow \nu) |
\mu\in B_\alpha, \nu \in I_\alpha\}}
 m_{\mu\rightarrow \nu}(\underline{x}_{\nu})
\biggr] .
\label{eq:frgbp}
\end{equation}

The free energy expression (\ref{eq:frgbp}) also appeared in
\cite{Yedidia-Freeman-Weiss-2005} as a basic assumption (both for
redundant and non-redundant region graphs). A theoretical challenge
for redundant region graphs, worth to be explored in future work,
is to derive  from the framework of partition function
expansion  an expression for $\tilde{F}_\alpha$
using only parent-to-child messages.

\section{Region Graph Expansion for the Grand Partition Function}
\label{sec:rsb}

If the region graph Bethe-Peierls free energy functional (\ref{eq:f0})
has only a single stationary point, a global minimum, then the rgBP equation
(\ref{eq:rgbp}) has only a single solution (fixed point). Reaching this unique
fixed point is computationally not hard. One may perform message-passing
iteration along the edges of the region graph $R$, or one
may work on the functional $F_0$ and find its unique minimum by gradient
descend methods. The situation is a little bit more complicated, for example,
if the $F_0$ functional has a single minimum and one or more saddle
points. In this situation, we expect that the rgBP equation still has only
one stable fixed point, which can be reached by direct iteration
(with some damping to accelerate convergence) or by minimizing $F_0$.

The real non-trivial situation occurs when the $F_0$ functional has
multiple minimal points
(see the right panel of figure \ref{fig:freelandscape}), and consequently
the rgBP equation (\ref{eq:rgbp}) has multiple stable fixed points.
We take $F_0$ as an approximate free energy landscape for the
system under study. Then each minimal point of $F_0$ can be regarded as
a metastable state $s$ of the system, it describes approximately the
system's collective behavior within certain timescale $\tau_s$.
The timescale $\tau_s$ is determined by the shape of the free energy landscape
surrounding metastable state $s$,
especially the relative height of the closest free energy saddle point.
For some mean-field models (such as
spin glasses on random graphs
\cite{Mezard-Parisi-2001}) $\tau_s$ is expected to
 approach infinity in the thermodynamic
limit $N \rightarrow \infty$ (see, e.g.,
\cite{Montanari-Semerjian-2006,Mezard-Montanari-2006}),
then a metastable state $s$ is a true thermodynamic Gibbs state (i.e.,
ergodicity is broken at $N\rightarrow \infty$). For finite-dimensional
systems, the timescale $\tau_s$ of a metastable state $s$ may be
finite even in the thermodynamic limit.

In this section, we regard each fixed point of the rgBP equation
(\ref{eq:rgbp}) as a {\em macrostate}, no matter whether it is a minimum of
$F_0$ or a maximum or saddle point. Viewing each macrostate as an
energy level, it is natural to define a grand
partition function $\Xi$ as
\cite{Monasson-1995,Mezard-Parisi-2001,Zhou-2007b,Xiao-Zhou-2011}
\begin{equation}
\label{eq:xi}
\Xi(y; \beta) =  \sum_{s} \exp\bigl[ -y F_{0}^{(s)}(\beta)\bigr] ,
\end{equation}
where $F_0^{(s)}(\beta)$ is the value of $F_0(\beta)$ at the rgBP
fixed point $s$; the parameter $y$ is the inverse
temperature at the level of macrostates.\footnote{Although (\ref{eq:xi})
sums over all the stationary points of $F_0$, we expect that
at sufficiently large values of $y$ the grand partition function
will be dominantly contributed by the free energy minimal points.}
The grand partition function $\Xi(y; \beta)$ is associated with
a grand free energy $G(y; \beta)$ through
\begin{equation}
G(y; \beta) \equiv -\frac{1}{y} \ln \Xi(y; \beta) .
\end{equation}
At a given inverse temperature $y$, the statistical weight of a macrostate
$s$ is
\begin{equation}
\mathcal{P}_B(s) = \frac{\exp\bigl[-y F_0^{(s)}(\beta)\bigr]}
{\sum\limits_{s} \exp\bigl[-y F_0^{(s)}(\beta)\bigr]} .
\end{equation}
From $\mathcal{P}_B(s)$ the mean value of free energy is
expressed as $ \langle F_0(\beta) \rangle_y \equiv
\sum_s \mathcal{P}_B(s) F_0^{(s)}(\beta)$. And
the entropy density $\Sigma(y; \beta)$ of macrostates is expressed as
\begin{equation}
\Sigma(y; \beta) \equiv - \frac{1}{N} \sum\limits_{s} \mathcal{P}_B(s)
\ln \mathcal{P}_B(s) .
\end{equation}
$\Sigma(y; \beta)$ is also called the complexity
in the spin glass literature \cite{Mezard-Parisi-2001}.
It is easy to verify that $G(y;\beta) = \langle F_0(\beta) \rangle_y
- \frac{N}{y} \Sigma(y; \beta)$.

We are interested in obtaining an approximate expression for
the grand free energy $G(y; \beta)$. For this purpose
we follow again the framework of partition function expansion
\cite{Xiao-Zhou-2011}. First notice that
\begin{equation}
 \Xi(y; \beta) = \prod\limits_{(\mu,\nu)\in R^\#}
\int {\rm D} p_{\mu\rightarrow \nu} \int
{\rm D} p_{\nu\rightarrow \mu}
\delta\bigl( p_{\mu\rightarrow \nu}-B_{\mu\rightarrow \nu}
\bigr) \delta\bigl( p_{\nu\rightarrow \mu}-B_{\nu\rightarrow \mu} \bigr)
e^{-y F_{0}} ,
\label{eq:Xi}
\end{equation}
where $\int {\rm D} p_{\mu\rightarrow \nu}$ means integrating over all
possible probability measures
$p_{\mu\rightarrow \nu}(\underline{x}_{\nu\#\mu}^{\nu})$ on the
edge $(\mu,\nu)$ of the region graph $R^\#$.
Because of the Dirac function
$\delta\bigl(p_{\mu\rightarrow \nu}-B_{\mu\rightarrow \nu}
(\{p_{\gamma\rightarrow
\mu} | \gamma \in \partial^\# \mu\backslash \nu\} ) \bigr)$, only the fixed
points of (\ref{eq:rgbp}) contribute to the grand partition function.

We then introduce on each edge $(\mu, \nu)$ of the region graph $R^\#$
two probability measures
$P_{\mu\rightarrow \nu}(p_{\mu\rightarrow \nu})$ and
$P_{\nu\rightarrow \mu}(p_{\nu\rightarrow \mu})$.
Using the expression (\ref{eq:f0}) for the free energy $F_0$,
the grand partition function is re-written as
\begin{eqnarray}
\Xi(y; \beta) =\prod\limits_{\alpha \in R}
 \prod_{\gamma\in \partial^\# \alpha} \int {\rm D}
p_{\gamma\rightarrow \alpha}
P_{\gamma\rightarrow \alpha}(p_{\gamma\rightarrow \alpha}) e^{-y f_\alpha}
 \prod\limits_{(\mu,\nu)\in R^\#}
\biggl[
\frac{ \delta(p_{\mu\rightarrow \nu}-B_{\mu\rightarrow \nu}
) \delta(p_{\nu\rightarrow \mu}-B_{\nu\rightarrow \mu})}
{ e^{-y f_{(\mu,\nu)}} P_{\mu\rightarrow \nu}(p_{\mu\rightarrow \nu})
P_{\nu\rightarrow \mu}(p_{\nu\rightarrow \mu})}
\biggr] .
\label{eq:Xi-exp-2}
\end{eqnarray}
This expression is very similar in form to (\ref{eq:Z-exp-3}), therefore the
method of partition function expansion
can be directly applied to (\ref{eq:Xi-exp-2}). As a result we obtain that
\begin{equation} \label{eq:gfeexpan}
G(y; \beta) = G_{0}  - \frac{1}{y}
\ln \biggl[1+ \sum\limits_{r^{loop} \subseteq R^\#}
L_{r^{loop}}^{(1)} \biggr] .
\end{equation}

The first term on the right of (\ref{eq:gfeexpan}) is expressed as
\begin{equation} \label{eq:gsp}
 G_{0}(y;\beta) = \sum\limits_{\gamma\in R} g_\gamma
 - \sum\limits_{(\mu,\nu)\in R^\#} g_{(\mu,\nu)} ,
\end{equation}
where
\begin{subequations}
\begin{eqnarray}
 g_\gamma & = &  - \frac{1}{y} \ln\biggl[
\prod_{\alpha \in \partial^\# \gamma}
\int {\rm D} p_{\alpha \rightarrow \gamma}
P_{\alpha \rightarrow \gamma}(p_{\alpha\rightarrow \gamma}) e^{-y f_\gamma}
\biggr] ,  \\
g_{(\mu,\nu)} & = & - \frac{1}{y}\ln\biggl[
\int \int {\rm D} p_{\mu\rightarrow \nu} {\rm D} p_{\nu\rightarrow \mu}
P_{\mu\rightarrow \nu}(p_{\mu\rightarrow \nu})
P_{\nu\rightarrow \mu}(p_{\nu\rightarrow \mu})
e^{-y f_{(\mu, \nu)}}
\biggr] .
\end{eqnarray}
\end{subequations}
The grand free energy $G_0$ as expressed by (\ref{eq:gsp}) can also
be regarded as a functional of the probability functionals
$\{P_{\mu\rightarrow \nu}(p_{\mu\rightarrow \nu}),
P_{\nu\rightarrow \mu}(p_{\nu\rightarrow \mu})\}$.

The correction contribution $L_r^{(1)}$ of a subgraph to the grand free energy
has a similar expression as (\ref{eq:correction}) \cite{Xiao-Zhou-2011}.
To ensure that any subgraph with at least one dangling edge has vanishing
correction contribution to the grand free energy, each probability function
$P_{\mu\rightarrow \nu} (p_{\mu\rightarrow \nu})$
needs to satisfy the following equation
\begin{equation}
 P_{\mu\rightarrow \nu}(p_{\mu\rightarrow \nu})
 =
\frac{
\prod\limits_{\gamma\in \partial^\# \mu\backslash \nu} \int
{\rm D} p_{\gamma\rightarrow \mu}
P_{\gamma\rightarrow \mu}(p_{\gamma \rightarrow \mu})
 e^{-y f_{\mu\rightarrow \nu}}
 \delta\Bigl(p_{\mu\rightarrow \nu}-B_{\mu\rightarrow \nu}
\bigl(\{p_{\gamma\rightarrow
\mu} | \gamma \in \partial^\# \mu\backslash \nu\} \bigr) \Bigr)}
{
\prod_{\gamma\in \partial^\# \mu\backslash \nu} \int
{\rm D} p_{\gamma\rightarrow \mu}
P_{\gamma\rightarrow \mu}(p_{\gamma\rightarrow \mu})
 e^{-y f_{\mu\rightarrow \nu}} } ,
 \label{eq:SPfix}
\end{equation}
where
\begin{equation}
\label{eq:fmutonu}
f_{\mu\rightarrow \nu} \equiv
- \frac{1}{\beta} \ln\biggl[
\sum_{\underline{x}_\mu}  \Psi_\mu (\underline{x}_\mu)
      \prod_{\gamma \in \partial^\# \mu \backslash \nu}
p_{\gamma \rightarrow \mu}
     (\underline{x}_{\mu\#\gamma}^\mu) \biggr] .
\end{equation}

Equation (\ref{eq:SPfix}) is called the region graph
survey-propagation equation, in correspondence to the 1RSB mean-field
theory of spin glasses \cite{Mezard-Parisi-2001,Mezard-etal-2002}. If
we neglect the loop correction contributions to $G(y; \beta)$, then
the grand free energy functional $G_0$ gives an approximate description of
the system's free energy landscape at the level of macrostates. It can
again be verified that, the first variation of $G_0$ with respect to
any of its arguments $P_{\mu\rightarrow \nu}(p_{\mu\rightarrow \nu})$ is
identically zero at a fixed point of (\ref{eq:SPfix}).

If we neglected all the loop correction contributions to $G(y; \beta)$
and approximate it with $G_0$, then an approximate expression for
the mean free energy of macrostates is
\begin{eqnarray}
\langle F_0(\beta) \rangle_y  \approx  \frac{\partial ( y G_0)}{\partial y}
& = & \sum\limits_{\gamma \in R}
\frac{\prod_{\alpha \in \partial^\# \gamma}
\int {\rm D} p_{\alpha \rightarrow \gamma}
P_{\alpha \rightarrow \gamma}(p_{\alpha\rightarrow \gamma})
f_\gamma e^{-y f_\gamma}}
{\prod_{\alpha \in \partial^\# \gamma}
\int {\rm D} p_{\alpha \rightarrow \gamma}
P_{\alpha \rightarrow \gamma}(p_{\alpha\rightarrow \gamma})
e^{-y f_\gamma}} \nonumber
\\
& & - \sum\limits_{(\mu,\nu)\in R^\#}
\frac{\int \int {\rm D} p_{\mu\rightarrow \nu}
{\rm D} p_{\nu\rightarrow \mu}
P_{\mu\rightarrow \nu}(p_{\mu\rightarrow \nu})
P_{\nu\rightarrow \mu}(p_{\nu\rightarrow \mu})
f_{(\mu,\nu)} e^{-y f_{(\mu, \nu)}}}
{\int \int {\rm D} p_{\mu\rightarrow \nu} {\rm D} p_{\nu\rightarrow \mu}
P_{\mu\rightarrow \nu}(p_{\mu\rightarrow \nu})
P_{\nu\rightarrow \mu}(p_{\nu\rightarrow \mu})
e^{-y f_{(\mu, \nu)}}} .
\label{eq:mf0app}
\end{eqnarray}
And the complexity $\Sigma(y; \beta)$ is expressed as
\begin{equation}
\Sigma(y; \beta)  \approx  \frac{ y \bigl( \langle F_0(\beta) \rangle_y
- G_0)}{N} .
\end{equation}
It is particularly interesting to determine the complexity value at
$y=\beta$, and to determine the maximal value of $y$ at which the
complexity becomes negative \cite{Mezard-Montanari-2009}.

It is not a simple task to solve numerically the
region graph survey-propagation equation (\ref{eq:SPfix}).
A direct approach is to iterate (\ref{eq:SPfix}) on the
region graph, with each
probability functional $P_{\mu\rightarrow \nu}(p_{\mu\rightarrow \nu})$
represented by a sample set of probability functions
$p_{\mu\rightarrow \nu}(\underline{x}_{\nu\#\mu}^{\nu})$.
A major complication is the reweighting factor
$e^{-y f_{\mu\rightarrow \nu}}$ in (\ref{eq:SPfix}). In the special situation
of $y=\beta$, the reweighting can be replaced by introducing new
auxiliary probability functions
\cite{Mezard-Montanari-2006,Krzakala-etal-PNAS-2007}. Several reweighting
tricks were discussed in the thesis of Zdeborov{\'{a}}
\cite{Zdeborova-2009}, which may be helpful for solving the
region-graph survey-propagation equation for general $y$ values.

The region-graph survey-propagation equation (\ref{eq:SPfix}) applies to
a general redundant or non-redundant region graph $R$.
Rizzo and co-authors \cite{Rizzo-etal-2010} derived a set of
generalized survey-propagation equations using the replica cluster
variation method, which are different from the region-graph
survey-propagation equation (\ref{eq:SPfix}). 
The connection between these two approaches needs to be further
studied.

\section{Numerical Results on the Two-Dimensional Ising
and  Edwards-Anderson Models}
\label{sec:2d}

We now apply the region graph belief-propagation equation to the 2D
Ising and Edwards-Anderson models, mainly for
the purpose of testing its performance.
Both models are described by the
energy function (\ref{eq:2DEA}). For the Ising model,
all edge coupling constants
$J_{i j}$ are  equal to a positive value $J$; for the EA model,
$J_{i j}$ are independent and identically distributed random
variables, taking value $J$ and $-J$ with equal probability.
All the external fields $h_i^0$ in (\ref{eq:2DEA})
are set to be zero for simplicity. In the numerical calculations,
$J$ and the Boltzmann constant $k_B$ are both set to be unity, so the
energy unit is $J$ and the temperature unit is $J/k_B$.

We consider $L\times L$ square lattices with $L$ coupling interactions in each
dimension. If we assume periodic boundary condition on both dimensions,
the total number of vertices (variable nodes) in the lattice is 
$N=L^2$; if open
boundary condition is used on both directions, the total
number of vertices is $N=(L+1)^2$.

The region graph for the square lattice is constructed to be non-redundant.
It has three types of regions, the `square' regions, the `stripe' regions,
and the `rod' regions. Each square region contains $n\times n$
vertices, each stripe region contains $n \times 2$ vertices,
and each rod region contains $n\times 1$ vertices.
All the coupling interactions between the vertices of a given region are
also included into this region.
Each square region is connected to four rod regions, each stripe region is
connected to two rod regions, and each rod region is connected to a square
region and a stripe region.
The region graph $R$ at $n=2$ is shown in figure~\ref{fig:rgfig}, where
a square region and a stripe region have the same shape.
For the general case
of $n\geq 2$, if we regard each square region as a
`giant vertex' and each stripe region plus its two connected rod regions
as  a `giant bond', then the region graph is again a
square lattice of giant vertices and giant bonds.
In this sense, our region graph
representation is a coarse-graining of the original lattice
that keeps its topology unchanged.

The partition function of the 2D Edwards-Anderson can be calculated
exactly by polynomial algorithms (see, for example,
Ref.~\cite{Thomas-Huse-Middleton-2011} and references therein).
Although the heuristic rgBP message-passing approach can only
obtain an approximate value for the free energy, it is very efficient
in estimating all the $N$ single-variable marginal probabilities
simultaneously.

\subsection{Region graph belief-propagation equations at $n=2$}

\begin{figure}
\begin{center}
\includegraphics[width=0.6\textwidth]{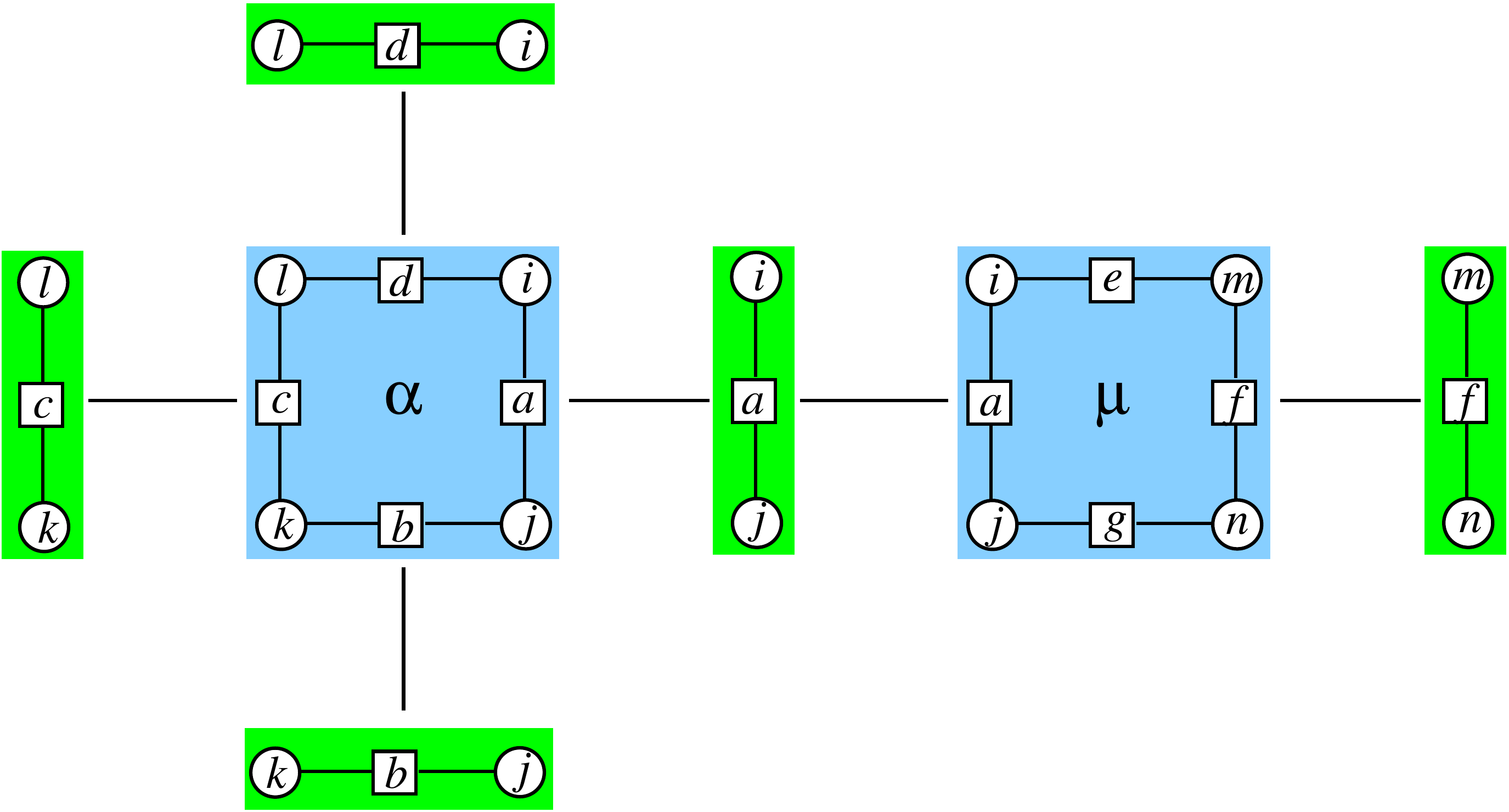}
\end{center}
\caption{ \label{fig:rgbp2D}
A local part of the region graph $R$ shown in
figure \ref{fig:rgfig}. The counting numbers
for a square region $\alpha$, a stripe region $\mu$, and a
rod region $a$ are, respectively, $c_\alpha=1$, $c_\mu=1$, $c_a=-1$.}
\end{figure}

A subgraph of the region graph $R$ at $n=2$
is plotted in figure \ref{fig:rgbp2D}. For notational simplicity we
denote the square regions and the stripe regions by Greek symbols
(such as $\alpha$ and $\mu$) and denote a rod region just by the index of
its function node (such as $a$ and $c$).
Consider a square region $\alpha$ and a rod region $a$ in
figure \ref{fig:rgbp2D}. The two probability distributions
between these regions can be parameterized as
%
\begin{subequations}
\begin{eqnarray}
p_{\alpha\rightarrow a}(\sigma_i, \sigma_j)
& \ \propto \ &  \exp\bigl( \beta h_{\alpha\rightarrow a}^{(i)} \sigma_i
+ \beta h_{\alpha\rightarrow a}^{(j)} \sigma_j
+ \beta J_{\alpha\rightarrow a}^{(i j)} \sigma_i \sigma_j\bigr)  ,
\label{eq:p11} \\
p_{a\rightarrow \alpha}(\sigma_i, \sigma_j)
& \ \propto \ & \exp\bigl( \beta h_{a\rightarrow \alpha}^{(i)} \sigma_i
+  \beta h_{a \rightarrow \alpha}^{(j)} \sigma_j
+ \beta J_{a \rightarrow \alpha}^{(i j)} \sigma_i \sigma_j \bigr) .
\label{eq:p12}
\end{eqnarray}
\end{subequations}
Similarly the two probability distributions between a stripe region $\mu$
and a rod region $a$ are expressed as
\begin{subequations}
\begin{eqnarray}
p_{\mu\rightarrow a}(\sigma_i, \sigma_j)
& \ \propto \ &  \exp\bigl( \beta h_{\mu\rightarrow a}^{(i)} \sigma_i
+ \beta h_{\mu\rightarrow a}^{(j)} \sigma_j +
\beta J_{\mu\rightarrow a}^{(i j)} \sigma_i \sigma_j\bigr)  ,
\label{eq:p21} \\
p_{a\rightarrow \mu}(\sigma_i, \sigma_j)
& \ \propto \ &
\exp\bigl( \beta h_{a\rightarrow \mu}^{(i)} \sigma_i
+  \beta h_{a \rightarrow \mu}^{(j)} \sigma_j +
\beta J_{a \rightarrow \mu}^{(i j)} \sigma_i \sigma_j \bigr) .
\label{eq:p22}
\end{eqnarray}
\end{subequations}
The self-consistent equations for these set of parameters
are derived from the rgBP equation (\ref{eq:rgbp}) and listed in
appendix \ref{app:rg}.

A trivial solution of the rgBP equation  is
the paramagnetic one with the fields in
the expressions (\ref{eq:p11})--(\ref{eq:p22}) all being identically zero,
\begin{eqnarray}
\label{eq:rg2para}
h_{\alpha\rightarrow a}^{(i)} = h_{\alpha\rightarrow a}^{(j)}=0 ,\quad
h_{a\rightarrow \alpha}^{(i)} = h_{a\rightarrow \alpha}^{(j)} = 0 ,
\quad
h_{\mu\rightarrow a}^{(i)} = h_{\mu\rightarrow a}^{(j)}=0 ,\quad
h_{a\rightarrow \mu}^{(i)} = h_{a\rightarrow \mu}^{(j)} = 0 .
\end{eqnarray}
The stability of the paramagnetic solution is analyzed through
a set of linearized rgBP iterative equations listed in
appendix \ref{app:stab}.  All the fields  such as
$h_{\alpha\rightarrow a}^{(i)}$ and $h_{\mu\rightarrow a}^{(i)}$
are randomly initialized, and their
values then evolve according to the linearized rgBP equations.
If all the fields finally decay to zero,
the paramagnetic solution is stable, otherwise
it is unstable and the original rgBP equation has other stable
fixed points.

The number ($2^n-1$) of parameters needed to completely characterize a
probability distribution of the rgBP equation
grows quickly with the number $n$ of vertices on a boundary line of
the square region. The paramagnetic solution and
its stability analysis have been worked out up to $n=10$ for
the Ising model. For the EA model we have only considered the simplest
case of $n=2$.

\subsection{The ferromagnetic Ising model}

At sufficiently high temperatures $T$
the paramagnetic solution (\ref{eq:rg2para}) is the
only solution for the rgBP equation at $n=2$. This trivial solution
becomes unstable at the critical temperature
$T_c^{(n=2)}=2.65635$ (periodic boundary conditions).
This value is higher than the
exact transition temperature
$T_c=2.26919$ \cite{Kramers-Wannier-1941a},
but a little bit lower than the value of $T_c^{(n=1)}=2.88539$
as obtained through the conventional belief-propagation equation
(i.e., $n=1$) \cite{Bethe-1935,Peierls-1936a}.

\begin{figure}
\begin{center}
\includegraphics[width=0.6\textwidth]{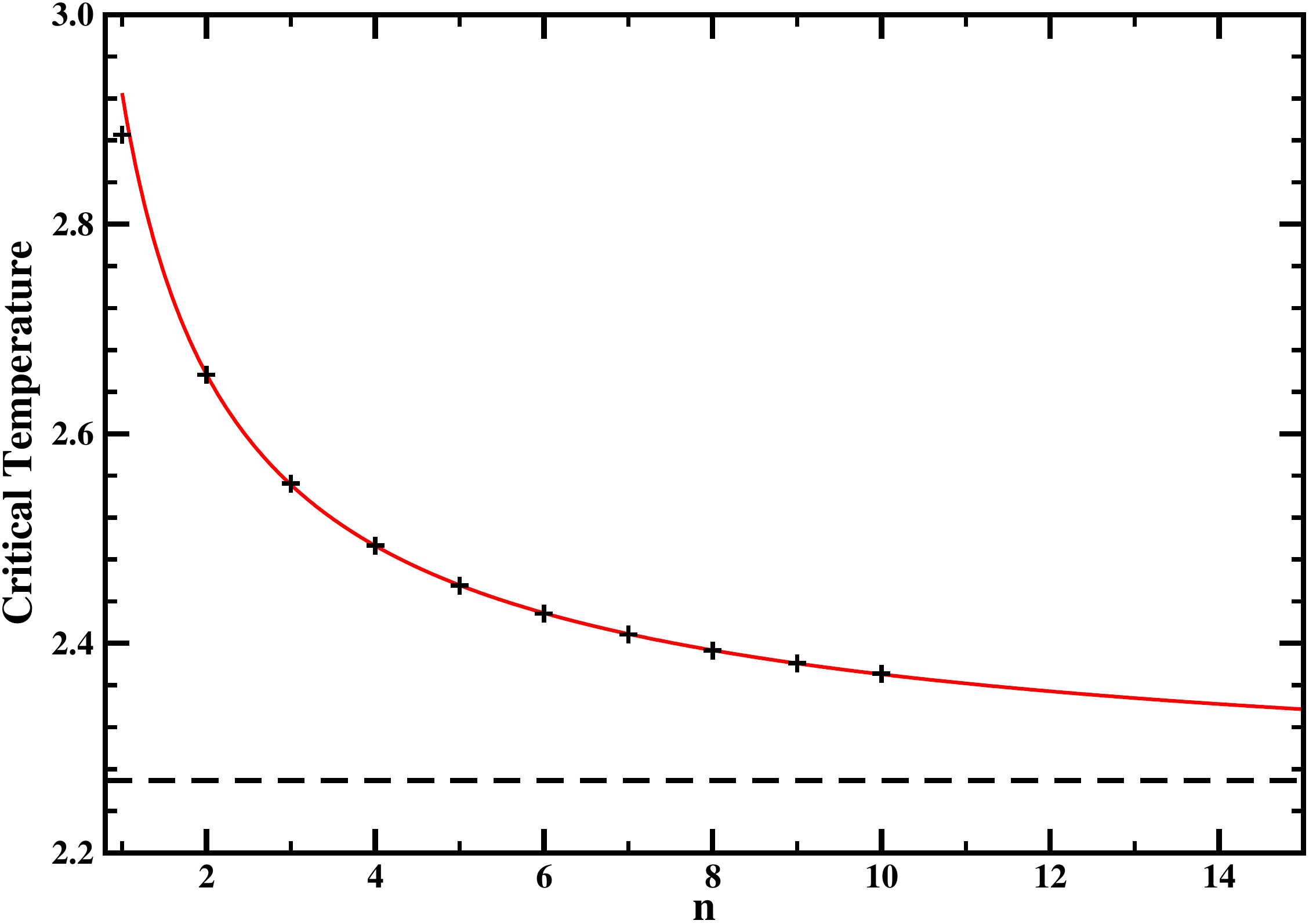}
\end{center}
\caption{
\label{fig:2disingTc}
The critical temperature $T$ at which the paramagnetic fixed-point of
the rgBP equation becomes unstable
(periodic boundary conditions). The integer
$n$ is the number of vertices on a boundary line of a square region.
The exactly known phase-transition temperature
$T_c= 2/\ln(1+\sqrt{2})\simeq 2.2692$ is indicated by
the horizontal dashed line. The red solid curve is a fitting function
$T = 2.2376 + 0.6875 n^{-0.7140}$ to the data with $n\geq 2$.}
\end{figure}

The paramagnet-ferromagnet transition temperature as predicted by the
rgBP equations decreases if larger square regions are used.
Figure~\ref{fig:2disingTc} demonstrates that the predicted
critical temperature $T_c^{(n)}$ can be fitted by
the following curve
\begin{equation}
T_c^{(n)} = T_c^\infty + \frac{a}{n^b} ,
\end{equation}
with $T_c^\infty = 2.2376 \pm 0.0037$,
$a=0.6875\pm 0.0017$, and $b=0.7140\pm 0.0108$. The fitted
value $T_c^\infty$ is slightly lower than the exact critical
point $T_c = 2.2692$. The data can also be fitted well by
$T_c^{(n)} = T_c + a^\prime / n^{b^\prime}$, with $a^\prime
= 0.688 \pm 0.007$ and $b^\prime = 0.8182 \pm 0.0078$.

\begin{figure}
\begin{center}
\includegraphics[width=0.495\textwidth]{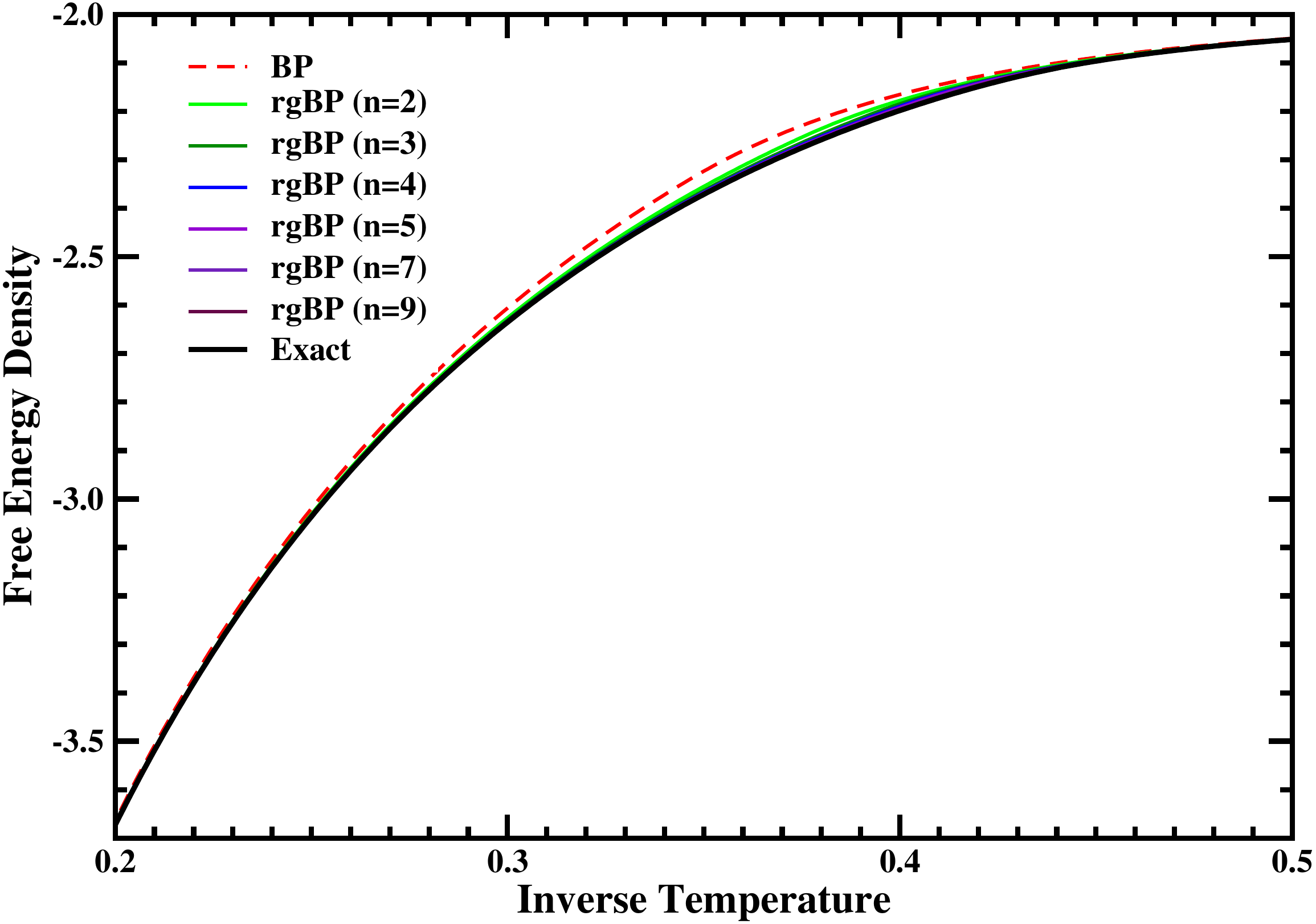}
\includegraphics[width=0.495\textwidth]{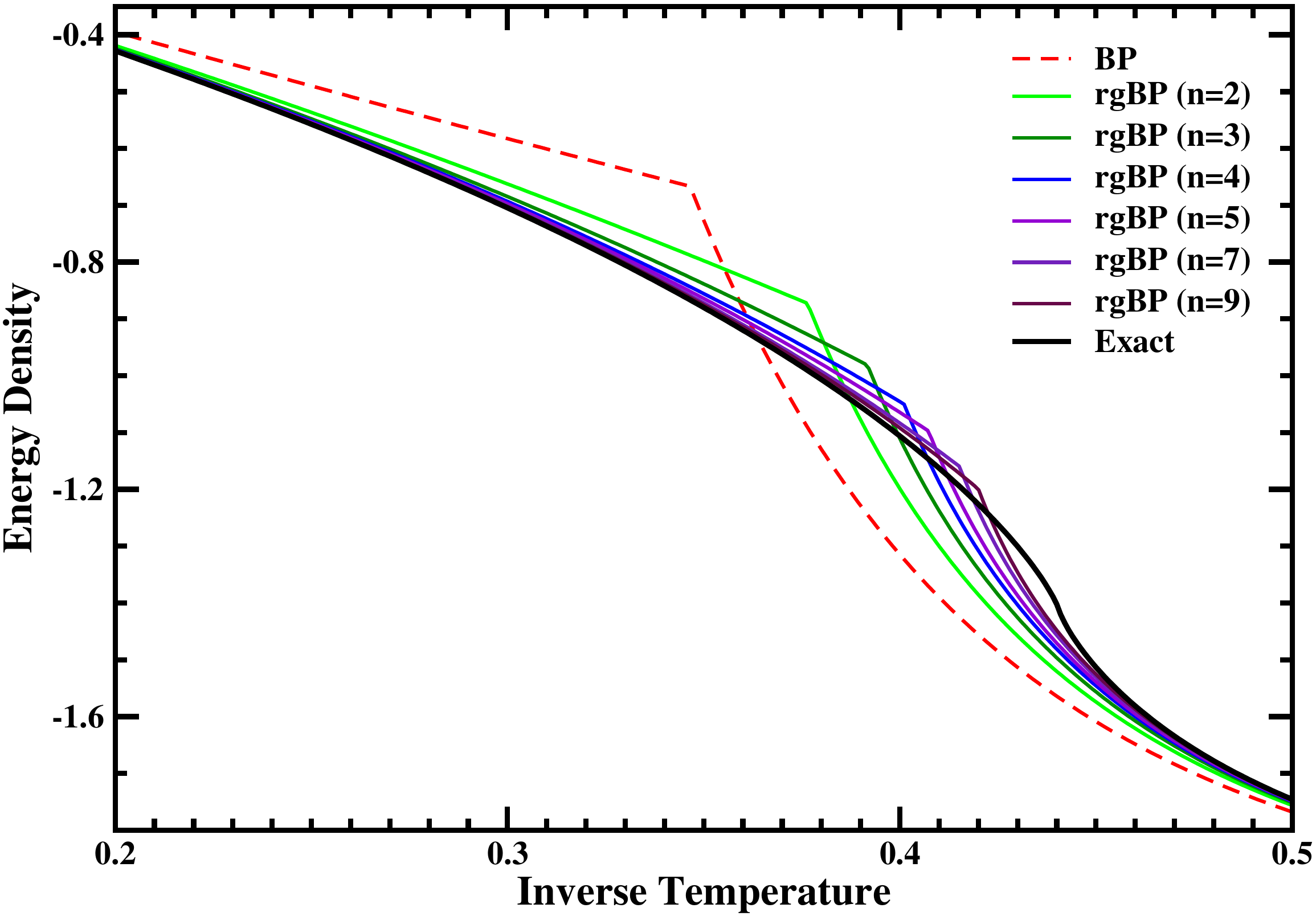}
\vskip 0.02\textwidth
\includegraphics[width=0.495\textwidth]{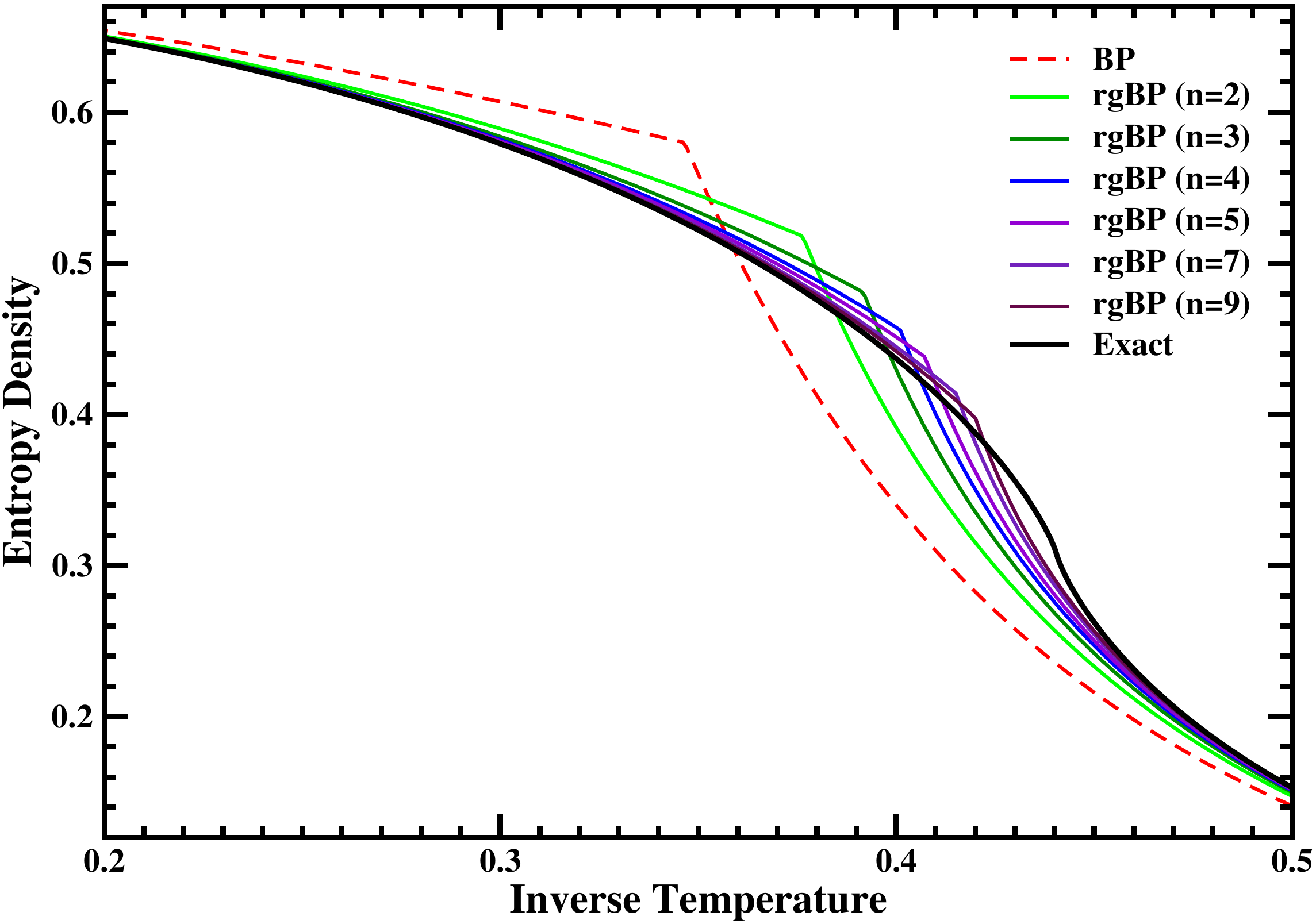}
\includegraphics[width=0.495\textwidth]{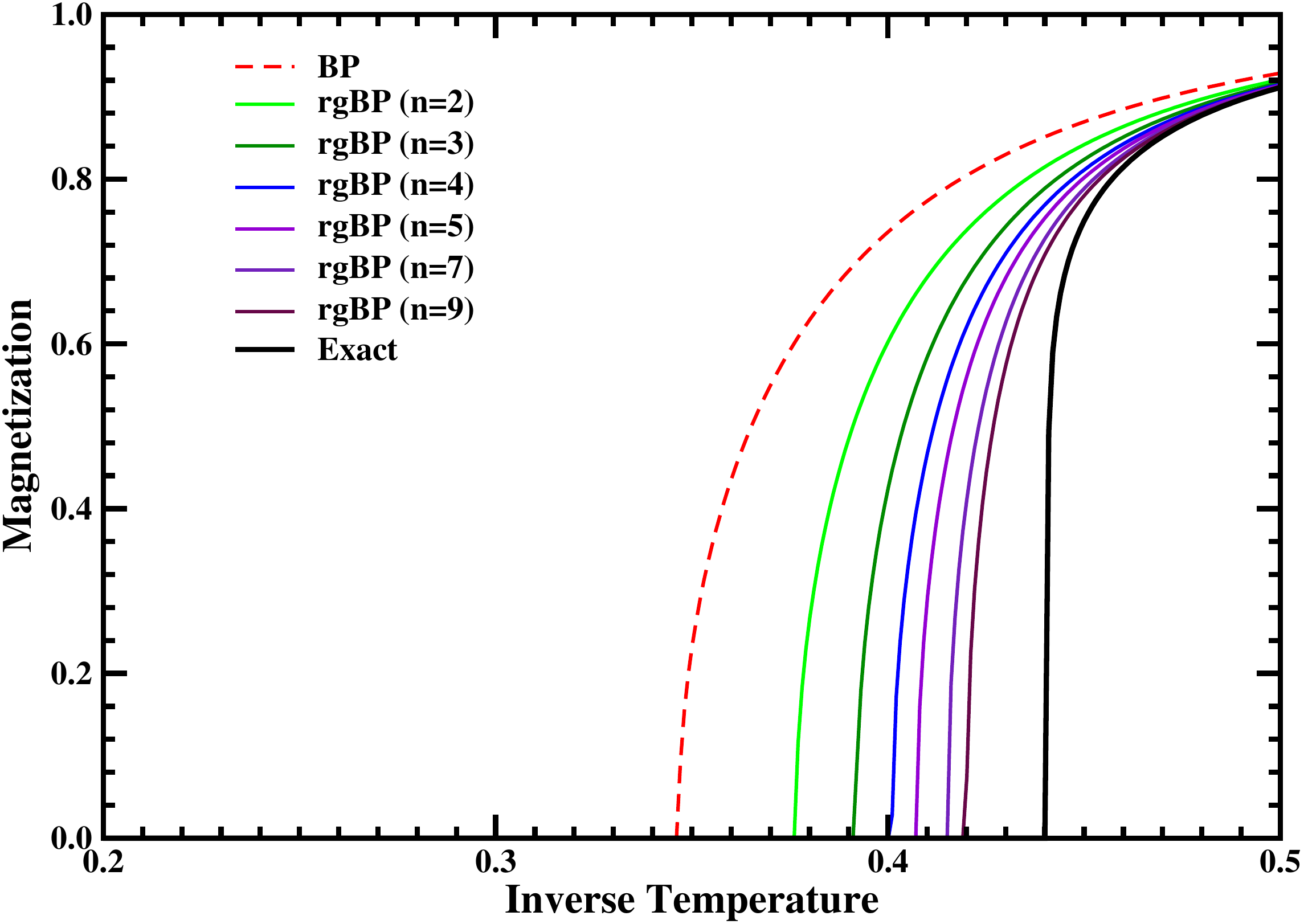}
\end{center}
\caption{\label{fig:2dIsing}
Results obtained on the ferromagnetic Ising model
(periodic boundary conditions) using the conventional
BP and the rgBP at various values of $n$. The results of Onsager's
exact solution are also shown for comparison.
(upper left) free energy density; (upper right) mean energy density;
(lower left) entropy density; (lower right) magnetization.
}
\end{figure}

The free energy density, mean energy density,
entropy density, and magnetization
as a function of temperature $T$ are shown in figure ~\ref{fig:2dIsing}
and compared with the exactly known results of Onsager \cite{Onsager-1944}.
As $n$ increases, the results are closer to the exact value.
The free energy as obtained by the rgBP equations is an upper bound to
the true free energy value of the system, but the difference is small at
$n\geq 2$. Both the energy density and the entropy density have a
kink at the rgBP critical point $T_c^{(n)}$, but this kink becomes more
and more weaker as $n$ increases (in the limit of $n\rightarrow \infty$
the exact Onsager solution should be reached). Using the theoretical
framework of coherent anomaly method \cite{Suzuki-etal-1995}, the results
obtained at different values of $n$ can be used to predict
the critical exponents of the 2D Ising model. The results of such an
exercise (to be carried out) will be reported elsewhere.

\begin{figure}
\begin{center}
\includegraphics[width=0.6\textwidth]{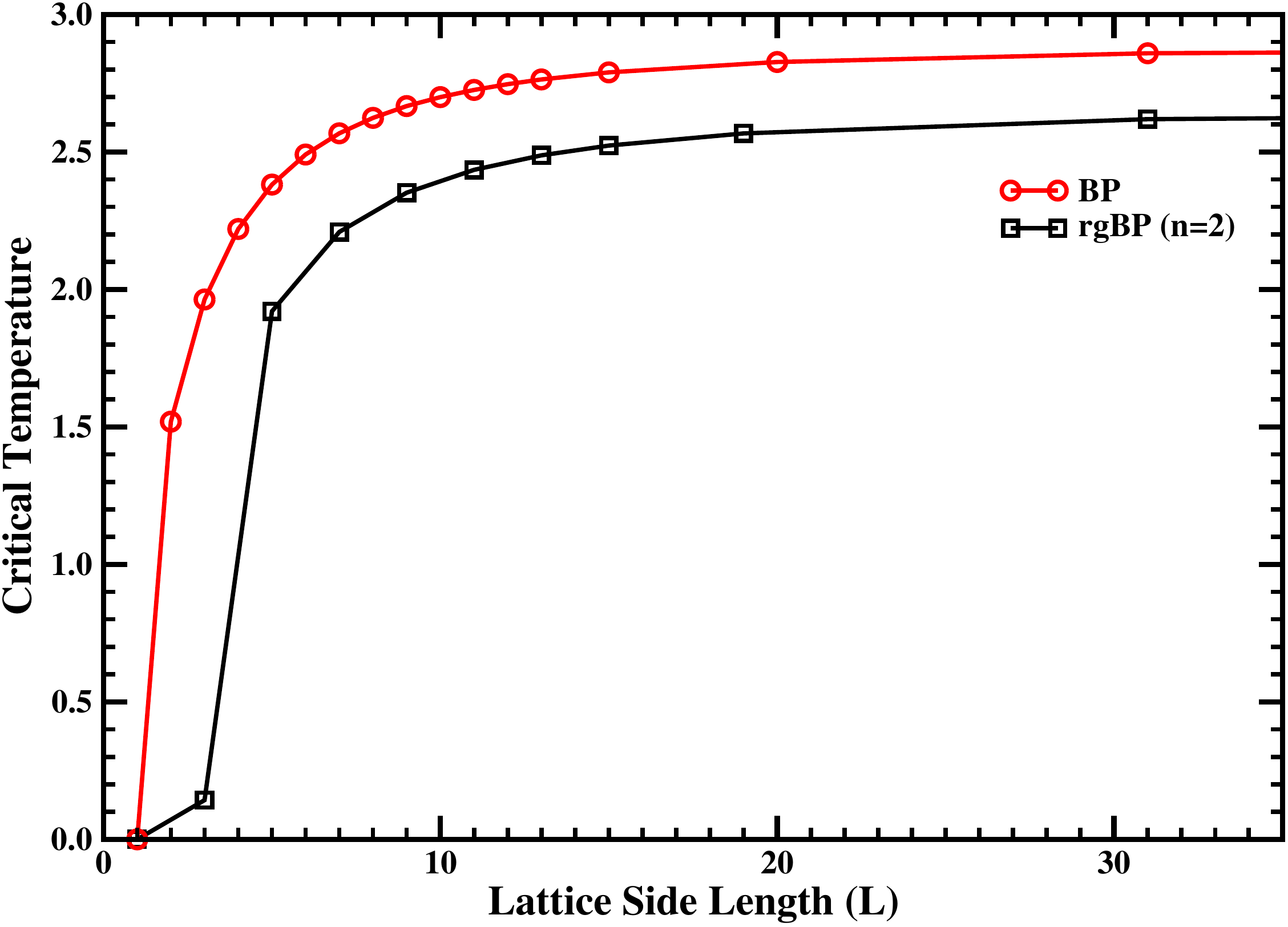}
\end{center}
\caption{
\label{fig:2dIsingOpen}
Threshold instability temperature of the paramagnetic
solution for the Ising model on a $L\times L$ square lattice
with open boundary condition. Circular points are obtained by
the BP approximation, square points by
the rgBP at $n=2$. At each value of $L$, the threshold temperature
as predicted by rgBP at $n=2$ is lower than that predicted by the
conventional BP.
}
\end{figure}

Under the periodic boundary condition, the instability temperature
of the paramagnetic
solution shown in figure~\ref{fig:2disingTc}
is independent of lattice side length $L$. But this is not the
case for the open boundary condition.
Under the open boundary condition, the shorter the lattice
side length $L$, the more stable the paramagnetic solution
is (see figure~\ref{fig:2dIsingOpen}). However, this difference in
threshold temperature between the
periodic and the open boundary conditions becomes very small
for $L> 100$. The rgBP equation at $n=2$ have two stable
ferromagnetic fixed points even for a very small $5\times 5$
square lattice (open boundary conditions) if the
temperature is lower than $1.92$ (see
figure~\ref{fig:2dIsingOpen}). For larger open square lattices,
the ferromagnetic solutions are stable at higher
temperatures. Of course for finite lattices there is no
real phase transition. The two low-temperature
ferromagnetic solutions are interpreted
as describing the two metastable states of
the square lattice. They are stable because the
correlation length in the system exceeds the length scale of
the maximal square region of the region graph. If $n$ exceeds the
correlation length of the system, the paramagnetic fixed point of
the rgBP equation will again be stable.

\subsection{The Edwards-Anderson model}

\begin{figure}
\begin{center}
\includegraphics[width=0.6\textwidth]{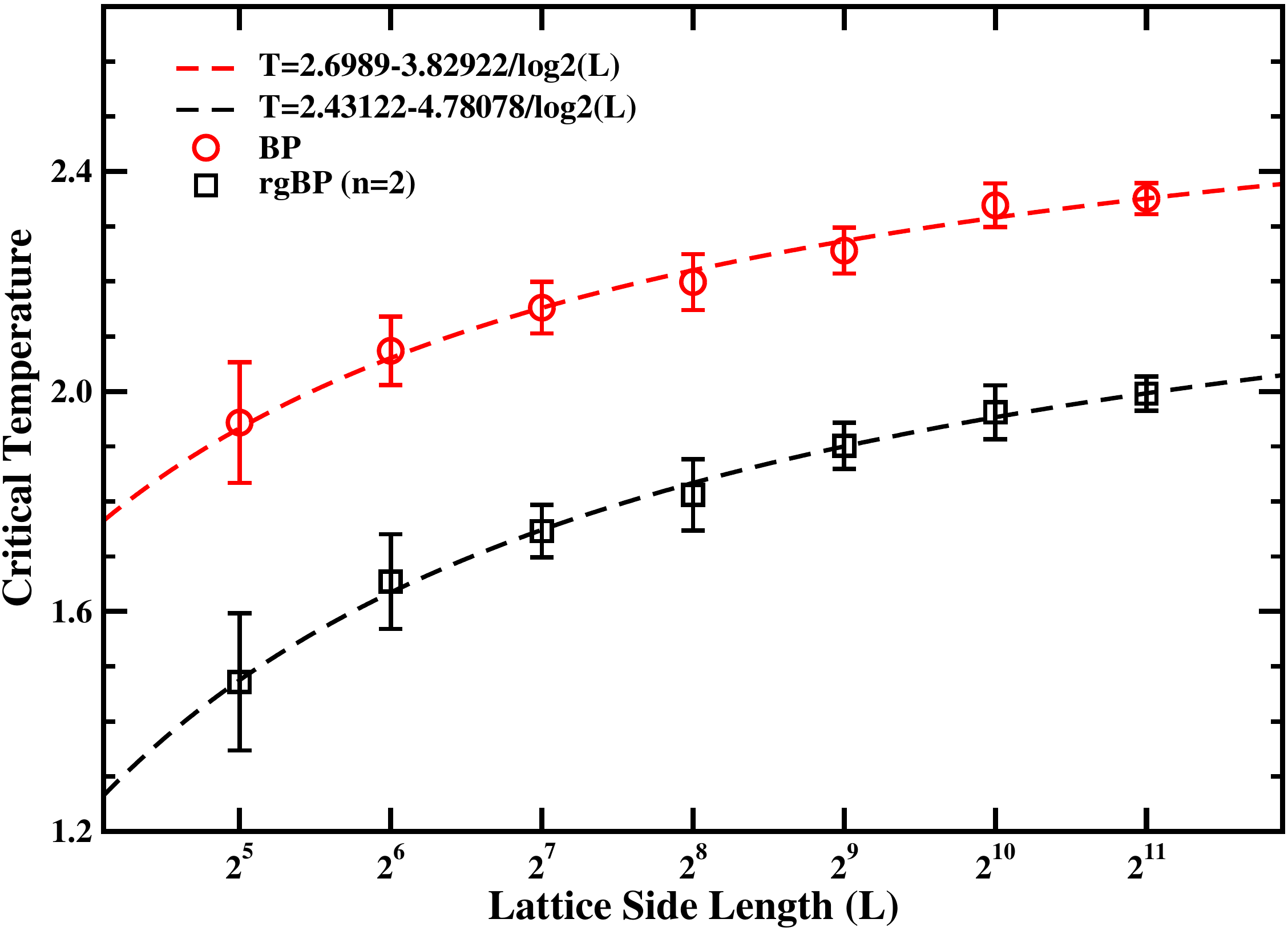}
\end{center}
\caption{
\label{fig:stabEA}
The threshold instability temperature of the
paramagnetic solution of the conventional BP and the
the  rgBP equation at $n=2$ for $L \times L$ periodic
square lattices. Each data point is obtained by averaging over
$10$ single instances of the Edwards-Anderson model. The dashed lines are
two fitting curves of the form $T=a-b/\log_2 L$, the two fitting parameters $a,b$
are shown in the figure.
}
\end{figure}

A set of single instances of the 2D EA model with $\pm J$ coupling constants
are randomly generated. For each instance the
threshold stability temperature $T_{c}^{(n=2)}$  of
the paramagnetic solution of the rgBP equation at $n=2$ is
determined numerically, and this value is compared with the
threshold value as obtained by the conventional BP approximation.
As shown in figure~\ref{fig:stabEA}, the performance of rgBP ($n=2$) is
better than that of BP, but  $T_c^{(n=2)}$ is still positive,
and it increases with the side length $L$ of the periodic square lattice.
Figure~\ref{fig:stabEA} also suggests that the finite-size corrections
to the  critical temperature
$T_c^{(n=2)}$ decrease as $1/\ln L$. This finite-size scaling
behavior is different from the observation of
\cite{Dominguez-etal-2011}.

The EA model on a square lattice has no real spin glass phase
at finite temperatures (see, for example,
\cite{Morgenstern-Binder-1980,Saul-Kardar-1993}). At any positive
temperature $T$ the system has only a paramagnetic phase, and the
magnetization on each vertex is equal to zero in the long time limit.
Apparently, the results of figure~\ref{fig:stabEA} with the
paramagnetic solution being unstable at positive temperatures are
in contradiction with the absence of finite-temperature spin
glass phase.

This apparent discrepancy can actually be removed.
The instability of the paramagnetic rgBP solution does not mean the
system is in a spin glass phase. It just signifies the
emergence of some collective domains in the square lattice, the
length scales of these collective domains exceed the length scale ($=n$)
of the region graph's maximal square region. The detailed arguments
go as follows.

\begin{figure}
\begin{center}
\includegraphics[height=0.405\textwidth]{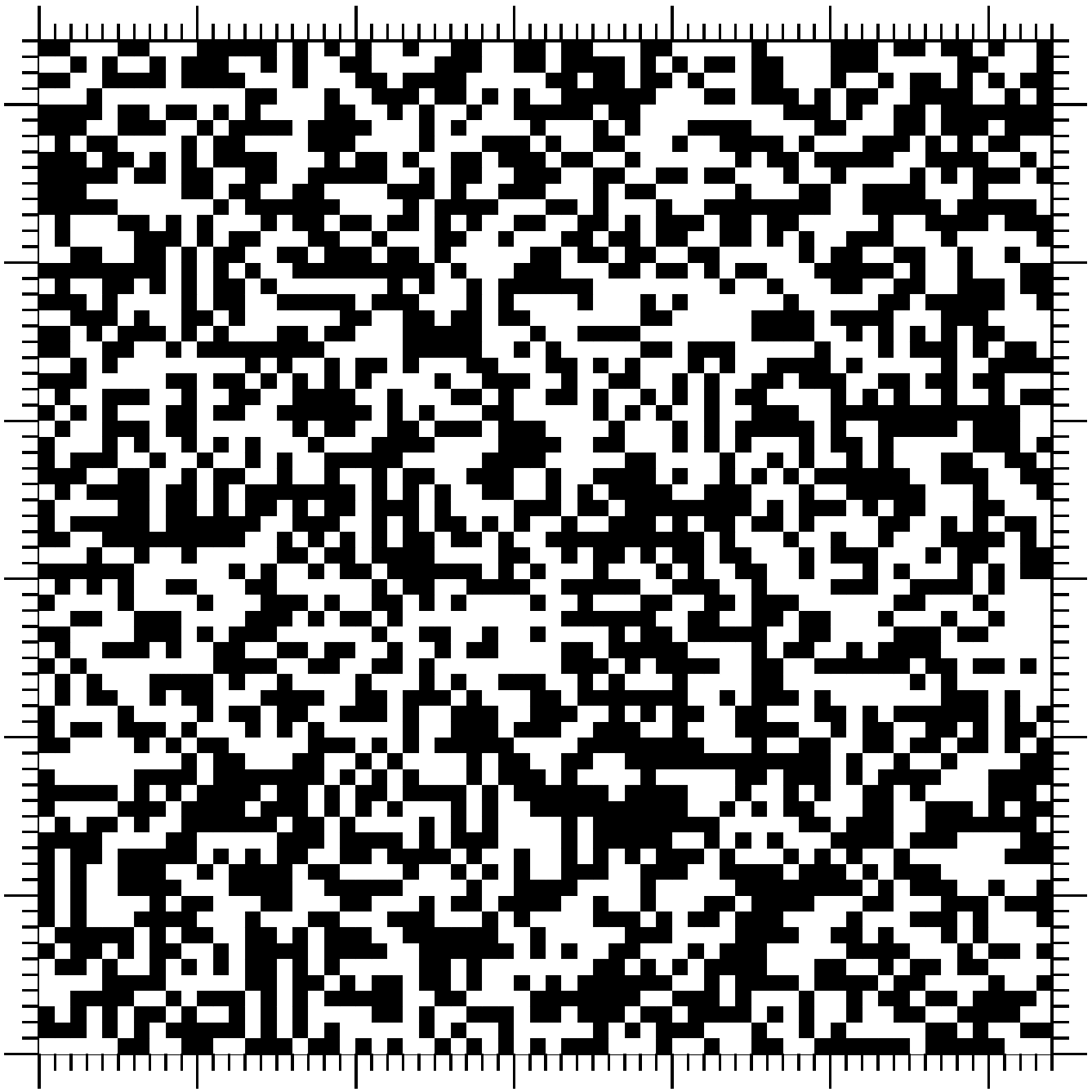}
\hspace {0.082\textwidth}
\includegraphics[height=0.405\textwidth]{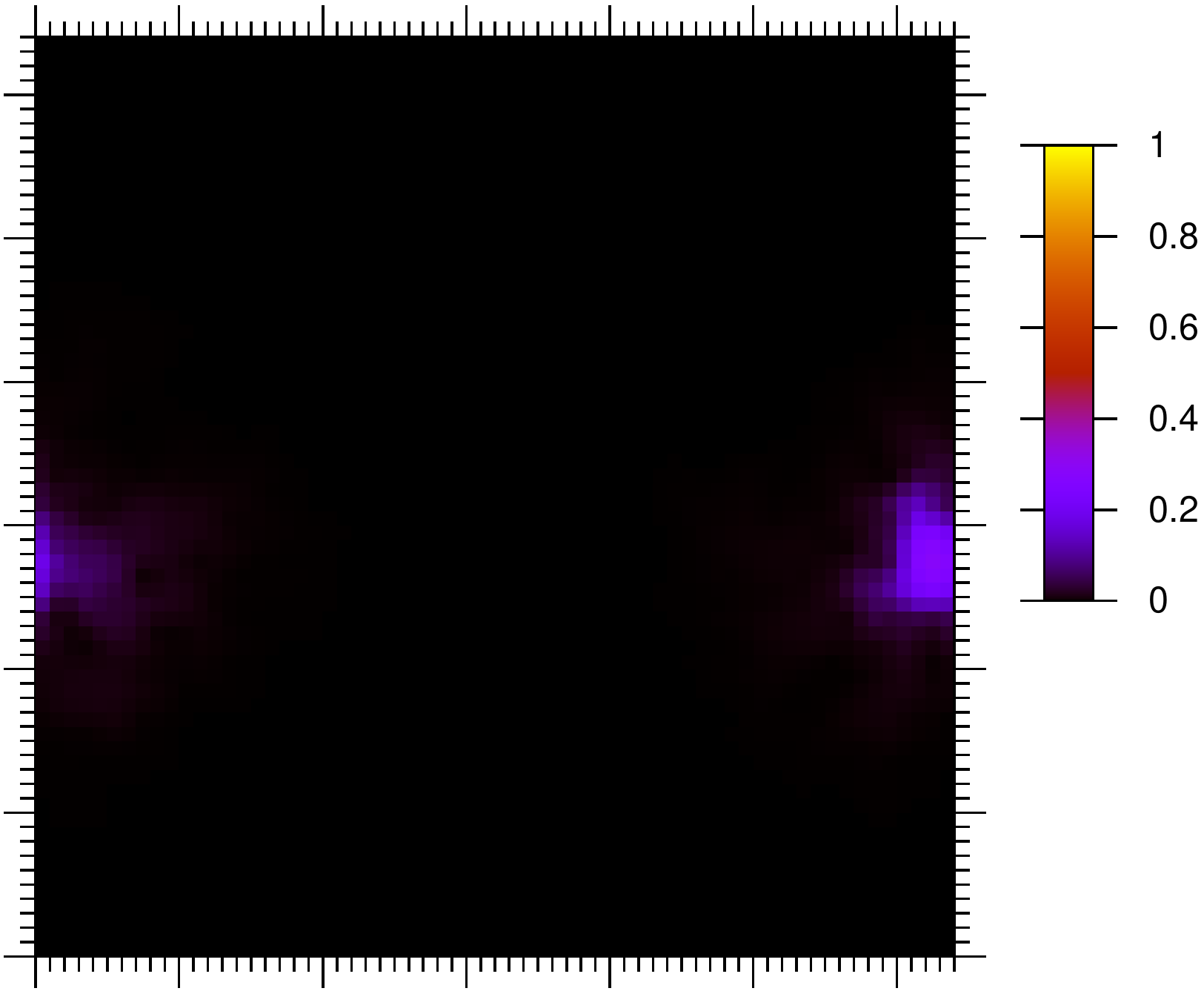}
\vskip 0.02 \textwidth
\includegraphics[height=0.405\textwidth]{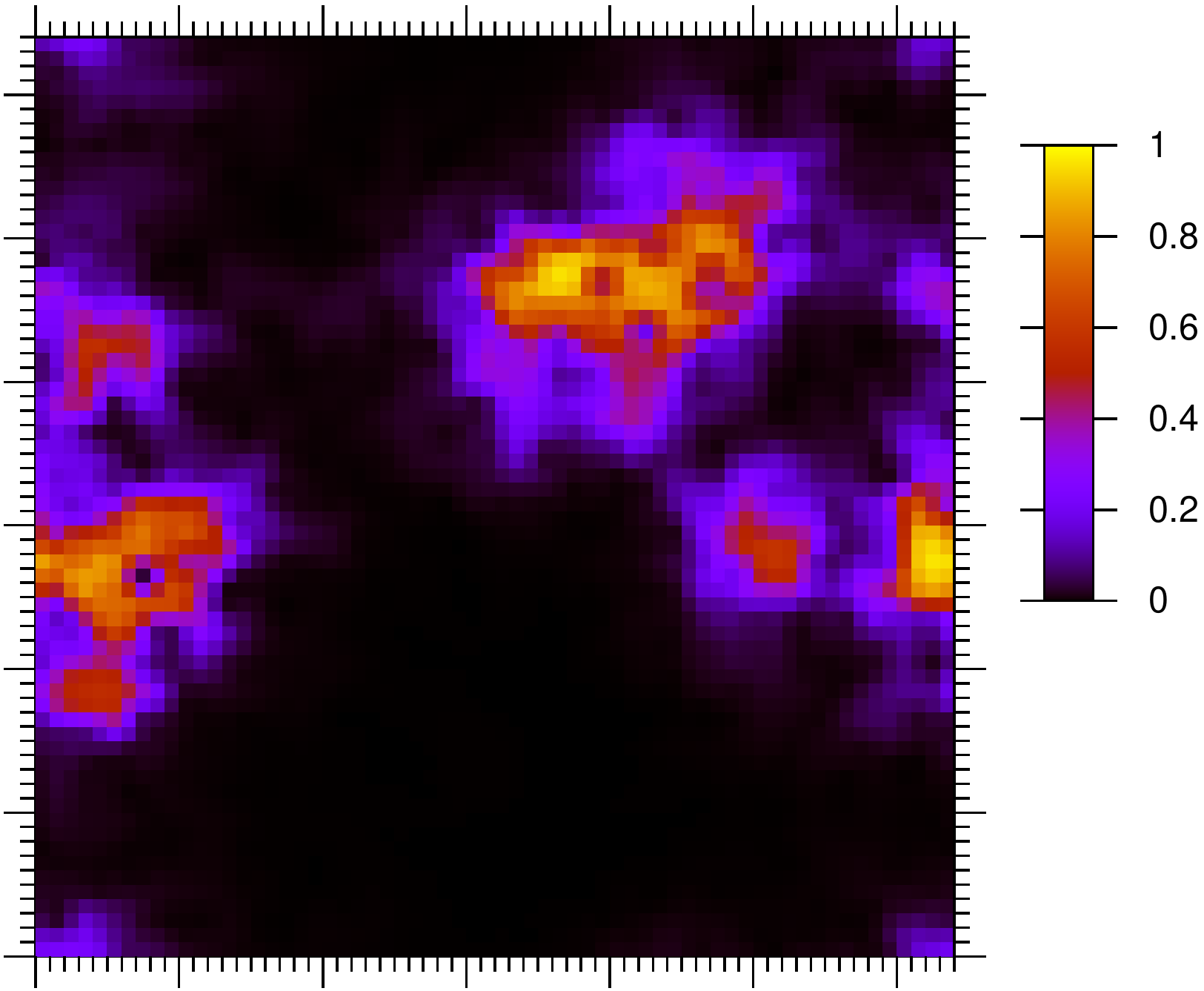}
\includegraphics[height=0.405\textwidth]{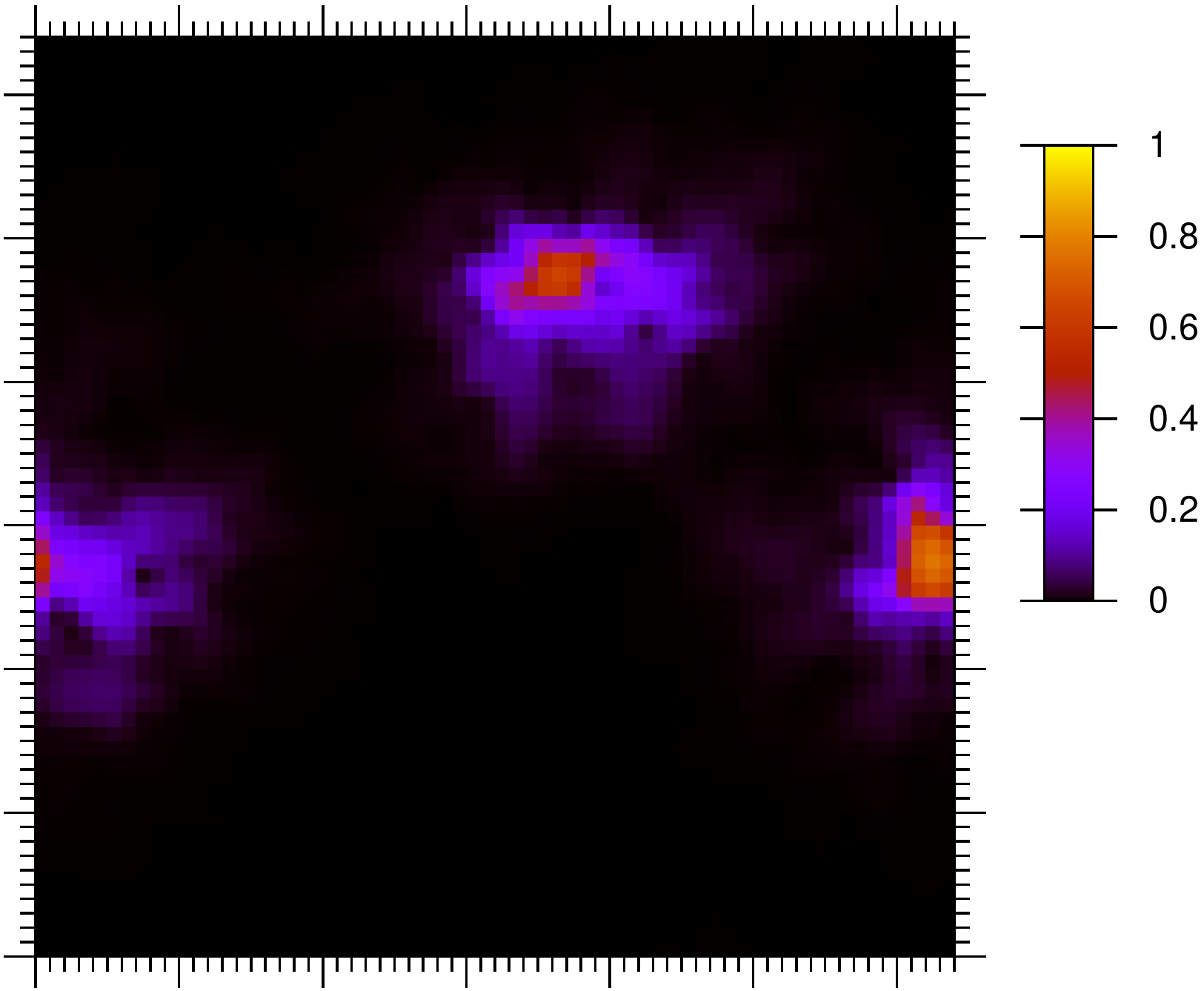}
\end{center}
\caption{
\label{fig:magevolution}
Emergence of domains with collective dynamics.
(upper left) Distribution of frustrated plaquettes in a single
instance of the 2D Edwards-Anderson model with periodic boundary
conditions. Each frustrated plaquette is shown as black, while each
non-frustrated plaquette is shown as white. For this instance, the
paramagnetic solution of the rgBP equation at $n=2$ becomes
unstable at $T_c^{(n=2)}=1.6618$ ($\beta =0.60177$).
(upper right) At temperature $T=1.6502$ ($\beta = 0.606$),
which is just slightly below $T_c^{(n=2)}$, one small collective
domain begins to emerge, as demonstrated by
non-zero  magnetizations for some plaquettes in a connected
cluster of plaquettes. The color of each plaquette encodes the mean
value of absolute magnetizations as averaged over its four vertices.
(lower right)  At a further  decreased temperature
$T= 1.5385$ ($\beta = 0.65$), another collective domain becomes
quite evident. As the temperature further decreases, more
collective domains form, and the formed collective
domains enlarge in size and their boundaries start to be in
contact. As an example, the mean absolute magnetization of
each plaquette is shown at $T=1.3333$ ($\beta = 0.75$) at the
lower left panel.
}
\end{figure}

An elementary square (including four coupling
interactions) of the  square lattice is called a
plaquette. There are two types of plaquettes, frustrated or
non-frustrated. A plaquette is said to be  frustrated
(non-frustrated)
if the product of the four edge coupling constants on its boundary
is negative (positive) \cite{Toulouse-1977}.
It is an obvious fact that
the four edge coupling energies of a frustrated
plaquette can not be {\em simultaneously} minimized.
For the EA model
studied in this paper, on average one-half of the plaquettes
in each problem instance are
frustrated, and these frustrated plaquettes are randomly
distributed on the 2D lattice (see the upper left panel of
figure~\ref{fig:magevolution} for a concrete example).

The abundance of frustrated plaquettes destroys the long-range
ferromagnetic correlations in the system. However, the local density
of frustrated plaquettes fluctuates considerably
at different parts of the lattice. There are
some small patches of the lattice that are mainly formed by
non-frustrated plaquettes. For example, in the $64 \times 64$
periodic square lattice of figure \ref{fig:magevolution} (upper
left panel), there is a $5 \times 5$ patch centered at position
$(62, 28)$ in which only $4$ of its $25$ plaquettes are
frustrated. The spin coupling interactions in such small patches
are essentially ferromagnetic in nature
(under a gauge transformation of the
spin variables and the coupling constants
\cite{Toulouse-1977}). Some of these ferromagnetic small
patches may considerably exceed the maximal $n\times n$ square region
in size, and at low enough temperatures, the ferromagnetic
correlation length will exceed $n$, making the rgBP fixed point to be
locally ferromagnetic (after the gauge transform) in these patches but
paramagnetic in the remaining parts of the square lattice. Given a
patch with a specified contour shape and a specified density of
unfrustrated plaquettes, the probability to discover such a patch is
higher in a square lattice with longer side length $L$.
Therefore, the observation that
$T_c^{(n=2)}$ increases with lattice size $L$
(figure~\ref{fig:stabEA}) is consistent with
figure~\ref{fig:2dIsingOpen}.

The marginal probability distribution of the two spins $\sigma_i$ and
$\sigma_j$ of a rod region $a$ (see figure~\ref{fig:rgbp2D}) is calculated as
\begin{eqnarray}
p_a(\sigma_i, \sigma_j) & \propto &
e^{\beta J_{i j} \sigma_i \sigma_j} p_{\alpha\rightarrow a}(\sigma_i,
\sigma_j)  p_{\mu\rightarrow a}(\sigma_i, \sigma_j)
\nonumber \\
& \propto & \exp\bigl[
\beta (h_{\alpha\rightarrow a}^{(i)} +
h_{\mu\rightarrow a}^{(i)} ) \sigma_i
+ \beta( h_{\alpha\rightarrow a}^{(j)} +
h_{\mu\rightarrow a}^{(j)}) \sigma_j
+ \beta (J_{\alpha\rightarrow a}^{(i j)}
+J_{\mu\rightarrow a}^{(i j)} - J_{i j} )
\sigma_i \sigma_j \bigr] .
\end{eqnarray}
Then we get the magnetization of vertex $i$ as
\begin{equation}
m_i   =
\frac{\tanh\bigl[\beta (h_{\alpha\rightarrow a}^{(i)} +
h_{\mu\rightarrow a}^{(i)})\bigr] +
\tanh\bigl[\beta (h_{\alpha\rightarrow a}^{(j)} +
h_{\mu\rightarrow a}^{(j)}) \bigr]
\tanh\bigl[\beta (J_{\alpha\rightarrow a}^{(i j)}
+J_{\mu\rightarrow a}^{(i j)} - J_{i j} )\bigr]}
{1+ \tanh\bigl[\beta (h_{\alpha\rightarrow a}^{(i)} +
h_{\mu\rightarrow a}^{(i)})\bigr]
\tanh\bigl[\beta (h_{\alpha\rightarrow a}^{(j)} +
h_{\mu\rightarrow a}^{(j)}) \bigr]
\tanh\bigl[\beta (J_{\alpha\rightarrow a}^{(i j)}
+J_{\mu\rightarrow a}^{(i j)} - J_{i j} )\bigr]
} .
\end{equation}
The magnetizations of all the vertices in the square lattice can
be obtained in a similar way. We define the mean absolute magnetization of
a plaquette as the average of the absolute magnetizations of its
four vertices.

Figure~\ref{fig:magevolution} shows the pattern of mean absolute
plaquette magnetizations at a given temperature $T$, for a
single instances of the EA model on a $64\times 64$
periodic square lattice. For $\beta < 0.60177$ the rgBP equation has only
the paramagnetic solution, and all the plaquette mean magnetizations are
zero. At $\beta =0.606$, the mean absolute magnetizations of a
small domain of the square lattice are nonzero
(figure~\ref{fig:magevolution}, upper right panel). As temperature
further decreases, this collective domain enlarges in area, and also
other collective domains start to form (figure~\ref{fig:magevolution},
lower right panel, $\beta=0.65$). As temperature further decreases,
different collective domains start to get into contact and they
compete for the boundary vertices (figure~\ref{fig:magevolution},
lower left panel, $\beta=0.75$)\footnote{At such low temperatures,
the rgBP equation starts to be difficult to converge, probably because of
the frustration effects due to domain competitions.}.

The heterogeneous patterns and its evolution shown in
figure~\ref{fig:magevolution} in some respect are similar to the
phenomenon of dynamical heterogeneity in structural glasses
\cite{Ediger-2000,Glotzer-2000}. This link deserves to be explored
more deeply. The existence of many collective domains and the
frustration effects between these domains very probably are
responsible for the glassy-like low-temperature dynamics of the
2D Edwards-Anderson model.

Figure~\ref{fig:magevolution} also suggests a way to improve the
rgBP prediction power. We infer that different local regions of the
square lattice have different correlation lengths.
To consider more precisely the correlations in the collective
domains, a conceptually easy way is to construct region graphs with
larger regions for these collective domains, while small regions are
used for the remaining parts to lower computational complexity.
This adaptive strategy needs to be implemented in future work.

We end this section by emphasizing that, the  BP and the rgBP 
iterative process are still able to 
converge when the paramagnetic fixed point becomes
unstable. For the 2D square-lattice Edwards-Anderson model 
(periodic boundary conditions), we found that 
BP converges as long as 
$T>1.52$, in agreement with earlier simulation results 
\cite{LageCastellanos-etal-2011,Dominguez-etal-2011}.
The rgBP at $n=2$ converges at even lower temperatures.

\section{Conclusion and outlook}
\label{sec:conclude}

In this paper we gave a detailed description of the region graph
partition function expansion approach (first introduced in
\cite{Zhou-etal-2011}), and
obtained approximate expresses for the free energy
and generalized free energies of a general graphical system with
an abundant number of short loops.
A series of message-passing equations (such as region graph
belief-propagation and region graph survey-propagation) were
derived in the expansion process. We have applied the rgBP equation
to the square lattice Ising model and Edwards-Anderson model and found
that it outperforms the conventional belief-propagation equation.
We also demonstrated that the fixed points of the rgBP equation
reveal the heterogeneous pattern of collective domains in the square
lattice. An adaptive strategy of improving the rgBP performance was
also suggested.

As discussed in the previous sections, there are many issues remain to be
explored. The adaptive rgBP scheme for the 2D Edwards-Anderson
model needs to be implemented. Another very interesting issue is
the effect of adding redundancy to the region-graph. A redundant
region-graph have been used for the 2D Ising model and
Edwards-Anderson model  in various papers within the
framework of cluster variation method
\cite{Kikuchi-1951,Morita-etal-1994,Yedidia-Freeman-Weiss-2005,Dominguez-etal-2011,LageCastellanos-etal-2012}.
The performance of rgBP needs to be tested on such a redundant region-graph.
We have performed some preliminary computations and found that the results of
rgBP depend on the particular way of removing redundancy. For example,
the region graph $R^\#$ shown in the right panel of 
figure~\ref{fig:reduntrgfig} actually is equivalent to the
non-redundant region graph $R$ of figure~\ref{fig:rgfig}. Some other 
ways of removing redundancy are able to make the paramagnetic
rgBP fixed point to be stable at even lower temperatures.

For a general redundant region graph $R$, the rgBP equation (\ref{eq:rgbp})
is not equivalent to the generalized belief-propagation equation 
(\ref{eq:ptoceq}) of \cite{Yedidia-Freeman-Weiss-2005}. We are working
on the issue of deriving the generalized belief-propagation equation from
the approach of partition function expansion.

The three-dimensional Edwards-Anderson model is believed to have a
true spin glass phase. This system will be studied using
rgBP and the region-graph survey-propagation equation in a future
work.

Particles in a dense liquid have translational degrees of freedom.
It remains to be seen whether a similar partition function
expansion scheme can be worked out for systems with mobile particles.
If approximate free energy landscapes can also be built for such
systems with the help of message-passing equations, it should be
very helpful for understanding structural glasses and supercooled
liquids.

A different theoretical approach (the replica cluster variation method)
of studying finite-dimensional spin glasses has been explored in
\cite{Rizzo-etal-2010,Dominguez-etal-2011,LageCastellanos-etal-2012}.
The message-passing process of this replica cluster variation method
appears to be much more complicated than the simple rgBP process.


\appendix

\section{Derivation of parent-to-child message-passing equation}
\label{app:proof}

Here we give a detailed derivation of equations (\ref{eq:pgamma4})
and (\ref{eq:miteration}). These equations
are valid for a non-redundant region graph $R$.
We assume $R$ to be non-redundant in this whole section.

First, the region subgraph formed by region $\gamma$ and all its
descendants is a connected tree (see the blue-shaded area of
figure~\ref{fig:rgptoc}). This property ensures the equivalence
of (\ref{eq:pgamma2}) with (\ref{eq:pgamma}). Notice that a
region $\mu\in B_\gamma$ may point to two or more regions of the
set $I_\gamma$.

Applying the definition (\ref{eq:Psialpha}) and then
the two identities (\ref{eq:cni}) and (\ref{eq:cna}), we obtain that
\begin{eqnarray}
\prod\limits_{\eta\in I_\gamma} \Psi_\eta(\underline{x}_\eta)
& = & \prod\limits_{a\in \gamma} \Bigl[ \psi_a(\underline{x}_{\partial a})
\Bigr]^{\sum\limits_{\eta \in I_\gamma \cap R_a} c_\eta}
\prod\limits_{i\in \gamma}
\Bigl[\psi_i(x_i) \Bigr]^{\sum\limits_{\eta \in  I_\gamma \cap R_i} c_\eta}
\nonumber \\
& = & \biggl[
\prod\limits_{a\in \gamma} \psi_a(\underline{x}_{\partial a})
\prod\limits_{i\in \gamma} \psi_i(x_i) \biggr]
\prod\limits_{b\in \gamma} \Bigl[ \psi_b(\underline{x}_{\partial b})
\Bigr]^{-\sum\limits_{\eta \in R_b \backslash I_\gamma} c_\eta}
\prod\limits_{j\in \gamma} \Bigl[ \psi_j(x_j)
\Bigr]^{-\sum\limits_{\eta \in  R_j \backslash I_\gamma} c_\eta} .
\label{eq:weight3}
\end{eqnarray}
The set  $R_b\backslash I_\gamma$ contains regions of subgraph $R_b$ except
those also belonging to set $I_\gamma$, and similarly for
$R_j \backslash I_\gamma$.

Because $R_b$ and $R_j$ are two connected
tree subgraphs, we have
\begin{equation}
\label{eq:identity4}
\sum\limits_{\eta \in R_b \backslash I_\gamma} c_\eta
= \sum\limits_{\nu\in R_b \cap I_\gamma}
\sum\limits_{\{(\mu\rightarrow \nu)| \mu\in B_\gamma\}}
\sum\limits_{\alpha \in R_b^{\mu\rightarrow \nu}} c_\alpha ,
\quad\quad
\sum\limits_{\eta \in R_j \backslash I_\gamma} c_\eta
= \sum\limits_{\nu\in R_j \cap I_\gamma}
\sum\limits_{\{(\mu\rightarrow \nu) | \mu \in B_\gamma\}}
\sum\limits_{\alpha \in R_j^{\mu\rightarrow \nu}} c_\alpha .
\end{equation}
In the above equation,
$R_{b}^{\mu\rightarrow \nu}$ denotes the branch of the
tree $R_b$ that is still connected with $\mu$ if the
directed edge $\mu\rightarrow \nu$ is removed; and
$R_j^{\mu\rightarrow \nu}$ has the same definition, i.e., it
is the branch of the tree $R_j$ that contains
region $\mu$ but not $\nu$. The possibility that a region
$\mu\in B_\gamma$ might point to two ore more regions in
$I_\gamma$ does not affect the validity of (\ref{eq:identity4}).
The reason is simple: if $\nu_1$ and $\nu_2$ are two children of
$\mu$ in $I_\gamma$, then $\nu_1$ and $\nu_2$ do not share
any function node nor any variable node in common.

Based on (\ref{eq:identity4}), (\ref{eq:weight3}) and (\ref{eq:pgamma2}),
we obtain the important expression (\ref{eq:pgamma4}). In that
equation, the parent-to-message
$m_{\mu\rightarrow \nu}(\underline{x}_\nu)$ is  defined as
\begin{equation}
\label{eq:mmutonu}
m_{\mu\rightarrow \nu}(\underline{x}_\nu)  \propto
p_{\mu\rightarrow \nu}(\underline{x}_\nu)
 \prod\limits_{b\in \nu}
\Bigl[ \psi_b(\underline{x}_{\partial b})
 \Bigr]^{-\sum\limits_{\alpha\in R_b^{\mu\rightarrow \nu}}
c_\alpha}
\prod\limits_{j\in \nu}
\Bigl[\psi_j(x_j) \Bigr]^{-\sum\limits_{\alpha\in R_j^{\mu\rightarrow \nu}}
 c_\alpha}
\end{equation}
up to a normalization constant (to be fixed by
$\sum_{\underline{x}_\nu} m_{\mu\rightarrow \nu}(\underline{x}_\nu)=1$).

Using the expression
(\ref{eq:rgbp}) for the probability distribution
$p_{\mu\rightarrow \nu}(\underline{x}_\nu)$, it is easy to show that
\begin{eqnarray}
m_{\mu\rightarrow \nu}(\underline{x}_\nu) & \propto &
\prod\limits_{b \in \nu} \bigl[ \psi_b(\underline{x}_{\partial b})
\bigr]^{-\sum\limits_{\alpha \in R_b^{\mu\rightarrow \nu}} c_\alpha}
\prod\limits_{j\in \nu}
\Bigl[\psi_j(x_j) \Bigr]^{-\sum\limits_{\alpha
\in R_j^{\mu\rightarrow \nu}} c_\alpha} \nonumber \\
& & \times
\sum\limits_{\underline{x}_{\mu}\backslash
\underline{x}_{\nu}} \biggl[
\prod\limits_{\eta\in I_\mu \backslash I_\nu}
\Psi_\eta (\underline{x}_{\eta}) \biggr]
\prod\limits_{\{(\alpha\rightarrow \lambda) |
\alpha \in B_\mu, \lambda \in I_\mu \backslash I_\nu\}}
p_{\alpha\rightarrow \lambda}(\underline{x}_\lambda) .
\end{eqnarray}
Notice that
\begin{eqnarray}
& \prod\limits_{\eta\in I_\mu \backslash I_\nu}
\Psi_\eta(\underline{x}_\eta)
=
\prod\limits_{a\in \mu} \Bigl[\psi_a(\underline{x}_{\partial a})
\Bigr]^{\sum\limits_{\eta \in R_a\cap (I_\mu \backslash I_\nu)} c_\eta}
\prod\limits_{i\in \mu} \Bigl[
\psi_i(x_i)\Bigr]^{\sum\limits_{\eta \in R_i\cap (I_\mu \backslash I_\nu)}
 c_\eta}  \\
& \quad =
\prod\limits_{a\in \mu\backslash \nu}
\Bigl[\psi_a \Bigr]^{1-\sum\limits_{\alpha\in R_a \backslash I_\mu} c_\alpha}
\prod\limits_{i\in \mu\backslash \nu}
 \Bigl[\psi_i \Bigr]^{1-\sum\limits_{\alpha\in R_i \backslash I_\mu} c_\alpha}
\prod\limits_{b\in\nu} \Bigl[\psi_b\Bigr]^{\sum\limits_{\eta\in R_b
\cap (I_\mu \backslash I_\nu)} c_\eta}
\prod\limits_{j\in \nu}\Bigl[\psi_j\Bigr]^{\sum\limits_{\eta\in R_j
\cap (I_\mu\backslash I_\nu)} c_\eta} . \nonumber
\end{eqnarray}
Then we have
\begin{eqnarray}
&  \prod\limits_{b\in \nu} \Bigl[\psi_b \Bigr]^{-
\sum\limits_{\alpha\in R_b^{\mu\rightarrow \nu}} c_\alpha}
\prod\limits_{j\in \nu} \Bigl[\psi_j\Bigr]^{-
\sum\limits_{\alpha\in R_j^{\mu\rightarrow \nu}} c_\alpha}
\prod\limits_{\eta\in I_\mu\backslash I_\nu} \Psi_\eta (\underline{x}_\eta)
\nonumber \\
& = \biggl[ \prod\limits_{a\in \mu\backslash \nu}
\psi_a \prod\limits_{i\in \mu\backslash \nu} \psi_i \biggr]
\prod\limits_{b\in \mu\backslash \nu}
\Bigl[\psi_b\Bigr]^{-\sum\limits_{\alpha\in R_b\backslash I_\mu} c_\alpha}
\prod\limits_{c\in \nu} \Bigl[\psi_c\Bigr]^{-
\sum\limits_{\alpha\in R_c^{\mu\rightarrow \nu}} c_\alpha+
\sum\limits_{\eta\in  R_c\cap (I_\mu\backslash I_\nu)} c_\eta}
\times \nonumber \\
& \prod\limits_{j\in \mu\backslash \nu} \Bigl[\psi_j\Bigr]^{-
\sum\limits_{\alpha\in R_j\backslash I_\mu} c_\alpha}
\prod\limits_{k\in \nu} \Bigl[\psi_k\Bigr]^{-
\sum\limits_{\alpha\in R_k^{\mu\rightarrow \nu}} c_\alpha +
\sum\limits_{\eta\in R_k \cap (I_\mu \backslash I_\nu)} c_\eta} .
\label{eq:product1}
\end{eqnarray}
It is easy to check that, for a function node $c\in \nu \in I_\mu$,
\begin{equation}
\sum\limits_{\alpha \in R_c^{\mu\rightarrow \nu}} c_\alpha
- \sum\limits_{\eta \in R_c \cap (I_\mu\backslash I_\nu)} c_\alpha
 = \sum\limits_{\alpha \in R_c^{\mu\rightarrow \nu} \backslash
I_\mu} c_\alpha ,
\end{equation}
where $R_c^{\mu\rightarrow \nu} \backslash I_\mu$ denotes the set formed
by all the regions in the region subtree
$R_c^{\mu\rightarrow \nu}$ except those
which are also members of the region set $I_\mu$. Similarly, for
a variable node $k\in \nu \in I_\mu$, we have
\begin{equation}
\sum\limits_{\alpha \in R_k^{\mu\rightarrow \nu}} c_\alpha
- \sum\limits_{\eta \in R_k \cap (I_\mu\backslash I_\nu)} c_\alpha
 = \sum\limits_{\alpha \in R_k^{\mu\rightarrow \nu} \backslash
I_\mu} c_\alpha ,
\end{equation}
with $R_k^{\mu\rightarrow \nu} \backslash I_\mu$ being the
the set formed  by all the regions in the region subtree
$R_k^{\mu\rightarrow \nu}$ except those
which are also members of the region set $I_\mu$.
With these two equalities, (\ref{eq:product1}) is re-written as
\begin{eqnarray}
&  \prod\limits_{b\in \nu} \Bigl[\psi_b \Bigr]^{-
\sum\limits_{\alpha\in R_b^{\mu\rightarrow \nu}} c_\alpha}
\prod\limits_{j\in \nu} \Bigl[\psi_j\Bigr]^{-
\sum\limits_{\alpha\in R_j^{\mu\rightarrow \nu}} c_\alpha}
\prod\limits_{\eta\in I_\mu\backslash I_\nu} \Psi_\eta (\underline{x}_\eta)
=  \biggl[ \prod\limits_{a\in \mu\backslash \nu}
\psi_a \prod\limits_{i\in \mu\backslash \nu} \psi_i \biggr] \nonumber \\
& \times
\prod\limits_{b\in \mu\backslash \nu}
\Bigl[\psi_b\Bigr]^{-\sum\limits_{\alpha\in R_b\backslash I_\mu} c_\alpha}
\prod\limits_{c\in \nu} \Bigl[\psi_c\Bigr]^{-
\sum\limits_{\alpha \in R_c^{\mu\rightarrow \nu} \backslash
I_\mu} c_\alpha}
\prod\limits_{j\in \mu\backslash \nu} \Bigl[\psi_j\Bigr]^{-
\sum\limits_{\alpha\in R_j\backslash I_\mu} c_\alpha}
\prod\limits_{k\in \nu} \Bigl[\psi_k\Bigr]^{-
\sum\limits_{\alpha \in R_k^{\mu\rightarrow \nu} \backslash
I_\mu} c_\alpha} .
\label{eq:product2}
\end{eqnarray}

Combining the above equation with (\ref{eq:mmutonu}) leads to
\begin{eqnarray}
& m_{\mu\rightarrow \nu}(\underline{x}_\nu)  \propto
\sum\limits_{\underline{x}_\mu \backslash \underline{x}_\nu}
\prod\limits_{a\in \mu\backslash \nu} \psi_a(\underline{x}_{\partial a})
\prod\limits_{i\in \mu\backslash \nu} \psi_i(x_i)
\prod\limits_{\{(\alpha \rightarrow \lambda) | \alpha\in B_\mu,
\lambda \in I_\mu\backslash I_\nu\}}
m_{\alpha\rightarrow \lambda}(\underline{x}_\lambda) \nonumber \\
\times
& \biggl\{
\prod\limits_{b\in \mu\backslash \nu} \Bigl[\psi_b\Bigr]^{-
\sum\limits_{\eta\in R_b\backslash I_\mu} c_\eta}
\prod\limits_{c\in \nu} \Bigl[\psi_c\Bigr]^{-
\sum\limits_{\alpha\in R_c^{\mu\rightarrow \nu}\backslash I_\mu} c_\alpha}
\prod\limits_{\{(\alpha\rightarrow \lambda) | \alpha\in B_\mu,
\lambda \in I_\mu\backslash I_\nu\}}
\prod\limits_{d\in \lambda} \Bigl[\psi_d\Bigr]^{
\sum\limits_{\eta \in R_d^{\alpha\rightarrow \lambda}} c_\eta} \nonumber \\
 & \prod\limits_{j\in \mu\backslash \nu} \Bigl[\psi_j\Bigr]^{-
\sum\limits_{\eta\in R_j\backslash I_\mu} c_\eta}
\prod\limits_{k\in \nu} \Bigl[\psi_k\Bigr]^{-
\sum\limits_{\alpha\in R_k^{\mu\rightarrow \nu}\backslash I_\mu} c_\alpha}
\prod\limits_{\{(\alpha\rightarrow \lambda) | \alpha\in B_\mu,
\lambda \in I_\mu\backslash I_\nu\}}
\prod\limits_{l\in \lambda} \Bigl[\psi_l\Bigr]^{\sum\limits_{\eta
 \in R_l^{\alpha\rightarrow \lambda}} c_\eta}
\biggr\} .
\end{eqnarray}
Using the the properties (\ref{eq:identity4}) for the region trees
$R_b$ (induced by each function node $b\in \mu$) and $R_k$ (induced by
each variable node $k\in \mu$), it is not difficult to verify that
the expression in the curly brackets of the above equation is equal to $1$.
Therefore we arrive at the message-passing equation (\ref{eq:miteration})
for $m_{\mu\rightarrow \nu}(\underline{x}_\nu)$.

\section{Derivation of the free energy expression (\ref{eq:frgbp})}
\label{app:fgbp}

We demonstrate that, for a non-redundant region graph $R$, the
region graph free energy $F_0$ can be expressed as
$F_0 = \sum_{\alpha \in R} c_\alpha \tilde{F}_\alpha$,
with $\tilde{F}_\alpha$ given by (\ref{eq:frgbp}).

First, we notice that $f_{(\mu,\nu)}$ as defined by (\ref{eq:fvande})
can be expressed as
\begin{equation}
f_{(\mu, \nu)} = f_\nu
+ \frac{1}{\beta} \ln \biggl[
\sum\limits_{\underline{x}_\nu} \Psi_\nu(\underline{x}_\nu)
\prod\limits_{\gamma \in \partial \nu \backslash \mu} p_{\gamma \rightarrow
\nu}(\underline{x}_{\nu\cap \gamma}^{\nu}) \biggr]
= f_\nu - f_{\nu\rightarrow \mu} ,
\label{eq:fmunu2}
\end{equation}
where $f_{\nu\rightarrow \mu}$ is defined through (\ref{eq:fmutonu}).
In writing down this equation, we have assumed that $\mu$ is a
parent of $\nu$. From equation (\ref{eq:fa3}) we then obtain that
\begin{equation}
\tilde{F}_\alpha = f_\alpha + \sum\limits_{\{(\mu\rightarrow \nu) | \mu \in
I_\alpha\}} f_{\nu\rightarrow \mu} .
\end{equation}
On the other hand, based on the definition (\ref{eq:fvande}) for $f_{\alpha}$
we derive that
\begin{equation}
f_\alpha  =  -\frac{1}{\beta} \ln \biggl[
\sum\limits_{\underline{x}_\alpha} \prod\limits_{\eta\in I_\alpha}
\Psi_\eta(\underline{x}_\eta)
\prod\limits_{\{(\mu\rightarrow \nu) | \mu \in B_\alpha, \nu \in I_\alpha\}}
p_{\mu\rightarrow \nu}(\underline{x}_\nu) \biggr]
- \sum\limits_{\{(\mu \rightarrow \nu) | \mu \in I_\alpha\}}
 f_{\nu\rightarrow \mu} .
\end{equation}
From the last two expressions we then get the following simple
formula for $\tilde{F}_\alpha$:
\begin{equation}
\label{eq:talpha2}
\tilde{F}_\alpha =
 -\frac{1}{\beta} \ln \biggl[
\sum\limits_{\underline{x}_\alpha} \prod\limits_{\eta\in I_\alpha}
\Psi_\eta(\underline{x}_\eta)
\prod\limits_{\{(\mu\rightarrow \nu) | \mu \in B_\alpha, \nu \in I_\alpha\}}
p_{\mu\rightarrow \nu}(\underline{x}_\nu) \biggr].
\end{equation}
This formula is very similar to (\ref{eq:frgbp}), but not yet identical.

Now we replace $p_{\mu\rightarrow \nu}(\underline{x}_\nu)$ of
(\ref{eq:talpha2}) by
$m_{\mu\rightarrow \nu}(\underline{x}_\nu)$ through the relation
(\ref{eq:mmutonu}), and obtain that
\begin{eqnarray}
\tilde{F}_\alpha & = &
-\frac{1}{\beta} \ln \biggl[
\sum\limits_{\underline{x}_\alpha} \prod\limits_{a \in \alpha} \psi_a
(\underline{x}_{\partial a}) \prod\limits_{i\in \alpha} \psi_i(x_i)
\prod\limits_{\{(\mu\rightarrow \nu) | \mu \in B_\alpha, \nu \in I_\alpha\}}
m_{\mu\rightarrow \nu}(\underline{x}_\nu) \biggr] \nonumber \\
& &
-\sum\limits_{\{(\mu\rightarrow \nu) | \mu \in B_\alpha, \nu \in I_\alpha\}}
\frac{1}{\beta} \ln \biggl[
\sum\limits_{\underline{x}_\nu}
 \prod\limits_{b\in \nu}
\Bigl[ \psi_b(\underline{x}_{\partial b})
 \Bigr]^{-\sum\limits_{\eta\in R_b^{\mu\rightarrow \nu}}
c_\eta}
\prod\limits_{j\in \nu}
\Bigl[\psi_j(x_j) \Bigr]^{-\sum\limits_{\eta \in R_j^{\mu\rightarrow \nu}}
 c_\eta} p_{\mu\rightarrow \nu}(\underline{x}_\nu)
\biggr].
\end{eqnarray}
We need to prove that
\begin{equation}
\sum\limits_{\alpha \in R} c_\alpha
\sum\limits_{\{(\mu\rightarrow \nu) | \mu \in B_\alpha, \nu \in I_\alpha\}}
\ln \biggl[
\sum\limits_{\underline{x}_\nu}
 \prod\limits_{b\in \nu}
\Bigl[ \psi_b(\underline{x}_{\partial b})
 \Bigr]^{-\sum\limits_{\eta \in R_b^{\mu\rightarrow \nu}}
c_\eta}
\prod\limits_{j\in \nu}
\Bigl[\psi_j(x_j) \Bigr]^{-\sum\limits_{\eta \in R_j^{\mu\rightarrow \nu}}
 c_\eta} p_{\mu\rightarrow \nu}(\underline{x}_\nu)
\biggr]
 = 0 .
\label{eq:talpha3}
\end{equation}
To prove this, it is first noticed that the left side of (\ref{eq:talpha3})
is equivalent to
\begin{equation}
\sum\limits_{\{(\mu\rightarrow \nu)\}}
\biggl\{ \sum\limits_{\{ \alpha | \mu \in B_\alpha,
\nu \in I_\alpha\}} c_\alpha \biggr\} \ln \biggl[
\sum\limits_{\underline{x}_\nu}
 \prod\limits_{b\in \nu}
\Bigl[ \psi_b(\underline{x}_{\partial b})
 \Bigr]^{-\sum\limits_{\eta \in R_b^{\mu\rightarrow \nu}}
c_\eta}
\prod\limits_{j\in \nu}
\Bigl[\psi_j(x_j) \Bigr]^{-\sum\limits_{\eta \in R_j^{\mu\rightarrow \nu}}
 c_\eta} p_{\mu\rightarrow \nu}(\underline{x}_\nu)
\biggr] .
\end{equation}
For a non-redundant region graph $R$, the sum in the curly brackets of
the above expression is identical to zero, i.e., for each directed
edge $\mu\rightarrow \nu$:
\begin{equation}
\sum\limits_{\{ \alpha | \mu \in B_\alpha,
\nu \in I_\alpha\}} c_\alpha = \sum\limits_{\alpha \geq \nu} c_\alpha
-\sum\limits_{\eta \geq \mu} c_\eta = 1- 1 =  0 .
\end{equation}

Combining equations (\ref{eq:f0cluster}), (\ref{eq:talpha2}) and
(\ref{eq:talpha3}), we obtain the objective equation:
\begin{equation}
F_0 = \sum\limits_{\alpha \in R}  c_\alpha
\left\{
 -\frac{1}{\beta} \ln \biggl[
\sum\limits_{\underline{x}_\alpha} \prod\limits_{a \in \alpha} \psi_a
(\underline{x}_{\partial a}) \prod\limits_{i\in \alpha} \psi_i(x_i)
\prod\limits_{\{(\mu\rightarrow \nu) | \mu \in B_\alpha, \nu \in I_\alpha\}}
m_{\mu\rightarrow \nu}(\underline{x}_\nu) \biggr]
\right\} .
\end{equation}

\section{Self-consistent equations at $n=2$}
\label{app:rg}

For the local structure shown in figure \ref{fig:rgbp2D},
the square-to-rod messages between
the regions $\alpha$ and $a$
are:
\begin{subequations}
\begin{eqnarray}
h_{\alpha \rightarrow a}^{(i)} & = &
 h_{d\rightarrow \alpha}^{(i)} +
\frac{1}{2 \beta}\ln \biggl[ \frac{\cosh[\beta (\tilde{h}_{l}+\tilde{J}_{d})]}
{\cosh[\beta (\tilde{h}_l-\tilde{J}_d)]}\biggr]
+\frac{1}{4 \beta }\ln\biggl[
\frac{1+\tanh [\beta \tilde{J}_c]\tanh[\beta (\tilde{h}_l+ \tilde{J}_d)]
\tanh [\beta (\tilde{h}_k+\tilde{J}_b)]}
{1+\tanh [\beta \tilde{J}_c]\tanh[\beta (\tilde{h}_l-\tilde{J}_d)]
\tanh[\beta (\tilde{h}_k
-\tilde{J}_b)]} \biggr] \nonumber \\
 & & \quad\quad\quad\quad\quad
  + \frac{1}{4 \beta}\ln\biggl[
\frac{1+\tanh [\beta \tilde{J}_c]\tanh[\beta (\tilde{h}_l+\tilde{J}_d)]
 \tanh [\beta (\tilde{h}_k-\tilde{J}_b)]}
{1+\tanh [\beta \tilde{J}_c] \tanh[\beta (\tilde{h}_l-\tilde{J}_d)]
 \tanh[\beta (\tilde{h}_k +\tilde{J}_b)]} \biggr] ,    \\
h_{\alpha\rightarrow a}^{(j)} & =&  h_{b\rightarrow \alpha}^{(j)} +
\frac{1}{2 \beta}\ln \biggl[ \frac{\cosh[\beta (\tilde{h}_{k}+\tilde{J}_{b})]}
{\cosh[\beta (\tilde{h}_k-\tilde{J}_b)]}\biggr]
+\frac{1}{4 \beta }\ln\biggl[ \frac{1+\tanh [\beta \tilde{J}_c]
\tanh[\beta (\tilde{h}_k+ \tilde{J}_b)] \tanh [\beta
(\tilde{h}_l+\tilde{J}_d)]}{1+\tanh [\beta \tilde{J}_c]
\tanh[\beta (\tilde{h}_k-\tilde{J}_b)] \tanh[\beta (\tilde{h}_l
-\tilde{J}_d)]} \biggr] \nonumber \\
& & \quad\quad\quad\quad\quad
+ \frac{1}{4 \beta}\ln\biggl[ \frac{1+\tanh [\beta \tilde{J}_c]
\tanh[\beta (\tilde{h}_k+\tilde{J}_b)]  \tanh [\beta
(\tilde{h}_l-\tilde{J}_d)]}{1+\tanh [\beta \tilde{J}_c]
\tanh[\beta (\tilde{h}_k-\tilde{J}_b)] \tanh[\beta (\tilde{h}_l
+\tilde{J}_d)]} \biggr] ,
\\
J_{\alpha\rightarrow a}^{(i j)} &=&  J_{i j} +
\frac{1}{4 \beta}\ln\biggl[ \frac{1+\tanh [\beta \tilde{J}_c]
\tanh[\beta (\tilde{h}_l+\tilde{J}_d)]  \tanh [\beta
(\tilde{h}_k +  \tilde{J}_b)]}{1+\tanh [\beta \tilde{J}_c]
\tanh[\beta (\tilde{h}_l+\tilde{J}_d)] \tanh[\beta (\tilde{h}_k-
\tilde{J}_b)]} \biggr] \nonumber \\
& & \quad\quad\quad\quad\quad
+ \frac{1}{4 \beta}\ln\biggl[ \frac{1+\tanh [\beta \tilde{J}_c]
 \tanh[\beta (\tilde{h}_l-\tilde{J}_d)]  \tanh [\beta
(\tilde{h}_k-\tilde{J}_b)]}{1+\tanh [\beta \tilde{J}_c]
\tanh[\beta (\tilde{h}_l-\tilde{J}_d)] \tanh[\beta (\tilde{h}_k+
\tilde{J}_b)]} \biggr] ,
\label{eq:Jalpha_a}
\end{eqnarray}
\end{subequations}
where we have introduced several shorthand notations
$$
\tilde{J}_{b} = J_{j k} + J_{b\rightarrow \alpha}^{(j k)} , \quad
\tilde{J}_c    =  J_{k l} + J_{c\rightarrow \alpha}^{(k l)} , \quad
\tilde{J}_d    =  J_{l i} + J_{d\rightarrow \alpha}^{(l i)} , \quad
\tilde{h}_{k}  =  h_{b\rightarrow \alpha}^{(k)}
+ h_{c\rightarrow \alpha}^{(k)} , \quad
\tilde{h}_{l}  = h_{d\rightarrow \alpha}^{(l)} +
 h_{c\rightarrow \alpha}^{(l)} .
$$
Similarly, the stripe-to-rod messages between
the regions $\mu$ and $a$ are:
\begin{subequations}
\label{eq:storms}
\begin{eqnarray}
h_{\mu \rightarrow a}^{(i)} & = & \frac{1}{2 \beta}\ln \biggl[
\frac{\cosh[\beta (h_{f\rightarrow \mu}^{(m)} + J_{i m})]}
{\cosh[\beta (h_{f\rightarrow \mu}^{(m)} -J_{i m})]}\biggr] \nonumber \\
& & \quad\quad
+\frac{1}{4 \beta }\ln\biggl[ \frac{1+\tanh [\beta (J_{m n} +
J_{f\rightarrow \mu}^{(m n)})]
\tanh[\beta (h_{f\rightarrow \mu}^{(m)} + J_{i m})]
\tanh [\beta (h_{f\rightarrow \mu}^{(n)}+J_{j n})]}
{1+\tanh [\beta (J_{m n} + J_{f\rightarrow \mu}^{(m n)})]
\tanh[\beta (h_{f\rightarrow \mu}^{(m)} -J_{i m})]
\tanh[\beta (h_{f\rightarrow \mu}^{(n)}-J_{j n}) ]} \biggr] \nonumber \\
& & \quad\quad
+ \frac{1}{4 \beta}\ln\biggl[ \frac{1+\tanh [\beta (J_{m n}
+ J_{f\rightarrow \mu}^{(m n)})]
\tanh[\beta (h_{f\rightarrow \mu}^{(m)} + J_{i m})]
\tanh [\beta (h_{f\rightarrow \mu}^{(n)}-J_{j n})]}{1+
\tanh [\beta (J_{m n} + J_{f\rightarrow \mu}^{(m n)} )]
\tanh[\beta (h_{f\rightarrow \mu}^{(m)} -J_{i m})]
\tanh[\beta (h_{f\rightarrow \mu}^{(n)}+J_{j n})]} \biggr] , \\
h_{\mu \rightarrow a}^{(j)} & = &  \frac{1}{2 \beta}\ln \biggl[ \frac{\cosh[\beta
(h_{f\rightarrow \mu}^{(n)} + J_{j n})]}
{\cosh[\beta (h_{f\rightarrow \mu}^{(n)} -J_{j n})]}\biggr] \nonumber \\
& & \quad\quad
+\frac{1}{4 \beta }\ln\biggl[ \frac{1+\tanh [\beta (J_{m n} +
J_{f\rightarrow \mu}^{(m n)})]
\tanh[\beta (h_{f\rightarrow \mu}^{(n)} + J_{j n})]
\tanh [\beta (h_{f\rightarrow \mu}^{(m)}+J_{i m})]}
{1+\tanh [\beta (J_{m n} + J_{f\rightarrow \mu}^{(m n)})]
\tanh[\beta (h_{f\rightarrow \mu}^{(n)} -J_{j n})]
\tanh[\beta (h_{f\rightarrow \mu}^{(m)}-J_{i m}) ]} \biggr] \nonumber \\
& & \quad\quad
+ \frac{1}{4 \beta}\ln\biggl[ \frac{1+\tanh [\beta (J_{m n} +
J_{f\rightarrow \mu}^{(m n)})]
\tanh[\beta (h_{f\rightarrow \mu}^{(n)} + J_{j n})]
\tanh [\beta (h_{f\rightarrow \mu}^{(m)}-J_{i m})]}{1+
\tanh [\beta (J_{m n} + J_{f\rightarrow \mu}^{(m n)} )]
\tanh[\beta (h_{f\rightarrow \mu}^{(n)} -J_{j n})]
\tanh[\beta (h_{f\rightarrow \mu}^{(m)}+J_{i m})]} \biggr] ,
\\
J_{\mu \rightarrow a}^{(i j)} &=&  J_{i j} +
\frac{1}{4 \beta} \ln\biggl[ \frac{1+\tanh [\beta (J_{m n}
+ J_{f\rightarrow \mu}^{(m n)})]
\tanh[\beta (h_{f\rightarrow \mu}^{(m)}+J_{i m} )]
\tanh [\beta (h_{f\rightarrow \mu}^{(n)} + J_{j n})]}{1+
\tanh [\beta( J_{m n} + J_{f\rightarrow \mu}^{(m n)}) ]
\tanh[\beta (h_{f\rightarrow \mu}^{(m)}+J_{i m})]
\tanh[\beta (h_{f\rightarrow \mu}^{(n)} -J_{j n})]} \biggr]  \nonumber \\
& & \quad\quad
+ \frac{1}{4 \beta} \ln\biggl[ \frac{1+\tanh [\beta (J_{m n} +
 J_{f\rightarrow \mu}^{(m n)})]
\tanh[\beta (h_{f\rightarrow \mu}^{(m)}-J_{i m} )]
\tanh [\beta (h_{f\rightarrow \mu}^{(n)} - J_{j n})]}{1+
\tanh [\beta( J_{m n} + J_{f\rightarrow \mu}^{(m n)}) ]
\tanh[\beta (h_{f\rightarrow \mu}^{(m)}-J_{i m})]
\tanh[\beta (h_{f\rightarrow \mu}^{(n)} +J_{j n})]} \biggr] ,
\label{eq:Jmu_a}
\end{eqnarray}
\end{subequations}

On the same edges
$(\alpha, a)$ and $(\mu, a)$, the rod-to-square and
rod-to-stripe messages are much simpler and are expressed as
\begin{subequations}
\begin{eqnarray}
& & h_{a\rightarrow \mu}^{(i)}   =  h_{\alpha \rightarrow a}^{(i)} , \quad\quad
h_{a\rightarrow \mu}^{(j)}   =  h_{\alpha \rightarrow a}^{(j)} , \quad\quad
J_{a\rightarrow \mu}^{(i j)} = J_{\alpha \rightarrow a}^{(i j)}- J_{i j}  ; \\
& & h_{a\rightarrow \alpha}^{(i)}   =  h_{\mu \rightarrow a}^{(i)} , \quad\quad
h_{a\rightarrow \alpha}^{(j)}   =  h_{\mu \rightarrow a}^{(j)} , \quad\quad
J_{a\rightarrow \alpha}^{(i j)} = J_{\mu \rightarrow a}^{(i j)}- J_{i j} .
\end{eqnarray}
\end{subequations}

\section{Stability analysis of the paramagnetic solution at $n=2$}
\label{app:stab}

At the paramagnetic fixed point
(\ref{eq:rg2para}), the effective couplings such as
$J_{\alpha\rightarrow a}^{(i j)}$ and $J_{\mu\rightarrow a}^{(i j)}$
are determined self-consistently
through (\ref{eq:Jalpha_a}) and (\ref{eq:Jmu_a}). Then the rgBP iteration
equations for the fields are linearized. The coefficients of the
linearized equations are obtained by the following expressions:

\begin{subequations}
\begin{eqnarray}
\frac{\partial h_{\mu\rightarrow f}^{(m)}}
{\partial h_{\alpha\rightarrow a}^{(i)}}
&=& \frac{\tanh[\beta J_{i m}] \bigl( 1- \tanh^2
[\beta J_{\alpha\rightarrow a}^{(i j)}] \tanh^2 [ \beta J_{j n}] \bigr)}
{1-\tanh^2 [\beta J_{\alpha\rightarrow a}^{(i j)}]
\tanh^2 [\beta J_{ i m}] \tanh^2 [\beta J_{j n}]} ,
\\
\frac{\partial h_{\mu\rightarrow f}^{(m)}}
{\partial h_{\alpha\rightarrow a}^{(j)}}
&=& \frac{\tanh[\beta J_{i m}] \tanh[\beta J_{\alpha\rightarrow a}^{(i j)}]
 \bigl( 1- \tanh^2 [ \beta J_{j n}] \bigr)}
{1-\tanh^2 [\beta J_{\alpha\rightarrow a}^{(i j)}]
\tanh^2 [\beta J_{ i m}] \tanh^2 [\beta J_{j n}]} ;
\\
\frac{\partial h_{\alpha\rightarrow d}^{(i)}}
{\partial h_{\mu\rightarrow a}^{(i)}} &=& 1 ,
\\
\frac{\partial h_{\alpha\rightarrow d}^{(i)}}
{\partial h_{\mu\rightarrow a}^{(j)}}
&=& \frac{\tanh[\beta \tilde{J}_a] \bigl( 1-
 \tanh^2 [\beta \tilde{J}_b] \tanh^2 [ \beta \tilde{J}_c] \bigr)}
{1-\tanh^2 [\beta \tilde{J}_{b} ]
\tanh^2 [\beta \tilde{J}_{a}] \tanh^2 [\beta \tilde{J}_{c}]} ,
\\
\frac{\partial h_{\alpha\rightarrow d}^{(l)}}
{\partial h_{\mu\rightarrow a}^{(i)}}
&=& 0,
\\
\frac{\partial h_{\alpha\rightarrow d}^{(l)}}
{\partial h_{\mu\rightarrow a}^{(j)}}
&=& \frac{\tanh[\beta \tilde{J}_c] \tanh [\beta \tilde{J}_b]
 \bigl( 1- \tanh^2 [\beta \tilde{J}_a] \bigr)}
{1-\tanh^2 [\beta \tilde{J}_{b} ]\tanh^2 [\beta \tilde{J}_{a}]
 \tanh^2 [\beta \tilde{J}_{c}]},
\\
\frac{\partial h_{\alpha\rightarrow c}^{(l)}}
{\partial h_{\mu\rightarrow a}^{(i)}}
&=& \frac{\tanh[\beta \tilde{J}_a] \bigl( 1-
\tanh^2 [\beta \tilde{J}_b] \tanh^2 [ \beta \tilde{J}_c] \bigr)}
{1-\tanh^2 [\beta \tilde{J}_{b} ]
\tanh^2 [\beta \tilde{J}_{a}] \tanh^2 [\beta \tilde{J}_{c}]} ,
\\
\frac{\partial h_{\alpha\rightarrow d}^{(l)}}
{\partial h_{\mu\rightarrow a}^{(j)}}
&=& \frac{\tanh[\beta \tilde{J}_a] \bigl( 1-
\tanh^2 [\beta \tilde{J}_b] \tanh^2 [ \beta \tilde{J}_c] \bigr)}
{1-\tanh^2 [\beta \tilde{J}_{b} ]
\tanh^2 [\beta \tilde{J}_{a}] \tanh^2 [\beta \tilde{J}_{c}]} .
\end{eqnarray}
\end{subequations}

The linearized rgBP equations for the fields are iterated  on a
given region graph. During each sweep of the iteration, the output
field messages of each square region of the region graph are
updated once, and the maximum among the absolute values of all
the updated fields is recorded. If this maximum decays to zero
with the iteration sweeps, the paramagnetic fixed point is then
declared as stable. When
the paramagnetic solution is unstable, this maximum value will eventually
increase with iteration sweeps (after a transient decreasing stage).



\bibliographystyle{spmpsci}      

\end{document}